\documentclass[aps,prd,preprint,preprintnumbers,groupedaddress,superscriptaddress,nofootinbib,showpacs,eqsecnum,draft]{revtex4-2}
\usepackage[T1]{fontenc}
\usepackage{graphicx}
\usepackage{amssymb}
\usepackage{amsmath}
\usepackage{bbm}
\usepackage{cancel}
\usepackage{epstopdf}
\usepackage{mathtools}
\usepackage{braket}
\usepackage[export]{adjustbox}
\usepackage[usenames,dvipsnames]{xcolor}
\usepackage{tikz}
\usepackage{tikz-feynman}
\usepackage{relsize}
\usepackage{titlesec}
\usepackage{setspace} 
\usepackage{xpatch}
\usepackage{xfrac}
\usepackage[final]{hyperref}
\hypersetup{
    colorlinks=true,
    linkcolor=blue,
    urlcolor=blue,
    linktoc=all
}

\usepackage{eso-pic}
\newcommand{\reportnum}[2]{
  \AddToShipoutPictureBG*{%
    \AtPageUpperLeft{%
      \hspace{0.75\paperwidth}%
      \raisebox{#1\baselineskip}{%
        \makebox[0pt][l]{\textnormal{#2}}
  }}}%
}

\renewcommand{\[}{\begin{equation}\begin{aligned}}
\renewcommand{\]}{\end{aligned}\end{equation}}

\renewcommand{\v}[1]{\boldsymbol{{#1}}}

\renewcommand{\Re}{\operatorname{Re}}
\usepackage[obeyDraft]{todonotes}
\makeatletter \def\@captype{figure} \makeatother 
\usetikzlibrary{calc}

\makeatletter

\pretocmd{\@affil@script}{\vspace*{2mm}}{}{}{}
\makeatother

\let\emptyset\varnothing

\tikzset{invclip/.style={clip,insert path={{[reset cm]
      (-16383.99999pt,-16383.99999pt) rectangle (16383.99999pt,16383.99999pt)
    }}}}

\def\dd{\hat d}
\def\del{\hat \delta}
\def\delp{\hat \delta^{(+)}}

\def\polv{\varepsilon}
\def\polvh#1{\varepsilon^{(#1)}}
\def\polvhconj#1{\varepsilon^{(#1)*}}

\def\Acl{A_\textrm{cl}}

\def\vpp{{\vphantom{\prime}}} 

\newcommand{\df}{d \Phi}
\def\Del{\del_\Phi}
\def\wn{\bar}
\definecolor{allOrderBlue}{rgb}{0.4,0.5,1}
\definecolor{patternBlue}{rgb}{0,0,1}
\definecolor{photonRed}{rgb}{1,0.2,0.2}

\def\qb{\bar q}
\def\Ord{\mathcal{O}}
\def\Lexp{\biggl\langle\!\!\!\biggl\langle}
\def\Rexp{\biggr\rangle\!\!\!\biggr\rangle}
\def\LexpT{\bigl\langle\mskip-5mu\bigl\langle}
\def\RexpT{\bigr\rangle\mskip-5mu\bigr\rangle}
\def\backdexpatch latexntA{\hspace*{-6mm}}
\def\nsand#1.#2.#3{%
        \left\langle\smash{#1}{\vphantom1}\right|{#2}%
        \left|\smash{#3}{\vphantom1}\right]}
\def\nsandaa#1.#2.#3{%
	\left\langle\smash{#1}{\vphantom1}\right|{#2}%
	\left|\smash{#3}{\vphantom1}\right\rangle}
\def\nsandbb#1.#2.#3{%
	\left[\smash{#1}{\vphantom1}\right|{#2}%
	\left|\smash{#3}{\vphantom1}\right]}
\def\nsandba#1.#2.#3{%
	\left[\smash{#1}{\vphantom1}\right|{#2}%
	\left|\smash{#3}{\vphantom1}\right\rangle}

\def\keta#1{%
        \left|\smash{#1}{\vphantom1}\right\rangle}
\def\ketb#1{%
        \left|\smash{#1}{\vphantom1}\right]}
\def\spa#1.#2{\left\langle#1\,#2\right\rangle}
\def\spb#1.#2{\left[#1\,#2\right]}
\def\spash#1.#2{\spa{\smash{#1}}.{\smash{#2}}}
\def\spbsh#1.#2{\spb{\smash{#1}}.{\smash{#2}}}

\DeclareMathOperator\arcsinh{arcsinh}
\def\bhn{{\bf \hat n}}

\def\sect#1{Sect.~{\ref{#1}}}
\def\Sect#1{Sect.~{\ref{#1}}}

\def\app#1{App.~{\ref{#1}}}

\def\fig#1{Fig.~{\ref{#1}}}

\def\eqn#1{eq.~(\ref{#1})}
\def\eqns#1#2{eqs.~(\ref{#1}) and (\ref{#2})}

\def\<{\langle}
\def\>{\rangl\noaffiliatione}
\def\expfrac#1#2{#1/#2}

\newbox\charbox
\newbox\slabox
\def\s#1{{      
        \setbox\charbox=\hbox{$#1$}
        \setbox\slabox=\hbox{$/$}
        \dimen\charbox=\ht\slabox
        \advance\dimen\charbox by -\dp\slabox
        \advance\dimen\charbox by -\ht\charbox
        \advance\dimen\charbox by \dp\charbox
        \divide\dimen\charbox by 2
        \raise-\dimen\charbox\hbox to \wd\charbox{\hss/\hss}
        \llap{$#1$}
}}
\def\cut#1{{      
        \setbox\charbox=\hbox{$#1$}
        \setbox\slabox=\hbox{$|$}
        \dimen\charbox=\ht\slabox
        \advance\dimen\charbox by -\dp\slabox
        \advance\dimen\charbox by -\ht\charbox
        \advance\dimen\charbox by \dp\charbox
        \divide\dimen\charbox by 2
        \raise-\dimen\charbox\hbox to \wd\charbox{\hss$|$\hss}
        \llap{$#1$}
}}

\def\finalk{r}

\def\pol{\varepsilon}
\def\timecomponentlabel{t}
\def\xcomponentlabel{x}
\def\ycomponentlabel{y}
\def\zcomponentlabel{z}

\reportnum{-2.5}{SAGEX 20-29-E}

\newcommand{\RadKer}{\mathcal{R}}
\newcommand{\RadKerCl}{\mathcal{R}^{(0)}}

\def\Rad{R}

\def\kb{\bar k}

\def\xferb{\overline{w}}

\def\pol{\varepsilon}

\def\beamsymbol{{%
\mathchoice{\scalebox{0.8}{$\odot$}}%
{\scalebox{0.8}{$\odot$}}%
{\scalebox{0.6}{$\odot$}}%
{\scalebox{0.4}{$\odot$}}}}

\def\kbbeam{\kb_{\beamsymbol}}
\def\beampolv{\polv_{\beamsymbol}}

\def\totalBeamMomentum{K_{\beamsymbol}}
\def\alphaNorm{\alpha_\beamsymbol}
\def\alphabNorm{\alphab_\beamsymbol}
\def\ANorm{A_\beamsymbol}
\def\hellabel{\eta}
\def\reference{\zeta}
\def\tb{\tilde b}
\def\tv{\tilde v}

\def\creation#1{a^\dagger_{#1}}
\def\annihilation#1{a^{\vphantom{\dagger}}_{#1}}
\def\creationh#1{\creation{(#1)}}
\def\annihilationh#1{\annihilation{(#1)}}

\def\operator{\mathbb}

\begin{document}

\newcommand{\Ampl}{\mathcal{A}}
\newcommand{\AmplB}{\mathcal{\bar A}}
\def\Int{\textrm{Int}}
\def\LIGOV{LIGO/Virgo}

\hfuzz=15 pt
\begin{minipage}{\textwidth}
\title{Waveforms from Amplitudes}

\def\squeezev{\vspace*{-2.5mm}}

\author{Andrea Cristofoli}
\affiliation{\looseness=-1%
\linespread{1}\selectfont%
Niels Bohr Institute, Blegdamsvej 17, 2100~K{\o}benhavn \O
\\ {\sf a.cristofoli@nbi.ku.dk}}

\author{Riccardo Gonzo}
\affiliation{\looseness=-1%
\linespread{1}\selectfont%
School of Mathematics, Trinity College, Dublin 2, Ireland
\\ {\sf gonzo@maths.tcd.ie}
}

\author{David~A.~Kosower}
\affiliation{\looseness=-1%
\linespread{1}\selectfont%
Institut de Physique Th\'eorique, CEA, CNRS, Universit\'e Paris--Saclay,
  F--91191 Gif-sur-Yvette cedex, France
  \\ {\sf David.Kosower@ipht.fr}
}

\author{Donal O'Connell}

\affiliation{\looseness=-1%
\linespread{1}\selectfont%
Higgs Centre for Theoretical Physics, School of Physics and Astronomy, 
The University of Edinburgh, Edinburgh EH9~3JZ, Scotland, UK%
\\ {\sf donal@ed.ac.uk}
}
\date\today

\begin{abstract}
We show how to compute classical
wave observables using
quantum scattering amplitudes.
We discuss observables both with incoming and with outgoing
waves.
The required classical limits are naturally described
by coherent states of massless bosons.  
We recompute the classic gravitational deflection of
light, and also show how to rederive Thomson
scattering.  We introduce a new class of local
observables, which includes the asymptotic electromagnetic
and gravitational Newman--Penrose scalars.  
As an example,
we compute a simple radiated waveform: the expectation of the
electromagnetic field in charged-particle
scattering.  At leading order, the waveform is
trivially related to the five-point scattering amplitude.
\end{abstract}

\pacs{\hspace{1cm}}

\maketitle

\end{minipage}

\newpage
\thispagestyle{empty}
{\baselineskip=19pt
\tableofcontents}
\vfill\eject

\section{Introduction}
\label{IntroductionSection}

Theoretical waveforms play an important role in the \LIGOV{} 
Collaboration's observational
program of gravitational-wave events from binary 
mergers~\cite{LIGO1,LIGO2}.
These waveforms provide templates that enable the detection of 
events against 
otherwise overwhelming noise backgrounds.
They also allow observers to extract the masses and spins of the 
binaries' constituents~\cite{LIGO3}. 
To date, theorists have computed waveforms (or equivalently, 
spectral functions for decaying binaries) 
using long-established effective-one-body (EOB) methods~\cite{EOB} and numerical-relativity approaches~\cite{NumericalRelativity},  in addition to
methods based on the `traditional' Arnowitt-Deser-Misner Hamiltonian formalism \cite{ADM},
direct post-Newtonian solutions in harmonic gauge \cite{Blanchet}, and computations in the effective-field theory approach pioneered by Goldberger and Rothstein \cite{EFT,EFT2}.

The start of the gravitational-wave observational era has spurred 
theorists to explore new approaches to computing classical 
observables for the two-body problem in gravity, 
in particular using quantum scattering amplitudes.  The connection 
between the quantum
$S$-matrix and observables in classical General Relativity (GR)
was first explored nearly fifty years ago by
Iwasaki~\cite{Iwasaki:1971vb}.  More recently, renewed
interest has been driven by modern on-shell techniques
for computing amplitudes and the double-copy
relation between Yang--Mills and gravitational 
amplitudes~\cite{Doublecopy1,Doublecopy2,Doublecopy3,Doublecopy4,%
Donal1,Donal2,Donal3,Donal4,
Donal5,Donal6,Donal7,Donal8,Doublecopy5,%
Doublecopy6,Doublecopy7,Doublecopy8,Doublecopy9,Doublecopy10,%
Doublecopy11,Doublecopy12,Doublecopy13,Doublecopy14,Doublecopy15,%
Doublecopy16,Burger:2021wss,Moynihan:2020ejh,Moynihan:2020gxj,%
Emond:2020lwi,Splitsignature}, as well
as the bounty of observations.  
Earlier investigations included applications to the two-body 
potential~\cite{EarlyTwoBody} and study of quantum corrections
to gravity~\cite{QuantumGravityCorrections}.

An important step was taken by
Cheung, Solon, and Rothstein~\cite{CRS}, who showed how to match 
effective field theories (EFTs) to scattering amplitudes
above threshold in order to extract a classical potential.  The classical potential can then
be used in the EOB or other frameworks to make predictions for bound-state quantities.
Bern, Cheung, Roiban, Shen, Solon, and Zeng used~\cite{Bern3G} this approach to 
compute the third-order corrections ($G^3$) to the conservative potential.  This milestone computation
went beyond what had been known from direct classical GR calculations, and provided 
the first concrete fulfillment of the promise of the scattering-amplitudes class of methods.
It used a two-loop scattering amplitude for massive particles, and was followed by many
new calculations using amplitude 
methods~\cite{Bern:2020buy,Bern:2020uwk,Bern:2021dqo,%
HPMRZ1,HPMRZ2,Cristofoli:2019neg,Bjerrum-Bohr:2019kec,%
Bjerrum-Bohr:2021vuf,Bjerrum-Bohr:2021din,%
Kosmopoulos:2021zoq,Huber:2020xny,Brandhuber:2019qpg,%
Chung:2020rrz,Guevara:2019fsj,Maybee:2019jus,Cheung:2020gyp,%
Arkani-Hamed:2019ymq,Bautista:2019tdr,Burger:2019wkq,Moynihan:2019bor,Emond:2019crr,Grignani:2020ahv}.
New EFT-based results have also 
emerged~\cite{Foffa:2019yfl,Kalin:2020mvi,Kalin:2020fhe,%
Kalin:2020lmz,Liu:2021zxr,Cho:2021mqw,Dlapa:2021npj,%
Foffa:2021pkg,Mougiakakos:2021ckm,Aoude:2020ygw,Foffa:2020nqe,Haddad:2020que,Goldberger:2020wbx,Levi:2020uwu,Levi:2020kvb,Aoude:2020onz,Levi:2019kgk,Kalin:2019inp,Blumlein:2021txj,Blumlein:2020pyo,Blumlein:2020znm,Blumlein:2019bqq,Kalin:2019rwq,Kalin:2019rwq,Damgaard:2019lfh,Foffa:2019eeb,Foffa:2019yfl,Blumlein:2019zku,Foffa:2019hrb,Loebbert:2020aos}.  In this context,  K\"alin and
Porto have pointed out 
an interesting analytic continuation from scattering
to bound-state observables~\cite{Kalin:2019rwq,Kalin:2019inp}.  
Several groups have pursued an eikonal 
approach~\cite{Heissenberg:2021tzo,DiVecchia:2021bdo,%
KoemansCollado:2020sxs,DiVecchia:2020ymx,KoemansCollado:2019ggb,%
DiVecchia:2021ndb,AccettulliHuber:2020oou}, and connections to 
it~\cite{Parra-Martinez:2020dzs}. 
Another approach which has seen recent
attention is the world-line formalism~\cite{EFT,Jakobsen:2021smu,Mogull:2020sak}. In the 
context of EFT, this world-line
approach is particularly important since
it makes immediate sense classically.
Treated as an effective quantum field
theory, this means that it organizes
quantum corrections particularly simply.
Finally, two
of the present authors have examined light-ray 
operators~\cite{Gonzo:2020xza}
and shock waves~\cite{Cristofoli:2020hnk}.  
Researchers working within a traditional
GR framework have also continued to produce new
results~\cite{He:2021cix,Damour:2020tta,Antonelli:2020ybz,%
Henry:2020pzq,Bini:2020hmy,Bini:2020nsb,Bini:2020zqy,Bini:2020wpo,%
Kuntz:2020gan,Bini:2020flp,Damour:2019lcq,Siemonsen:2019dsu,%
Barack:2019agd,Bini:2019nra,Antonelli:2019ytb}. 

In a previous paper~\cite{KMOC},  two of the present authors and 
Maybee outlined an 
observables-based approach to computing classical quantities.
It starts with an observable
in the quantum theory, expressing it in terms of scattering amplitudes;
and then uses an efficient and controlled method for taking the 
classical limit.  In this approach, rather than trying to compute intermediate quantities such as the conservative potential, we write down a formal expression for an observable of interest --- for example, the total change in the momentum of one of two scattered particles, \textit{aka\/} its impulse --- in the quantum theory.
With an appropriate wavefunction for the initial state, we can express the chosen observable in terms of quantum scattering amplitudes.  We further restore powers of $\hbar$ via dimensional analysis.  At this stage, the $\hbar$ scaling is naively bad, as the observable may be seemingly divergent in the classical, $\hbar\rightarrow 0$ limit, and loop corrections appear to be increasingly divergent with increasing order.

The original paper~\cite{KMOC} focused on scattering two massive 
particles. Appropriate wavefunctions were necessary to localize each 
incoming particle. This localization will sharpen in the classical 
limit, when we are focusing on point particles.  The localization will in turn lead us to retain momenta for the scattering particles in the expression for the observable,  but to use \textit{wavenumbers\/} for exchanged, emitted, or virtual massless particles (photons or gravitons). The change of variables from momenta to wavenumbers for the latter reveals additional powers of $\hbar$ that then yield a finite classical limit at each perturbative order.
Herrmann, Parra-Martinez, Ruf, and Zeng~\cite{HPMRZ1,HPMRZ2}, and 
separately Bautista and Guevara~\cite{Bautista:2019tdr} 
have applied this approach in their calculations.

Ref.~\cite{KMOC} did not discuss massless bosonic particles, in 
particular in the initial state.  We remedy that in this article. Furthermore, ref.~\cite{KMOC} focused only on global
observables, which require surrounding
an event with a detector of $4\pi$ coverage. We remedy this as well with a discussion of local observables, such as
electromagnetic and gravitational waveforms. 
Newman--Penrose~\cite{Newman:1961qr} scalars provide a natural
language for these quantities. We will
introduce these two principal topics of our article in the remainder of this introduction.

Let us begin with the question of initial-state massless bosons.
In the classical limit, 
one describes \emph{massive} particles as superpositions of 
single-particle states.  They ultimately appear as point-like particles
or extended bodies.
In contrast, \emph{massless} bosons appear as waves or wave packets.
It is no longer possible to describe them as superpositions of
single-particle states.  Instead, we shall see that they
emerge most naturally from coherent states of the corresponding
quantum fields. 
Such states are inherently superpositions of multiparticle states. 

The significance of coherent states was emphasized by Glauber 
from 1963 on.  He proved that every 
quantum state of radiation --- that is, every density matrix --- 
can be described as a suitable superposition of coherent 
states~\cite{Glauber:1963fi,Glauber:1963tx}. In particular, in 
the classical limit one can describe these density matrices 
using the so-called Glauber--Sudarshan 
$P$-representation~\cite{Glauber:1963tx,Sudarshan:1963ts}.
In this representation, there is a classical probability 
distribution in the space of coherent states. The application of 
coherent states to the classical limit of quantum scattering 
amplitudes started soon afterwards in the work of Frantz, Kibble, 
and Brown~\cite{Frantz:1965,Kibble:1965zza,Brown:1969nb}, but a 
systematic analysis of the question was still 
lacking~\cite{Birula1977}. Most calculations were limited to the 
solvable model of the linear interactions of a current (or 
a stress tensor) with the associated 
field~\cite{coherent-textbook}: in this case the $S$-matrix is 
solvable to all orders in perturbation theory, and its structure 
is exactly equivalent to a coherent state. Yaffe later 
showed~\cite{Yaffe:1981vf} that coherent states are very 
convenient for understanding the emergence of the classical 
approximation from quantum physics quite generally. Concrete 
applications are nonetheless rare in the literature, especially 
outside the case of a single particle interacting with a fixed 
coherent background (see ref.~\cite{Ilderton:2017xbj} and 
references therein)\footnote{Ref.~\cite{Endlich:2016jgc} offers
a notable exception in the context of the superradiance problem.}.
Coherent states have a close connection to soft limits
and infrared divergences, which provide a natural arena for 
their emergence in the late-time dynamics
of QED and linearized gravity~\cite{Bloch:1937pw,Kulish:1970ut,%
Nordstrom:1970zz,Jakob:1990zi,Ware:2013zja,Addazi:2019mjh}. 

Let us turn next to the question of local observables. In ref.~\cite{KMOC}, the authors studied time-integrated 
observables, in the context of scalar electrodynamics, and 
validated the amplitude-based approach through comparisons with 
direct calculations in classical electromagnetism.  What is of 
more direct interest to observers, however, are 
time-\textit{dependent\/} observables such as radiation 
waveforms. These are examples of a 
class of observables which
are \emph{local} in the sense that
they describe a measurement
at one spacetime point (or in a small
region of spacetime). The time-integrated
observables of~\cite{KMOC} in principle
require an apparatus which completely
surrounds a scattering event, so that
(for example) the impulse of any incoming particle can be measured. We describe this class of observables as \emph{global} as a result.

In this article, we establish
a direct connection between local observables, such as waveforms,
and scattering amplitudes. 
We validate our approach with a 
calculation of a simple 
waveform, arising from the scattering of two charged particles in 
scalar QED.  
We will see that
waveforms are effectively amplitudes for detecting massless particles, or waves in the classical limit.  We show how to write appropriate quantum observables, and how to take their limits.  Finally, we provide
a direct connection between the 
celebrated Newman--Penrose formalism~\cite{Newman:1961qr} and scattering amplitudes. 

As our work has progressed, we have become aware of a parallel line of investigation by Bautista, Guevara, Kavanagh and Vines~\cite{BGKV}. Their work is broadly
complementary to ours, but touches on some
of the same themes: the connection between
the Compton amplitude and classical wave
scattering, for example, and the close
connection between the Newman-Penrose
scalars and helicity amplitudes.

Our article is organized as follows.  We begin in the next section with a review of the formalism of ref.~\cite{KMOC}.  
In \sect{MasslessClassicalLimitSection}, we review 
coherent states for the electromagnetic field, show how
they correspond to classical fields, and give a simple example
of a light beam built from them.  
In \sect{ScatteringLightSection}, we discuss global observables with
massless waves in the initial state, 
concentrating on the impulse in this context. As examples, we discuss Thomson
scattering and its relation to the Compton
amplitude, and
we examine the calculation of the gravitational deflection
of light within our formalism. We turn to
the second major topic of our article in \sect{ObserversSection}
with a discussion of the general form of local observables far from some event. \Sect{SpectralFunctionSection} follows with an
introduction to spectral aspects
of local observables, leading to
the Newman--Penrose projection formalism.
In \sect{ThomsonScatteringSection}, we pause the general development to 
give example of a local observable: 
the scattered radiation field  
in Thomson scattering.  In \sect{EmissionSection}, we present the 
general form of the emission waveform when
two massive particles scatter, and in 
\sect{ExplicitWaveformSection} we give
explicit results for electromagnetic emission in charged-particle
scattering to leading order.  We discuss the connection between the 
waveform and the total radiated momentum 
in \sect{RadiatedMomentumSection},
and end with concluding remarks in \sect{ConclusionsSection}.

\section{Review of Formalism}
\label{ReviewSection}

We use relativistic units, retaining $c=1$, even as we restore $\hbar$ explicitly.
This means that we must distinguish units of energy and length, which we denote
by $[M]$ and $[L]$ respectively.  In this article, we will use a different
normalization than the conventions of ref.~\cite{KMOC} (which are also the
conventions of Peskin and Schroeder~\cite{PeskinSchroeder}).
Here, we normalize the annihilation and creation operators such that,
\begin{equation}
[\annihilation{p}, \creation{p'}] = (2\pi)^3 2E_p \delta^{(3)}(\v{p} - \v{p}')\,.
\label{CreationAnnihilationNormalization}
\end{equation}
(Bold symbols denote spatial three-vectors.)
Accordingly, $n$-point scattering amplitudes continue to have dimension $[M]^{4-n}$.

We keep $[M]^{-1}$ as the dimension of single-particle states $\keta{p}$, 
\begin{equation}
|p \rangle \equiv \, \creation{p} |0\rangle\,,
\label{MassiveSingleParticleState}
\end{equation}
with the
vacuum state being dimensionless.  We define $n$-particle plane-wave states as simply
the tensor product of normalized single-particle states.  (The normalization of the
single-particle states is the same as in ref.~\cite{KMOC}.)  The state $|p\rangle$
represents a particle of momentum $p$ and positive energy, while $\langle p|=\langle0|\annihilation{p}$
is the conjugate state.

We find it convenient to define an $n$-fold Dirac $\delta$ distribution with normalization
absorbing $2\pi$s,
\begin{equation}
\del^{(n)}(p) \equiv (2\pi)^n \delta^{(n)}(p)\,.
\label{delDefinition}
\end{equation}

The scattering matrix $S$ and the transition matrix $T$ are both dimensionless.
Scattering amplitudes are matrix elements of the latter between plane-wave states,
\begin{equation}
\langle p'_1 \cdots p'_m | T | p_1 \cdots p_n \rangle = 
\Ampl(p_1 \cdots p_n \rightarrow p'_1 \cdots p'_m) 
\del^{(4)}(p_1 + \cdots p_n - p'_1 - \cdots - p'_m)\,.
\end{equation}
As our formalism encompasses both QED and gravity, as well as other theories with massless
force carriers, we denote the coupling by $g$.  In electrodynamics, it corresponds
to $e$, while in gravity to $\kappa=\sqrt{32\pi G}$.  It is not dimensionless once
we have restored the factors of $\hbar$; rather, it is $g/\sqrt{\hbar}$ that is
the dimensionless coupling.

We start by taking the momenta of all particles as the primary variables; 
but as explained in the introduction, for most massless momenta, 
wavenumbers are the variables of interest.  We introduce a notation for the
wavenumber $\wn p$ associated to the momentum $p$,
\begin{equation}
\wn p \equiv p / \hbar\,.
\label{WavenumberNotation}
\end{equation}

\def\lcomp{\ell_c}
\def\lpack{\ell_w}
\def\lscatt{\ell_s}

\def\phlwave{\ell_\lambda}
\def\phlperp{\ell_\perp}
\def\phlpara{\ell_\parallel}

\def\vpt{{\vphantom{2}}}
We use the notation of ref.~\cite{KMOC} for the on-shell phase-space measure,
\begin{equation}
\df(p_i^\vpt) \equiv \dd^4 p_i^\vpt \, \delp(p_i^2-m_i^2)\,.
\label{dfDefinition}
\end{equation}
We will leave the mass implicit, along with the designation
of the integration variable as the first summand when the argument is a sum.
The notation for the measure again absorbs factors of $2\pi$,
\begin{equation}
\dd^4 p \equiv \frac{d^4 p}{(2\pi)^4}\,,
\end{equation}
and as usual,
\begin{equation}
\delta^{(+)}(p^2-m^2) = \Theta(p^\timecomponentlabel)\,\delta(p^2-m^2)\,,
\end{equation}
so that,
\begin{equation}
\delp(p^2-m^2) = 2\pi\Theta(p^\timecomponentlabel)\,\delta(p^2-m^2)\,.
\end{equation}
($p^\timecomponentlabel$ is the energy component of the four-vector.)

Given our convention for normalizing single-particle states, their inner product is,
\begin{equation}
\langle p' | p \rangle = 2 E_p \del^{(3)} (\v{p}-\v{p}').
\label{MomentumStateNormalization}
\end{equation}
The expression on the right-hand side
is the appropriately normalized delta function for the on-shell 
measure, which is convenient to express in more compact notation,
\begin{equation}
\Del(p_1^\vpp-p_1') \equiv  
2 E_{p_1} \del^{(3)} (\v{p}_1^\vpp-\v{p}_1')\,.
\end{equation}
We should understand the argument on the left-hand side as a function of four-vectors.
In this notation, \eqn{MomentumStateNormalization} is simply,
\begin{equation}
\langle p' | p \rangle = \Del(p-p')\,.
\end{equation}
With this notation, we can also rewrite the normalization of creation and annihilation
operations~(\ref{CreationAnnihilationNormalization}) in a natural form,
\begin{equation}
[\annihilation{p\vphantom{p'}}, \creation{p'}] = \Del(p - p')\,.
\label{CompactCreationAnnihilationNormalization}
\end{equation}
We will also employ the notation $\annihilation{}(k)\equiv \annihilation{k}$ and
$\creation{}(k)\equiv \creation{k}$ to allow for additional indices.

Ref.~\cite{KMOC} exclusively considered the scattering of two massive point-like particles.
In this article we go beyond that discussion to consider initial 
states which may involve massless radiation. 
However, when appropriate
we will continue to use the notation of ref.~\cite{KMOC} for 
initial states involving only massive particles:
we take the initial momenta to be $p_1$ and $p_2$, initially 
separated by a transverse impact parameter $b$. The latter is
transverse in that $p_1\cdot b = 0 = p_2\cdot b$.  

In the quantum theory, the system
of massive particles is described by wave functions, which we build out of plane waves.
In the classical limit, these wave functions must localize the two point-like particles, and must separate them clearly.
We describe the incoming particles in the far past
by wave functions $\phi_i(p_i)$, which we take to have 
reasonably well-defined positions and
momenta. We will review the requirements on the wave packets,
discussed in detail in sect.~4 of ref.~\cite{KMOC}, below.

We express the initial state in terms of plane waves
$| p_1\, p_2 \rangle_\textrm{in}$,
\begin{equation}
\begin{aligned}
| \psi \rangle_\textrm{in} &= \int \! \dd^4 p_1 \dd^4 p_2 \, \delp(p_1^2 - m_1^2) 
\delp(p_2^2 - m_2^2) \, \phi_1(p_1) \phi_2(p_2) \, e^{i b \cdot p_1/\hbar} | p_1\, p_2 
\rangle_\textrm{in}
\\ &= 
\int \! \df(p_1) \df(p_2)\;
  \phi_1(p_1) \phi_2(p_2) \, e^{i b \cdot p_1/\hbar} | p_1\, p_2 \rangle_\textrm{in}\,.
\end{aligned}
\label{InitialState}
\end{equation}

We require each wave function $\phi_i$ to be normalized to unity,
\begin{equation}
\int \! \df(p_1)\; |\phi_1(p_1)|^2 = 1\,,
\label{WavefunctionNormalization}
\end{equation}
so that our incoming state is also normalized to unity,
\begin{equation}
\begin{aligned}
{}_{\textrm{in}}\langle \psi | \psi \rangle_{\textrm{in}} 
&= \int \! \df(p_1) \df(p_2) \df(p'_1) \df(p'_2) e^{i b\cdot(p_1-p_1')/\hbar}
\\&\hspace*{15mm}\times \phi_1(p_1) \phi_1^*(p_1') \, \phi_2(p_2) \phi_2^*(p_2') 
\, \Del(p_1 - p_1') \, \Del (p_2 - p_2') \\
&= \int \! \df(p_1)\df(p_2)\; |\phi_1(p_1)|^2 |\phi_2(p_2)|^2
\\ &= 1\,.
\end{aligned}
\end{equation}

Finally, we turn to a review of the classical limit.
As discussed in ref.~\cite{KMOC}, there are three scales we must consider in the context of massive particle scattering: the
Compton wavelengths of the particles, $\lcomp^{(i)} \equiv \hbar/m_i$;
the intrinsic spread of the two particles' wavepackets, given by $\lpack$;
and the scattering length, $\lscatt$.  Taking the classical
limit requires that we impose the `Goldilocks' conditions,
\begin{equation}
\lcomp^{(i)} \ll \lpack \ll \lscatt\,.
\label{Goldilocks}
\end{equation}
The calculation of the scattering reveals that
$\lscatt\sim\sqrt{-b^2}$.

In order to expand in the $\hbar\rightarrow 0$ limit and extract the leading, classical,
term for any observable, as mentioned above we must make the powers of $\hbar$ explicit.
These arise from two sources: powers ordinarily hidden inside electromagnetic or
gravitational couplings; and powers arising from keeping the wavenumbers of massless
particles fixed rather than their momenta.  This is true both for emitted and
virtual particles, when considering quantities such as the total emitted radiation.

\section{Classical Limit for Massless Particles}
\label{MasslessClassicalLimitSection}

We are now ready to address the first major topic of this article: how to include initial-state massless classical waves in the formalism of ref.~\cite{KMOC}.
A naive extension of the considerations of 
ref.~\cite{KMOC} to massless particles is clearly impossible.  
A particle's Compton wavelength diverges when its mass goes to 
zero, making it impossible to satisfy the required conditions~(\ref{Goldilocks}). 
It doesn't make sense to treat messengers (photons or gravitons) 
as point-like particles.
Indeed, Newton and Wigner~\cite{Newton:1949cq} and Wightman~\cite{Wightman:1962sk}
proved rigorously long ago that a strict localization of 
known massless particles in position space 
is impossible\footnote{The proof holds for vector bosons and gravitons.}.
A proper treatment instead relies on coherent states.  
We begin such a treatment in the following subsection
by discussing general aspects of coherent states, 
focusing on the electromagnetic case.  We then describe the 
kind of coherent states of interest to us.

\subsection{Coherent States of the Electromagnetic Field}

\def\Aexpand{%
\begin{equation}
\opA_\mu(x) = \frac1{\sqrt{\hbar}}\sum_\hellabel \int \df(k) \, 
\bigl[ \annihilationh{\hellabel}(k) \polvhconj{\hellabel}_\mu(k)\, e^{-i k\cdot x/\hbar} 
 + \creationh{\hellabel}(k) \polvh{\hellabel}_\mu(k)\, e^{+ik \cdot x/\hbar} \bigr] \,,
\end{equation}}

\def\opA{\operator{A}}
\def\opN{\operator{N}}
\def\opK{\operator{K}}
\def\Foutdn{\langle F^\textrm{out}_{\mu\nu}(x)\rangle}
\def\vmu{{\vphantom{\mu}}}
\def\Fopdn{\operator{F}_{\mu\nu}}
\def\Eop{\operator{E}}
\def\Top{\operator{T}}
We can write the electromagnetic field operator as,
\begin{equation}
\opA_\mu(x) = \frac1{\sqrt{\hbar}}\sum_\hellabel \int \df(k) \, 
\bigl[ \annihilationh{\hellabel}(k) \polvhconj{\hellabel}_\mu(k)\, e^{-i k\cdot x/\hbar} 
 + \creationh{\hellabel}(k) \polvh{\hellabel}_\mu(k)\, e^{+ik \cdot x/\hbar} \bigr] \,,
\label{eq:aField}
\end{equation}
where $\hellabel = \pm$ labels the helicity, and the 
polarization vectors satisfy,
\begin{equation}
\bigl[\polvh{\hellabel}_\mu(k) \bigr]^* = 
\polvh{-\hellabel}_\mu(k)\,.
\end{equation}
We follow the usual amplitudes convention of representing 
an outgoing positive-helicity
photon of momentum $k$ by $\polvh{+}_\mu(k)$, which 
also corresponds to an incoming
negative-helicity photon of the opposite momentum. To understand 
the helicity flip for an incoming
state, note that we can analytically continue an incoming 
momentum $k$ to an outgoing momentum $k' = -k$.
The energy component $k^{\prime\timecomponentlabel}$ of the 
outgoing momentum is now negative. Thus, in an all-outgoing 
convention,
positive-helicity photons of momentum $k$ with 
$k^\timecomponentlabel>0$ are represented by the polarization 
vector $\polvh{+}_\mu(k)$,
while positive-helicity photons of momentum $k$ with 
$k^\timecomponentlabel<0$ are represented by the polarization 
vector $\polvh{-}_\mu(k)$.

More generally, 
$\creationh{\hellabel}(k)$ creates a
single-messenger state of momentum $k$ and helicity $\hellabel$, while
$\annihilationh{\hellabel}(k)$ destroys such a state.  Equivalently,
the latter operator creates a conjugate state of momentum $k$ and helicity $\hellabel$.
 
The commutation relations are
\begin{equation}
\bigl[ \annihilationh{\hellabel}(k), \creationh{\hellabel'}(k') \bigr] = 
  \delta_{\hellabel, \hellabel'} \Del(k-k') \,.
\end{equation}
For example, a  single-particle positive-helicity state is
\begin{equation}
|k^+ \rangle \equiv \creationh{+}(k) |0\rangle = 
\bigl[\annihilationh{+}(k)\bigr]^\dagger |0 \rangle \,.
\end{equation}
The conjugate state is $\langle k^+|$.

Using the form of the electromagnetic field in \eqn{eq:aField}, the electromagnetic field strength operator is,
\[
\Fopdn(x) = -\frac{2i}{\hbar^{3/2}}\sum_\hellabel \int \df(k) \, 
\bigl[ \annihilationh{\hellabel}(k) \,k_{[\mu}^\vmu \polvhconj{\hellabel}_{\nu]}(k)\, e^{-i k\cdot x/\hbar} 
 - \creationh{\hellabel}(k) \,k_{[\mu}^\vmu \polvh{\hellabel}_{\nu]}(k)\, e^{+ik \cdot x/\hbar} \bigr] \,,
 \label{eqn:fieldstrengthdn}
\]
where as usual the subscripted brackets denote
antisymmetrization, 
\begin{equation}
A_{[\mu}B_{\nu]} = \scalebox{0.9}{$\displaystyle\frac{1}{2}$}
(A_\mu B_\nu-A_\nu B_\mu) \,.
\end{equation}

\def\coherent#1{\mathbb{C}_{#1}}
\def\normcoherent#1{\mathcal{N}_{#1}}
\def\coherentbra#1{\langle #1|}
\def\coherentket#1{| #1\rangle}
\def\coherentinner#1{\langle #1| #1\rangle}

Introduce the coherent-state operator,
\begin{equation}
\coherent{\alpha,(\eta)} \equiv 
\normcoherent{\alpha} \exp \biggl[ \int \df(k) \, \alpha(k) 
\creation{(\eta)}(k) \biggr]\,,
\end{equation}
where the normalization $\normcoherent{\alpha}$ will be given 
below.  
We can build coherent states of the electromagnetic field using 
this operator,
such as a positive-helicity one,
\begin{equation}
\coherentket{\alpha^{+}} =  \coherent{\alpha,(+)} |0\rangle \,.
\label{CoherentStateDefinition}
\end{equation}
More generally, we could consider coherent states containing both 
helicities. Since coherent-state operators for different 
helicities commute and every polarization vector can be 
decomposed in the 
helicity basis, there is no loss of generality in making a 
specific helicity choice
for the coherent states we consider. The coherent state operators 
are unitary,
\begin{equation}
(\coherent{\alpha,(\hellabel)})^{\dagger} = 
(\coherent{\alpha,(\hellabel)})^{-1} \,.
\end{equation}

The normalization factor $\normcoherent{\alpha}$ is determined by the condition
$\langle \alpha^{+}|\alpha^{+}\rangle = 1$, that is,
\begin{equation}
\normcoherent{\alpha} = \exp\left[ -\frac12 \int \df(k) \, |\alpha(k)|^2 \right] \,,
\end{equation}
as can be seen by using the Baker--Campbell--Hausdorff formula. 
 
At this stage,
the function $\alpha(k)$ is quite general, however in specific
examples, we may take it to be real. 
We will see below that it is subject
to certain restrictions in the classical limit.  We will also see that its functional form will determine 
the physical shape of the corresponding state, so we will call it the `waveshape' function.

The coherent-state creation operator acting on the vacuum can be rewritten
using the Baker-Campbell-Hausdorff identity as a displacement operator~\cite{Frantz:1965,Kibble:1965zza} yielding 
\begin{align}
\coherent{\alpha,(\eta)} |0\rangle =\exp\biggl[\int \df(k) \alpha(k)
  (\creationh{\eta}(k) - \annihilationh{\eta}(k)) \biggr] |0\rangle \,.
\label{CoherentStateDefinitionII}
\end{align}
\def\phdag{{\vphantom{\dagger}}}
Its action on creation and annihilation operators is given by,
\begin{equation}
\begin{aligned}
\coherent{\alpha,(\eta)}^{\dagger} \annihilationh{\rho}(k) 
\coherent{\alpha,(\eta)}^\phdag &= \annihilationh{\rho}(k) + 
\delta_{\eta \rho}^\phdag\, \alpha(k)\,,
\\ 
\coherent{\alpha,(\eta)}^{\dagger} \creationh{\rho}(k) 
\coherent{\alpha,(\eta)}^\phdag &= 
\creationh{\rho}(k) + \delta_{\eta \rho}^\phdag\,\alpha^*(k) \,.
\end{aligned}
\end{equation}

To interpret the state, let us compute 
$\langle \alpha^+ | \opA^\mu(x) | \alpha^+ \rangle$. 
It is useful to note,
\begin{equation}
\begin{aligned}
\annihilationh{+}(k) \coherentket{\alpha^+} &= \alpha(k) \coherentket{\alpha^+} \,,\\
\annihilationh{-}(k) \coherentket{\alpha^+} &= 0\,,\\
\coherentbra{\alpha^+} \creationh{+}(k) &= \coherentbra{\alpha^+} \alpha^*(k) \,,\\
\coherentbra{\alpha^+} \creationh{-}(k) &= 0 \,,
\end{aligned}
\label{CoherentStateProperties}
\end{equation}
which incidentally imply that the dimension of $\alpha(k)$ is the same as the dimension of the annihilation operator: inverse mass. It is then easy to see that,
\def\alphab{\bar\alpha}
\begin{equation}
\begin{aligned}
\coherentbra{\alpha^+} \opA_\mu(x) \coherentket{\alpha^+} &= 
\frac1{\sqrt{\hbar}} \int \df(k) \, 
\bigl[ \alpha(k) \polvhconj{+}_\mu(k) e^{-i k\cdot x/\hbar} 
 + \alpha^*(k) \polvh{+}_\mu(k) e^{+ik \cdot x/\hbar} \bigr]
 \\&= 
\int \df(\kb) \, 
\bigl[ \alphab(\kb) \polvhconj{+}_\mu(\kb) e^{-i \kb\cdot x} 
 + \alphab^*(\kb) \polvh{+}_\mu(\kb) e^{+i \kb \cdot x} \bigr] \\
&\equiv A_{\textrm{cl} \, \mu}(x)\,,
\end{aligned}
\label{Aexpectation}
\end{equation}
where we define 
\begin{equation}
\alphab(\kb) \equiv \hbar^{\expfrac32}\alpha(k) \,.
\end{equation}
Additional constraints on $\alphab$ will emerge below from the 
consideration of correlators in the classical limit.
Note that the polarization vector is invariant under the rescaling 
from a momentum to a wavevector:
$\polvh{\hellabel}(\kb) = \polvh{\hellabel}(k)$ is independent of $\hbar$.

\def\fourA{\widetilde{A}}
Now, the most general solution of the classical Maxwell equation in empty space is,
\begin{equation}
\sum_\hellabel \Acl^{(\hellabel) \, \mu}(x) = \sum_\hellabel \int \df(\kb)  
\bigl[ \fourA_\hellabel(\kb) \polvhconj{\hellabel}{}^\mu(\kb) e^{-i \kb \cdot x} 
+ \fourA_\hellabel^*(\kb) \polvh{\hellabel}{}^\mu(\kb) e^{+i \kb \cdot x} \bigr] \,,
\label{ClassicalField}
\end{equation}
in terms of Fourier coefficients $\fourA_\hellabel(\kb)$,
which we can identify as $\alphab(\kb)$. 
Evidently our state $|\alpha^+\rangle$ contributes only the terms of 
positive  helicity 
($\hellabel = +$); a more general coherent state involving creation 
operators of both helicities 
would generate this most general solution of the free 
Maxwell equations. 
In examples we will consider, 
the simpler state $|\alpha^+\rangle$ will suffice. 

To further illuminate the meaning of coherent states, we may consider 
scattering amplitudes in the presence of a 
non-trivial background field $\Acl(x)$. The scattering matrix in the 
presence of this background field
depends on it. We denote this dependence by $S(\Acl)$.
Using the properties of the coherent state operator it can be shown 
that,
\begin{equation}
\begin{aligned}
\coherent{\alpha,(\hellabel)}^{\dagger} S(A) 
\coherent{\alpha,(\hellabel)} = S(A + \Acl^{(\hellabel)}) \,.
\label{eqn:S_matrix_coherent}
\end{aligned}
\end{equation}
Coherent states thus allow us to capture the physics of a specific 
background field based on vacuum scattering amplitudes:
\begin{equation}
\coherent{\alpha,(\hellabel)}^{\dagger} S(0)
\coherent{\alpha,(\hellabel)} = S(\Acl^{(\hellabel)}) \,.
\end{equation}
The formulation of the perturbation theory in a fixed background is particularly convenient when the Feynman rules --- or the scattering amplitudes --- in the background are known exactly \cite{FixedBackground}.

\subsection{Classical Coherent States}

The coherence of a state does not suffice for it to behave classically. We must also
require factorization of expectation values,
\begin{equation}
\label{eq:classicalFactorization}
\coherentbra{\alpha^+} \opA^\mu(x) \opA^\nu(y) \coherentket{\alpha^+} \simeq 
\coherentbra{\alpha^+} \opA^\mu(x) \coherentket{\alpha^+} 
\coherentbra{\alpha^+} \opA^\nu(y) \coherentket{\alpha^+} \,.
\end{equation}
A straightforward calculation in a light-cone gauge defined by a light-like vector $q$ shows that,
\begin{equation}
\begin{aligned}
&\coherentbra{\alpha^+} \opA^\mu(x) \opA^\nu(y) \coherentket{\alpha^+} = 
\\ &\hskip 10mm
\coherentbra{\alpha^+} \opA^\mu(x) \coherentket{\alpha^+} 
\coherentbra{\alpha^+} \opA^\nu(y) \coherentket{\alpha^+}
 \hphantom{\langle} + \frac1\hbar \int \df(k) \biggl[ \eta^{\mu\nu} 
   - \frac{k^\mu q^\nu + k^\nu q^\mu}{k \cdot q + i \delta}\biggr] e^{-i k \cdot (x-y) /\hbar}  
\\ &\hskip 10mm =
\coherentbra{\alpha^+} \opA^\mu(x) \coherentket{\alpha^+} 
\coherentbra{\alpha^+} \opA^\nu(y) \coherentket{\alpha^+}
 \hphantom{\langle} + \hbar \int \df(\kb) \biggl[ \eta^{\mu\nu} 
   - \frac{\kb^\mu q^\nu + \kb^\nu q^\mu}{\kb \cdot q + i \delta}\biggr] e^{-i \kb \cdot (x-y)}  \,.
\end{aligned}
\end{equation}
Similarly for the field strengths, in a gauge independent way using~\eqn{eqn:fieldstrengthdn}, we obtain
\begin{equation}
\begin{aligned}
\coherentbra{\alpha^+} \mathbb{F}^{\mu\nu}(x) \mathbb{F}^{\rho\sigma}(y) \coherentket{\alpha^+} =\ 
&\coherentbra{\alpha^+} \mathbb{F}^{\mu\nu}(x) \coherentket{\alpha^+} 
\coherentbra{\alpha^+} \mathbb{F}^{\rho\sigma}(y) \coherentket{\alpha^+} \\
& + 4\hbar  
\partial^{[\mu} \eta^{\nu][\sigma}\partial^{\rho]}  
\, \int \df(\kb) \, e^{-i \kb \cdot (x-y)} \,.
\end{aligned}
\label{eq:towardsClassicalFactorization}
\end{equation}

For classical behavior, the second term on the right-hand side of \eqn{eq:towardsClassicalFactorization} must be negligible compared to the
first term.  Writing $F_\textrm{cl}^{\mu\nu}(x) \equiv \coherentbra{\alpha^+} \mathbb{F}^{\mu\nu}(x) \coherentket{\alpha^+}$, 
the right-hand side becomes,
\begin{equation}
F_\textrm{cl}^{\mu\nu}(x) F_\textrm{cl}^{\rho\sigma}(y)
+ \frac{\hbar}{\pi^2}
\partial^{[\mu} \eta^{\nu][\sigma}\partial^{\rho]} 
 \frac{1}{(\mathbf{x} - \mathbf{y})^2 - (x^0-y^0 - i \delta)^2} \,.
\end{equation}
The first term has a nontrivial limit as $\hbar\rightarrow 0$, whereas the second term goes to zero 
in the limit, consistent with our expectations. For $\hbar \neq 0$, it is not possible to satisfy 
the inequality in the full spacetime region due to the divergence on the light-cone 
$(x^0 - y^0)^2 = |\v{x} - \v{y}|^2$ of the massless photon propagator: causally
connected measurements cannot be disentangled. 
We expect these contributions to fade away in the
classical limit of a physical observable~\cite{Kibble:1965zza}.
The factorization condition, which is trivial in the classical
limit, has been dubbed the ``complete coherence condition'' in the literature\footnote{In the
quantum optics literature the normal-ordered correlator of the electric field at different 
spatial locations
can have various degrees of coherence \cite{coherence}.}, a term coined by 
Glauber~\cite{Glauber:1963tx}.

\def\Nphoton{N_\gamma}
As usual, we define the operator measuring the number of photons to be,
\[
\opN_\gamma = \sum_\hellabel \int \df(k) \, \creationh{\hellabel}(k) \annihilationh{\hellabel}(k) \,.
\]
A short computation shows that the expectation number $\Nphoton$  of photons in our coherent state is,
\begin{equation}
\begin{aligned}
\Nphoton &= \coherentbra{\alpha^+}\opN_\gamma\coherentket{\alpha^+}
\\ &= \int \df(k) |\alpha(k)|^2
\\ &= \frac1{\hbar}\int \df(\kb) |\alphab(\kb)|^2\,.
\end{aligned}
\label{eq:nphotons}
\end{equation}
The classical limit $\hbar\rightarrow 0$ thus corresponds to the limit of
a large number of photons, that is a limit of large occupation
number~\cite{Yaffe:1981vf}.  The desired factorization property~\eqn{eq:classicalFactorization} will thus hold when,
\begin{equation}
\Nphoton \gg 1 \,.
\end{equation}
We must choose the waveshape $\alpha$ such that the integral in
the last line of \eqn{eq:nphotons} is not parametrically small as $\hbar \rightarrow 0$. A simple way to do so is to chose
$\alphab$ independent of $\hbar$.

Similarly, the momentum carried by the coherent state is,
\begin{equation}
\begin{aligned}
\totalBeamMomentum^\mu &= \coherentbra{\alpha^+}\opK^\mu\coherentket{\alpha^+}
\\ &= \int \df(k) |\alpha(k)|^2\,k^\mu
\\ &= \int \df(\kb) |\alphab(\kb)|^2\,\kb^\mu\,.
\end{aligned}
\end{equation}
This quantity (``K beam'') is finite in the classical limit, as expected. 

We emphasize that this coherent-state construction and its connection to classical states generalizes to any
massless particle, including gravitons.
Finally, it is worth remarking on the important and familiar case of geometric optics. This is a purely classical approximation
to wave phenomena, valid in situations where the wavelength is negligible in comparison to other physical scales. An important
example, which we discuss below, is 
of the gravitational bending of light. 

\subsection{Localized Beams of Light}
\label{LocalizedBeams}

In this paper, one of our foci will be on phenomena associated with scattering light from a point-like object. For problems of this 
type to be well-defined, the incoming wave must be spatially separated from the incoming particle in the far past. Consequently,
we need to understand how to
describe a localized incoming beam of light. We can choose the beam to be moving in the $z$ direction, 
localized around the origin of the $x$--$y$ plane. To see how to do this, let's consider some 
examples. 

The simplest option for the waveshape is,
\begin{equation}
\alpha(k) = \alphaNorm \Del(k - \hbar \kbbeam)\,,
\label{eq:simplewave}
\end{equation}
where $\kbbeam$ (``k-bar beam'') is the overall wavevector of the wave,
and $\alphaNorm$ (``$\alpha$ beam'')
is a constant which scales like $\sqrt{\hbar}$.  
Defining $\alphabNorm = \hbar^{-1/2}\alphaNorm$, this choice implies that,
\begin{equation}
\alphab(\kb) = \alphabNorm \Del(\kb-\kbbeam)\,,
\end{equation}
and that the classical field takes the form, 
\begin{equation}
\Acl^\mu(x) = 2\Re \alphabNorm \,
\beampolv^{*\mu}(\kbbeam) e^{-i \kbbeam \cdot x} \,.
\end{equation}
It is worth pointing out here that the expectation value of the gauge potential between coherent states is always a real quantity: a physical field which can be measured.
We can choose
\begin{equation}
\begin{aligned}
\kbbeam^\mu &= (\omega, 0, 0, \omega) \\
\beampolv^\mu &= \frac{1}{\sqrt{2}} (0, 1, i, 0)\,, \\
\end{aligned}
\end{equation}
to provide an explicit example.
If we pick the normalization of $\alphab$ to be given by $\alphabNorm = \ANorm / \sqrt{2}$ with $\ANorm$ real,
the classical field for this example is,
\begin{equation}
\Acl^\mu(x)= \ANorm\, (0, \cos \omega(t-z), -\sin \omega (t-z), 0) \,,
\end{equation}
which is a plane wave of circular 
polarization\footnote{The wave $\langle \alpha^- | \opA^\mu | \alpha^- \rangle$ is 
circularly polarized in the opposite sense.} 
moving in the $z$-direction with angular frequency $\omega$. 
This wave is completely delocalized, which is a disadvantage for our purposes: we wish to have a clean
separation between the incoming wave and point-like particle states.

\def\deltasmF#1{\delta_{#1}}
\def\deltasm#1#2{\deltasmF{#1}(#2)}
\def\deltasmp#1#2#3{\deltasmF{#1}^{#3}(#2)}
\def\deltasmalt#1#2{\delta_{\textrm{sm}}(#1;#2)}

To localize the wave, we may ``broaden'' the delta function in \eqn{eq:simplewave}. 
Define,
\begin{equation}
\deltasm{\sigma}{\kb} \equiv 
\frac{1}{\sigma \sqrt\pi}    \exp \biggl[ - \frac{\kb^2}{\sigma^2}\biggr] \,,
\end{equation}
which is normalized so that
\begin{equation}
\int_{-\infty}^\infty d\kb \, \deltasm{\sigma}{\kb} = 1 \,.
\end{equation}
The peak width is controlled by the parameter $\sigma$.  As $\kb$ is a wavenumber,  
$\sigma$ has dimensions of inverse length. We may choose our incoming wave, moving along 
the $z$-axis, to be symmetric under a rotation about that axis. Consider the choice,
\begin{equation}
\begin{aligned}
\alpha(k) &= \frac1{\hbar^3} |\v k| (2\pi)^3 \ANorm \sqrt{2\hbar} \, 
\deltasm{\sigma_\parallel}{\omega - k^\zcomponentlabel/ \hbar} 
\deltasm{\sigma_\perp}{k^\xcomponentlabel / \hbar}
\deltasm{\sigma_\perp}{k^\ycomponentlabel/ \hbar}\,;
\end{aligned}
\label{eq:beamAlphaI}
\end{equation}
or equivalently,
\begin{equation}
\begin{aligned}
\alphab(\kb) &= {\sqrt2} |\v \kb| (2\pi)^3 \ANorm \, 
\deltasm{\sigma_\parallel}{\omega - \kb^\zcomponentlabel} 
\deltasm{\sigma_\perp}{\kb^\xcomponentlabel} 
\deltasm{\sigma_\perp}{\kb^\ycomponentlabel} \,,
\end{aligned}
\label{eq:beamAlphaII}
\end{equation}
with $\ANorm$ real.  (We use the superscripts $\timecomponentlabel$,
$\xcomponentlabel$, $\ycomponentlabel$, and $\zcomponentlabel$ 
to denote the corresponding components of $\kb$.)
We have introduced  two measures of beam spread, $\sigma_\parallel$
and $\sigma_\perp$, along and transverse to the wave 
direction respectively.
The corresponding classical field is,
\begin{equation}
\begin{aligned}
\Acl^\mu(x) &=  {\sqrt{2}} \ANorm  
\Re \int d^3 \kb\,\beampolv^{*\mu}(\kb)  
\, \deltasm{\sigma_\parallel}{\omega - \kb^\zcomponentlabel} 
\\&\hskip 35mm\times \deltasm{\sigma_\perp}{\kb^\xcomponentlabel}
\deltasm{\sigma_\perp}{\kb^\ycomponentlabel} e^{-i \kb \cdot x}\Big|_{\kb^\timecomponentlabel=\sqrt{(\kb^\xcomponentlabel)^2
           +(\kb^\ycomponentlabel)^2+(\kb^\zcomponentlabel)^2}}\,.
\end{aligned}
\end{equation}
We emphasize that other choices of wave shape are available in the classical theory: the only
constraint is that $N_\gamma$ must be large.

Let us further refine our example by taking $\sigma_\parallel$ 
to be very small compared to the other two scales, $\sigma_\perp$
and $\omega=\kbbeam^{\timecomponentlabel}$.  We are thus 
considering a monochromatic beam, for which we can replace 
$\deltasmF{\sigma_\parallel}$ by a Dirac delta distribution.  
Doing so, we obtain,
\begin{equation}
\begin{aligned}
\Acl^\mu(x) =    
\sqrt{2}{\ANorm}
\Re \int d^2 \kb_\perp \,\beampolv^{*\mu}(\kb)  \,
\delta_{\sigma_\perp}( \kb^\xcomponentlabel) 
\delta_{\sigma_\perp}( \kb^\ycomponentlabel) \,
e^{-i t\sqrt{\omega^2 + (\kb^\xcomponentlabel)^2 
   + (\kb^\ycomponentlabel)^2}} e^{i \omega z} 
e^{i \kb^\xcomponentlabel \, x}e^{i \kb^\ycomponentlabel \, y} \,.
\end{aligned}
\label{WavesFieldStepI}
\end{equation}
We can simplify this expression with the following considerations. For the beam to be moving in 
the $z$-direction, the photons in the beam should dominantly have their momenta, or equivalently
their wavenumbers, aligned in the 
$z$-direction. However, the broadened distribution $\delta_{\sigma_\perp}$ 
does allow small components 
of momentum in the $x$ and $y$ directions. These components should be subdominant. 
The corresponding $x$ and $y$ 
wavenumbers are of order $\sigma_\perp$ while the wavenumber in the $z$ direction is of 
order $\omega$.  Let us define the (reduced) wavelength $\lambda \equiv \omega^{-1}$.
We must thus require,
\begin{equation}
{\lambda^{-1}} \gg  \sigma_\perp \,.
\end{equation}
We can also define a transverse size of the beam, 
\begin{equation}
\phlperp = {\sigma_\perp^{-1}} \,,
\end{equation}
along with a `pulse length',
\begin{equation}
\phlpara = {\sigma_\parallel^{-1}} \,.
\end{equation}
We see that we must require,
\begin{equation}
\lambda \ll \phlperp \,.
\label{PhotonConditionI}
\end{equation}
In other words, a collimated monochromatic beam must have a transverse size which is large in 
units of the beam's wavelength.  The requirement~(\ref{PhotonConditionI}) is in some respects
analogous to the first part of the `Goldilocks' condition~(\ref{Goldilocks}). However, we
emphasize that \eqn{PhotonConditionI} arises from our desire to localize the wave in
the far past. In particular, waves violating the requirement~\eqref{PhotonConditionI} may
still be classical.

Turning back to \eqn{WavesFieldStepI}, we may now simplify the time-dependent 
exponential factor.
The broadened delta distribution $\delta_{\sigma_\perp}$ forces,
\begin{equation}
(\kb^\xcomponentlabel)^2 + (\kb^\ycomponentlabel)^2 \lesssim
\sigma_\perp^2 = \phlperp^{-2} \,,
\end{equation}
so that,
\begin{equation}
\sqrt{\omega^2 + (\kb^x)^2 + (\kb^y)^2} \lesssim 
 \sqrt{\omega^2 + \phlperp^{-2}} \simeq \omega + \Ord(\phlperp^{-2}\omega^{-2}) \simeq \omega\,.
\label{BroadenedDeltaFunctionExpansionLO}
\end{equation}
For the wave's field, we obtain, in this approximation,
\begin{equation}
\begin{aligned}
\Acl^\mu(x) &=  \sqrt{2}\ANorm \Re \biggl\{e^{-i \omega(t-z)} 
\int d^2 \kb_\perp \,\beampolv^{*\mu}(\kb)  
\,\deltasm{\sigma_\perp}{\kb^\xcomponentlabel} 
\deltasm{\sigma_\perp}{\kb^\ycomponentlabel} 
e^{i \kb^\xcomponentlabel\, x}
 e^{i \kb^\ycomponentlabel\, y}\biggr\} \\
&=  \sqrt{2}{\ANorm}  \Re \biggl\{
e^{-i \omega(t-z)}\,\beampolv^{*\mu}(\kbbeam^\vmu) 
\int d^2 \kb_\perp \, 
\deltasm{\sigma_\perp}{\kb^\xcomponentlabel}
\deltasm{\sigma_\perp}{\kb^\ycomponentlabel} 
e^{i \kb^\xcomponentlabel \, x}
e^{i \kb^\ycomponentlabel \, y}\biggr\} \,,
\end{aligned}
\label{WavesFieldStepII}
\end{equation}
where we can replace $\beampolv^\mu(\kb)$ 
by $\beampolv^\mu(\kbbeam^\vmu)$
because of the smallness of the transverse 
components of $\kb$. (Recall that 
$\kbbeam^\mu = (\omega,0,0,\omega)$.)
To continue, we may note that the integral,
\begin{equation}
\int_{-\infty}^\infty {d\qb}\, e^{i \qb x}\deltasm{\sigma}{\qb} =  
e^{-x^2 \sigma^2 /4} \,,
\label{eq:supportSuppression}
\end{equation}
so that we finally obtain,
\begin{equation}
\Acl^\mu(x) =  \sqrt{2} \ANorm  \Re 
\Bigl[ e^{-i \omega(t-z)}\beampolv^{*\mu}(\kbbeam^\vmu) \, 
e^{-(x^2 + y^2)/(4\phlperp^2)} \Bigr]\,.
\label{WavesFieldLO}
\end{equation}
This is indeed a wave of circular polarization along the $z$-axis, with finite size in the 
$x$--$y$ plane. 

Our approximation that $\sigma_\parallel$ is negligible gives us 
a beam of \textit{infinite\/}
spatial extent along the direction of propagation (here, the 
$z$ axis). 
Were we to stop short of the $\sigma_\parallel \rightarrow 0$ 
limit, we would find a finite size in this 
direction too.  The occupation number, which is divergent for 
infinite extent in the $z$-direction, would also become finite 
for nonvanishing $\sigma_\parallel$.

The classical field in \eqn{WavesFieldLO} describes a beam of 
light that does not
spread in the transverse direction, in apparent contradiction to the 
non-zero transverse momenta
the integral contains.  This seeming contradiction is lifted when we 
compute the field of \eqn{WavesFieldStepI} 
to the next order in $1/(\omega\phlperp)$ and $t/ \phlperp$, as described in 
\app{BeamSpreadAppendix}. The result for short enough times is,
\begin{equation}
\begin{aligned}
\Acl^\mu(x) &=  \sqrt{2} \ANorm \Re  \biggl\{
\frac{\exp[-i \omega(t-z)]}{1+i \frac{t}{2\omega\phlperp^2}}\beampolv^{*\mu}(\kbbeam^\vmu) \, 
\exp\left[-\frac{(x^2 + y^2)}
    {4\phlperp^2 [1+i t/(2\omega\phlperp^2)]}\right]\biggr\} \\
&\phantom{=}+ \frac{\ANorm }{\sqrt{2} }  \Re \biggl\{
\exp[-i \omega(t-z)]
\left[i \frac{x}{\phlperp^2} \partial_{\kb^\xcomponentlabel} 
\beampolv^{*\mu}(\kb) \Big|_{\kb=\kbbeam} 
+ i \frac{y}{\phlperp^2} \partial_{\kb^\ycomponentlabel} 
\beampolv^{*\mu}(\kb) \Big|_{\kb=\kbbeam}\right] 
\\&\hspace*{20mm}\times
  \exp\left[-\frac{(x^2 + y^2)}{4\phlperp^2}\right]\biggr\}
  + \cdots \,.
\end{aligned}
\label{WavesFieldNLO}
\end{equation}

\section{Global Observables with Incoming Radiation}
\label{ScatteringLightSection}
In the previous section, we examined the use of coherent states to 
describe waves built up of massless messengers (photons or 
gravitons), and understood that the classical
limit emerges in the limit of large occupation number.  In this section, we turn to dynamics:
we will consider the scattering of a messenger wave and a scalar point particle. Real-life examples 
are the classical scattering of a light beam off a charged point particle;
a light beam scattering gravitationally off a point particle; or
a gravitational wave scattering off a point particle. 

Our focus in this section will be on \emph{global} observables, obtained by surrounding the scattering event
with a distant sphere of detectors. These detectors can register the total change in momentum (or impulse) of the particle,
or of the wave, during scattering. 
These are the same kinds of observables considered in
ref.~\cite{KMOC}.
The main novelty in
this section will be the computation of global observables 
for scattering with incoming classical radiation, which
we will describe using the coherent states discussed in the 
previous section.
In the following sections we will discuss local observables. 

Two examples will allow us to explore different aspects of the 
dynamics: the electromagnetic impulse on a charge
in a spatially localized beam of light (Thomson scattering); and 
the General-Relativistic deflection of light in the 
geometric-optics limit. 
We begin by discussing the details of the requirements imposed by the dynamics in the classical limit,
and the nature of the initial state.

\subsection{Setup}

In the classical limit, the Compton wavelength $\lcomp$ of a 
point-like particle must be unobservably small.
However, there is (in general) no need for the wavelength of
massless waves to be small. On the contrary, finite-wavelength 
classical waves are quotidian phenomena, and
propagate along the pages of many classical-physics textbooks.

In the scattering of two point-like particles, this requirement on $\lcomp$ would be violated if
the particles approach at distances smaller than (or of order of) their Compton wavelength, because then the underlying
wave nature of the particles becomes important. Thus we arrive at 
the conclusion that classical scattering of two particles obtains only when the impact parameter $b \neq 0$. 

In contrast, for a wave of wavelength $\lambda$ interacting with a 
particle, we simply require that $\lambda$ be much larger than the 
Compton wavelength $\lcomp$ of the particle. When this is the 
case, the messengers comprising the wave cannot resolve
the quantum structure of the particle. For the classical 
point-particle approximation to be valid, we further
require that $\lambda$ should be large compared to the finite size $\lpack$ of the particle's wave
packet. We thus have the requirement,
\[
\lcomp \ll \lpack \ll \lambda \,,
\]
for classical interactions of a wave with a particle of Compton wavelength $\lcomp$. There is no 
\textit{a priori\/} constraint on the impact parameter $b$.

\graphicspath{ {images/} }
\begin{figure}[!htb]
    \centering
    \begin{minipage}{.5\textwidth}
        \centering
        \includegraphics[draft=false,width=0.5\textwidth]{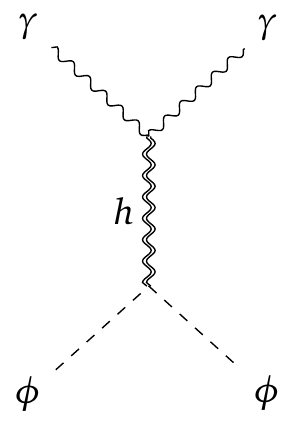}
    \end{minipage}%
    \begin{minipage}{0.5\textwidth}
        \centering
        \includegraphics[draft=false,width=0.7\textwidth]{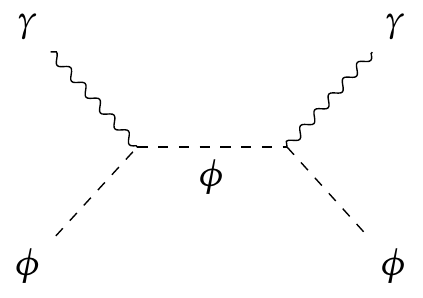}
    \end{minipage}
\caption{While the t-channel graviton exchange contribution exists for a photon interacting gravitationally with a scalar, this is not true in electromagnetic case}
\label{fig:MessengerScattering}
\end{figure}
As exemplified in \fig{fig:MessengerScattering}, in the
electromagnetic scattering of a photon off a charged particle,
there is no $t$-channel contribution. Correspondingly 
we
are primarily interested in the $b\simeq0$ case.
(More precisely, we are interested in $b$ smaller than the
transverse size of the beam.)
We will explore this in more detail below.
In contrast, in
the gravitational scattering of a photon off a neutral particle,
there are both $s$- and $t$-channel contributions.  In
this case, we are interested in general $b$. 

The interaction between our particle and our wave introduces 
another length scale to consider, 
namely the scattering length $\lscatt$. Let $q = \hbar \qb$ be a 
characteristic momentum exchange
associated with the interaction; then the scattering length is 
defined to be,
\begin{equation}
\lscatt = \frac{1}{\sqrt{|\qb^2|}} \,.
\end{equation}
The value of the scattering length depends on the details of the 
scattering process. In the case where
two point-like particles scatter, for instance, one finds that 
$\lscatt \sim b$. In the case at hand
where a particle interacts with a wave this need not be the case. 
Indeed for an $s$ channel processes
it is more natural to expect $\lscatt$ to be determined by the 
off-shellness of intermediate propagators
such as $s -m^2$. For definiteness let us take the momentum of the 
incoming particle to be $p_1 = m_1 u_1$ 
while the incoming wave has characteristic wavenumber
$\kbbeam$. Then $s - m_1^2 = 2 \hbar \kbbeam \cdot p_1$, so 
that the scattering length should be,
\begin{equation}
\lscatt \sim \frac{1}{\kbbeam \cdot u_1} \,.
\end{equation}
This is simply of the order of the wavelength of the incoming wave.

We turn next to the construction of the incoming state.  As in ref.~\cite{KMOC} and 
in \eqn{InitialState}, we write the point particle as a superposition of plane-wave
states weighted by a wavefunction $\phi(p)$.  Following the discussion in the
previous section, we write the messenger wave as a coherent state of helicity $\eta$
characterized by
the waveshape $\alpha(k)$.  We start with a basis of states constructed out of
coherent states~(\ref{CoherentStateDefinition}) of definite helicity
$|\alpha^{\eta}\rangle$ for the messenger and plane-wave states for the massive particle,
\begin{equation}
|p_1\,\alpha_2^{\eta}\rangle_{\textrm{in}} = |p_1\rangle |\alpha_2^{\eta}\rangle\,.
\end{equation}
Our initial state then takes the form,
\begin{equation}
\begin{aligned}
| \psi_w \rangle_\textrm{in} &= 
\int \! \df(p_1)\;
  \phi_1(p_1)  \, e^{i b \cdot p_1/\hbar} | p_1\, \alpha_2^{\eta} \rangle_\textrm{in}\,.
\end{aligned}
\label{WaveInitialState}
\end{equation}
The impact parameter $b$ now separates the particle from the center of the beam in the far past.
As in the earlier discussion, the state is normalized to unity, 
${}_\textrm{in}\langle\psi_w|\psi_w\rangle_\textrm{in} = 1$.
(We will leave the `in' subscript implicit going forward.)

Information about the classical four-velocity of the point particle is hidden inside
$\phi(p)$.  The explicit example studied in ref.~\cite{KMOC} made use of a linear exponential
(which slightly counter-intuitively reduces to a Gaussian in the nonrelativistic limit).
In the same way, the information about the overall momentum
$\totalBeamMomentum$ 
of the messenger wave is
hidden inside $\alpha(k)$.  

In the following, we will make use of the coherent wave shape $\alpha(k)$ chosen in \eqn{eq:beamAlphaI} 
corresponding to the choice of $\bar\alpha(k)$ of \eqn{eq:beamAlphaII}, independent of $\hbar$ as desired.
We will elucidate inequalities between the various parameters defining the beam below, where relevant.

\subsection{General Expression for the Impulse}

Before we discuss the details of specific examples, let us investigate the general structure of the impulse, $\langle\Delta p_1\rangle$, on a massive particle during a scattering event with a classical wave.
We can carry over the
expression from ref.~\cite{KMOC},
\def\WI#1{I^\mu_{w(#1)}}
\def\WIo#1#2{I^{\mu,#2}_{w(#1)}}
\def\MI#1{I^\mu_{(#1)}}
\begin{equation}
\begin{aligned}
\langle \Delta p_1^\mu \rangle 
&= \langle \psi_w | \, i [ \operator P_1^\mu, T ] \, | \psi_w \rangle + \langle \psi_w | \, T^\dagger [ \operator P_1^\mu, T] \, |\psi_w \rangle
\\ &= \WI1 + \WI2 \,.
\end{aligned}
\label{ImpulseMaster}
\end{equation}
Compared to ref.~\cite{KMOC}, only the initial state is different.

Before studying the expansion of this expression,
we remark that there is an equivalent formulation in 
terms of the background field,
\begin{equation}
\begin{aligned}
\langle \Delta p_1^\mu \rangle &= \int \df(p_1) \df(p^{\prime}_1) \, \phi_1(p_1) \phi^{*}_1(p_1^{\prime}) e^{-ib\cdot(p_1'-p_1)/\hbar} \bra{p_1^{\prime}} \coherent{\alpha,(\eta)}^{\dagger} i [\mathbb{P}_1^{\mu}, T] \coherent{\alpha,(\eta)}^\phdag \ket{p_1} \\ 
&\hphantom{=}\, +  \int \df(p_1) \df(p^{\prime}_1) \, \phi_1(p_1) \phi^{*}_1(p_1^{\prime}) e^{-ib\cdot(p_1'-p_1)/\hbar} \bra{p^{\prime}_1} \coherent{\alpha,(\eta)}^{\dagger}  T^{\dagger} [\mathbb{P}_1^{\mu}, T ] \coherent{\alpha,(\eta)}^\phdag
\ket{p_1} \\ 
&=\int \df(p_1) \df(p^{\prime}_1) \, \phi_1(p_1) \phi^{*}_1(p_1^{\prime}) e^{-ib\cdot(p_1'-p_1)/\hbar} \bra{p_1^{\prime}} i [\mathbb{P}_1^{\mu}, T(\Acl^{(\eta)})] \ket{p_1}\\
&\hphantom{=}\, +  \int \df(p_1) \df(p^{\prime}_1) \, 
\phi_1(p_1) \phi^{*}_1 (p_1^{\prime})  e^{-ib\cdot(p_1'-p_1)/\hbar}  \bra{p^{\prime}_1} T^{\dagger}(\Acl^{(\eta)}) [\mathbb{P}_1^{\mu}, T(\Acl^{(\eta)}) ] \ket{p_1} \,,
\end{aligned}
\end{equation}
where the scattering matrix computed from the background 
$\Acl^{(\eta)}$ is denoted by $T(\Acl^{(\eta)})$, and we have used 
the relation $\coherent{\alpha,(\eta)}^{\dagger} \coherent{\alpha,(\eta)}^\phdag = \mathbbm{1}$. While we will focus on the formulation~(\ref{ImpulseMaster}), it is intriguing to notice the linear term of the impulse $\WI1$ is closely related to the two-point function of the massive scalar field in the coherent state background. 
As a consequence, we should expect a resummation of all higher-order results.

Returning to \eqn{ImpulseMaster}, we note that --- just as in the scattering of two massive particles --- only the first term
contributes at leading order (LO) in the generic coupling $g$. This LO contribution
arises at $\Ord(g^2)$; the second term only contributes starting at $\Ord(g^4)$.
Let us focus on the $\WI1$ term, and write out the details of
the wavefunction~(\ref{WaveInitialState}),
\begin{equation}
\begin{aligned}
\WI1 &= \int \df(p_1^\vpp)\df(p_1')\; e^{-i b\cdot(p_1'-p_1^\vpp)/\hbar} \phi_1(p_1^\vpp)\phi_1^*(p_1')
\,
i (p_1'-p_1^\vpp)^\mu 
\langle p_1'\, \alpha_2^{\eta}|T|p_1^\vpp\,\alpha_2^{\eta}\rangle\,.
\end{aligned}
\label{ImpulseStep1}
\end{equation}
\def\innerh{\zeta}

The matrix elements of coherent states are not of definite order in 
perturbation theory.
In order to isolate the contributions at each order, 
one would ordinarily introduce a complete set of 
states of definite particle number on each side of the $T$ matrix,
\begin{equation}
\begin{aligned}
\WI1 &= \sum_{X,X'}\sum_{\innerh,\innerh'=\pm}
\int \df(p_1^\vpp)\df(p_1')\df(r_1^\vpp)\df(r_1')
 \df(k_2^\vpp)\df(k_2')\; 
\\[-3mm] &\hspace*{30mm} \times 
e^{-i b\cdot(p_1'-p_1^\vpp)/\hbar} 
  \phi_1(p_1^\vpp)\phi_1^*(p_1')
\,i (p_1'-p_1^\vpp)^\mu 
\\ &\hspace*{30mm} \times 
\langle p_1'\, \alpha_2^{\eta}|r_1'\,k_2^{\prime\innerh'}\,X'\rangle
\langle r_1'\,k_2^{\prime \innerh'}\,X'| T
|r_1^\vpp\,k_2^{\innerh}\,X\rangle
\langle r_1^\vpp\,k_2^{\innerh}\,X|p_1^\vpp\,\alpha_2^{\eta}\rangle\,.
\end{aligned}
\label{ImpulseStep2}
\end{equation}
The sums over $X$ and $X'$ are over different numbers of messengers, including none, and
include the phase-space integrals over their momenta.  Charge conservation implies that each
intermediate state must contain one net massive-particle number; we drop additional 
particle--antiparticle pairs as their effects will disappear in the 
classical limit, and we denote
the massive-particle momenta by $r_1^\vpp$ and $r_1'$.  
Moreover, in order to satisfy on-shell
conditions of the $T$ matrix element, each intermediate state must 
contain at least one messenger, whose momenta are denoted 
by $k_2^\vpp$ and $k'_2$.

The LO contribution to $\WI1$ is the simplest.  One
may be tempted to believe that it arises from terms with
$X=X'=\emptyset$, but this is not quite right: that would omit disconnected parts of the $S$-matrix. In the situation at hand, a great many photons are present in the initial state; the dominant contribution to the interaction occurs when most photons pass directly from the initial to the final state. Thus rather than taking $X = X' = \emptyset$, we instead
need to sum over additional messengers in the coherent
states.  These sums over non-interacting
messengers, contributing disconnected $S$-matrix terms, are necessary to recover the correct normalization.

\def\trn{\tilde r^\vpp}
\def\trp{\tilde r'}
One can carry out these sums explicitly, but it is
convenient instead to introduce an alternate representation
for the $T$ matrix in terms of creation and
annihilation operators.  As the incoming state
$|\psi_w\rangle$ given in \eqn{WaveInitialState} contains one 
massive particle and
an arbitrary number of photons (or messengers more generally),
we must consider terms with a pair of massive-particle
annihilation and creation operators, and an arbitrary
nonzero number of messenger annihilation and creation
operators (not necessarily paired).  That representation
has the form,
\begin{equation}
\begin{aligned}
T = \sum_{\tilde\hellabel, \tilde\hellabel'} 
\int \df(\trn_1, \trp_1,\tilde k_2^\vpp, \tilde k_2') \; 
\langle \trp_1\tilde k_2^{\prime\tilde\hellabel'}  | T 
| \trn_1 \tilde k_2^{\tilde\hellabel} \rangle \;
\creationh{\tilde\hellabel'}(\tilde k_2') \creation{}(\trp_1)\, 
\annihilation{}(\trn_1)\annihilationh{\tilde\hellabel}(\tilde k_2^\vpp) 
 + \cdots \,,
\end{aligned}
\label{AlternateTRepresentation}
\end{equation}
where the ellipsis indicates higher order terms in the coupling $g$ as well
as amplitudes which do not contribute in the classical limit.
We will summarily drop all these terms in the following, retaining only the
explicit $\Ord(g^2)$ term.
The measure here is a shorthand,
\begin{equation}
    \df(\trn_1, \trp_1,\tilde k_2^\vpp, \tilde k_2')
    =  \df(\trn_1) \df(\trp_1)\df(\tilde k_2^\vpp)\df(\tilde k_2')\,.
\end{equation}
The advantage of the 
representation~(\ref{AlternateTRepresentation}) is that the 
creation
and annihilation operators act simply on coherent states,
yielding factors of $\alpha(k_2^\vpp)$ and $\alpha^*(k_2')$, and
taking care of the normalization for us.  Each term
within this representation contains an ordinary
(connected) amplitude with a definite number of external
messengers.

\def\dakNoteA#1{\textsl{#1}}
\def\class{\textrm{cl}}
\def\beam{B}
\def\qbperp{\qb_\perp}

The required matrix element for the integrand term
in \eqn{AlternateTRepresentation} can be computed easily,
\begin{equation}
\begin{aligned}
\langle p_1'\, \alpha_2^{\eta}| 
T|p_1\,\alpha_2^{\eta}\rangle &=
\langle \trp_1 \tilde k_2^{\prime\tilde\hellabel'} | T 
|\trn_1  \tilde k_2^{\tilde\hellabel} \rangle \;
\langle p_1'\, \alpha_2^{\eta}|
\creationh{\tilde\hellabel'}(\tilde k_2') \creation{}(\tilde p')\, 
\annihilationh{\tilde\hellabel}(\tilde k_2^\vpp) 
\annihilation{}(\tilde p) |p_1\,\alpha_2^{\eta}\rangle
\\ &= \Del(\trn_1-p_1)\,\Del(\trp_1-p'_1)\,
      \delta_{\tilde\hellabel,\hellabel}
      \delta_{\tilde\hellabel',\hellabel}
      \alpha_2(\tilde k_2^\vpp)\alpha_2^*(\tilde k_2')
      \,\langle \trp_1 \tilde k_2'^{\tilde\hellabel'} | T 
| \trn_1 \tilde k_2^{\tilde\hellabel} \rangle\,,
\end{aligned}
\label{IntegrandMatrixElement}
\end{equation}
where we neglected all the terms in
the ellipsis of \eqn{AlternateTRepresentation}.
Notice that we encountered the matrix element $\langle \alpha_2^{\hellabel} | \alpha_2^{\hellabel}\rangle = 1$: this conveniently takes care of all the disconnected diagrams.
The remaining matrix element introduces the desired
scattering amplitude,
\begin{equation}
\begin{aligned}
\langle \trp_1\,\tilde k_2^{\prime\hellabel'}| T
|\trn_1\,\tilde k_2^{\tilde\hellabel}\rangle
&= \Ampl(\trn_1\,\tilde k_2^{\tilde\hellabel}\rightarrow \trp_1\,k_2^{\prime\tilde\hellabel'})\,
\del^4(\trn_1+\tilde k_2^\vpp-\trp_1-\tilde k_2')\,.
\end{aligned}
\label{ComptonMatrixElement}
\end{equation}
As usual, the superscripts on the messenger momenta denote the corresponding physical helicity.
To write it in the usual amplitudes convention, $A(0 \rightarrow p_1, p_2, \ldots)$, we must
cross the momenta to the other side. This flips the helicity of incoming messengers.

Using the results of \eqns{IntegrandMatrixElement}{ComptonMatrixElement}
in \eqn{ImpulseStep1} and carrying out the sums over
$\tilde\hellabel,\tilde\hellabel'$, we obtain,
\begin{equation}
\begin{aligned}
\WI1 &= \int \df(p_1)\df(p_1')\df(k_2^\vpp)\df(k_2')\; 
\phi_1(p_1^\vpp)\phi_1^*(p_1')
     \alpha_2(k_2^\vpp)\alpha_2^*(k_2')
 \\[-3mm] &\hspace*{20mm} \times 
 e^{-i b\cdot(p_1'-p_1^\vpp)/\hbar} 
\,i (p_1'-p_1^\vpp)^\mu 
\\ &\hspace*{20mm} \times 
\Ampl(p_1^\vpp\,k_2^{\eta}\rightarrow p_1'\,k_2^{\prime \eta})\,\del^4(p_1^\vpp+k_2^\vpp-p_1'-k_2')
\,,
\end{aligned}
\label{LOImpulseStep2}
\end{equation}
where we have dropped the tildes on $k_2$ and $k_2'$.

If we make the usual change of variables to the momentum mismatches $q_{1,2}$,
\begin{equation}
\begin{aligned}
q_1 &= p'_1-p_1\,, \\
q_2 &= k'_2-k_2\,;
\end{aligned}
\label{MomentumMismatches}
\end{equation}
use the delta function to integrate over $q_2$; and drop the subscript on $q_1$, we find,
\begin{equation}
\begin{aligned}
\WI1 &= \int \df(p_1)\df(k_2)\dd^4 q\;\del(2q\cdot p_1+q^2)\del(2q\cdot k_2-q^2)
     \Theta(p_1^t+q^t)\Theta(k_2^t-q^t)\; 
\\[-3mm] &\hspace*{30mm} \times 
\phi_1(p_1)\phi_1^*(p_1+q)\,\alpha_2^*(k_2-q)\alpha_2(k_2)
\\ &\hspace*{30mm} \times 
 e^{-i b\cdot q/\hbar} 
\,i q^\mu \,
\Ampl(p_1\,k_2^{\eta}\rightarrow p_1+q,\,(k_2-q)^{\eta})
\,.
\end{aligned}
\label{LOImpulseStep3}
\end{equation}

The analysis of the classical limit as
far as the $\phi_1(p_1)\phi_1^*(p_1+q)$ factor is concerned is the same as in ref.~\cite{KMOC}.
It requires us to take the wavenumber mismatch as our integration variable in lieu
of the momentum mismatch. 
At leading order, we do not have to worry about terms singular in $\hbar$, so the evaluation
as far as the massive particle is concerned will take,
\begin{equation}
\begin{aligned}
\del(2q\cdot p_1+q^2) &\rightarrow \hbar^{-1}\del(2\qb\cdot p_1)\,,
\\ \phi(p_1+q) &\rightarrow \phi(p_1)\,.
\end{aligned}
\label{MassiveParticleLimit}
\end{equation}
Removing the coupling from inside the scattering amplitude (as in ref.~\cite{KMOC},
the reduced amplitude is denoted by $\AmplB$), we find for the classical limit,
\begin{equation}
\begin{aligned}
\WIo1{\class} &= 
 g^2 \Lexp \int \df(\kb_2)\dd^4 \qb\;\del(2\qb\cdot p_1)\del(2\qb\cdot \kb_2-\qb^2) \, \Theta(\kb_2^t - \qb^t)\;
\alphab_2^*(\kb_2-\qb)\alphab_2(\kb_2)
\\[-3mm] &\hspace*{45mm} \times 
\, e^{-ib\cdot \qb} 
\,i \qb^\mu \,
\AmplB(p_1\,\hbar \kb_2^{\eta}\rightarrow p_1+\hbar \qb,\,\hbar (\kb_2-\qb)^{\eta})
\Rexp\,.
\end{aligned}
\label{LOImpulseClassical}
\end{equation}
As in ref.~\cite{KMOC}, the double-angle brackets indicate an average over the wave function of the point-like particle. Classically, this is a function of the momentum $p_1$ with a very sharp peak at $p_1 = m_1 u_1$ where $u_1$ is the classical (proper) velocity and $m_1$ is the particle's mass.

We can now apply this general result in a variety of specific cases. We shall describe two examples in detail: Thomson scattering of a charge by a wave, with $b \simeq 0$, and gravitational scattering of light by a mass in the geometric-optics limit. 

\subsection{Impulse in Thomson Scattering}
\label{sec:thomsonImpulse}

\graphicspath{ {images/} }
\begin{figure}[tb]
\centering\includegraphics[draft=false,width=0.7\textwidth]{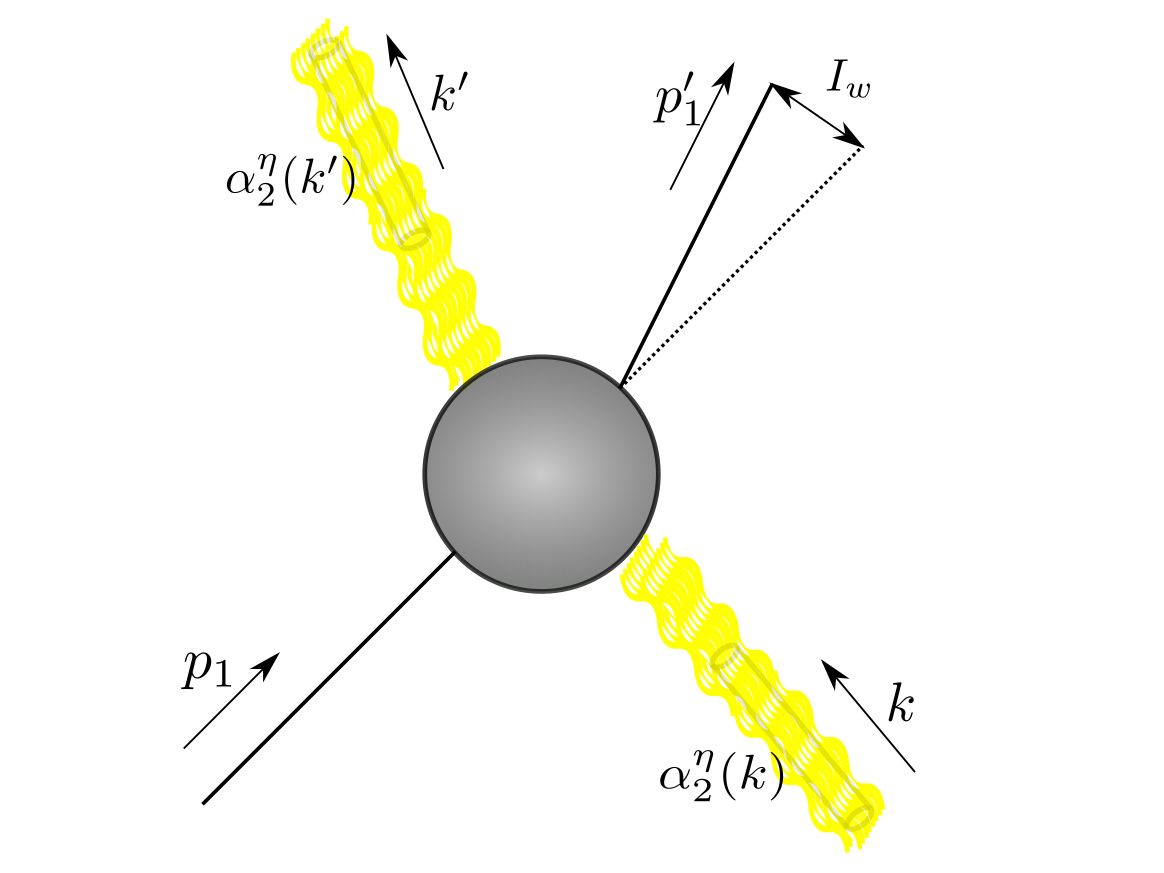}
\caption{Impulse in scattering of a massive object off a coherent state background.}
\label{fig:impulse}
\end{figure}

Our first application is to Thomson scattering,
of a particle of charge $Qe$ and mass $m$,
by a collimated beam of light.
We take the light beam to have positive helicity,
corresponding to the coherent state $\coherentket{\alpha^{+}}$. We need the four-point tree Compton amplitude in scalar QED,
\[
\mathcal{\AmplB}(p_1^\vpp, k_2^{\hellabel} \rightarrow p'_1, k_2'^{\hellabel'}) &= 2 Q^2\, 
\polvhconj{\hellabel}(k_2^\vpp) \cdot \polvh{\hellabel'}(k'_2) \\
&= 2 Q^2\, \polvh{-\hellabel}(k_2^\vpp) 
           \cdot \polvh{\hellabel'}(k'_2) \,,
\label{eq:comptonGeneral}
\]
where we have chosen the gauge,
\begin{equation}
\polvh{\hellabel} \cdot p_1 = 0 \,,
\end{equation}
for both photons.
Alternatively, in spinor variables, we 
have a gauge-invariant expression for the helicity amplitude, 
namely
\begin{equation}
\mathcal{\AmplB}(p_1, k_2^+ \rightarrow p'_1, k_2'^+) = -\frac{Q^2}{2} \, \frac{\langle k_2^\vpp | p_1 |k_2' ]^2}{k_2 \cdot p_1 \, k'_2 \cdot p_1} \,.
\label{eq:comptonExplicit}
\end{equation}
This form of the amplitude is manifestly gauge independent, but it 
depends explicitly on spinors $|k_2' \rangle$ and $|k_2^\vpp]$ 
associated with photon momenta. As usual,
in the classical limit we prefer to work with photon wavenumbers. We therefore introduce rescaled spinors,
\[
|\kb_2' \rangle &\equiv \hbar^{-1/2} \keta{k_2'} \,,\\
|\kb_2^\vpp] &\equiv \hbar^{-1/2} \ketb{k_2^\vpp} \,,
\]
which are directly associated with the photon wavenumbers. The amplitude then has the expression,
\[
\mathcal{\AmplB}(p_1, k_2^+ \rightarrow p'_1, k_2'^+) = -\frac{Q^2}{2} \, \frac{\langle \kb_2 | p_1 |\kb_2' ]^2}{\kb_2 \cdot p_1 \, \kb'_2 \cdot p_1} \,.
\label{eq:comptonExplicitWN}
\]

Choosing $b = 0$, and for a more symmetric presentation, writing $k = k_2$ and $k' = k_2 - q$, 
the impulse \eqn{LOImpulseClassical} takes the form,
\begin{equation}
\langle \Delta p^\mu \rangle =  \frac{Q^2 e^2}{2} \int \df(\kb) \df(\kb') \, \del(2 p \cdot (\kb - \kb')) \, \alphab^*(\kb') \alphab(\kb) \, i(\kb' - \kb)^\mu  \frac{\langle \kb | p | \kb']^2}{(\kb \cdot p)^2} \,.
\end{equation}
This expression may be compared with the classical electromagnetic result, obtained by iterating the classical Lorentz force twice. Thus we see in an explicit example
that a vanishing impact parameter is perfectly acceptable in the classical scattering of waves off matter, in contrast to the situation for two massive particles scattering.

It is interesting that the Compton amplitude appears at tree level in the classical physics of wave scattering off massive particles. This amplitude is also
relevant~\cite{Guevara:2018wpp} for purely massive particle scattering, though at one loop order. While the amplitude is very simple for spinless particles, 
it is considerably more complicated~\cite{Chung:2018kqs} for particles with large spins. Currently we do not have a clear understanding of the appropriate
Compton amplitude for the Kerr black hole, or of what principle we could use to determine it. This is an important area for further research. Our work
suggests one angle of attack: information about the classical part of the Compton amplitude could be extracted by a purely classical analysis of the impulse on
a massive spinning object in scattering off a messenger wave. This is one topic under independent study in ref.~\cite{BGKV}.

\subsection{Light Deflection in Gravitational Scattering}
\label{LightBendingSection}
A second interesting application of the formulas derived in the previous section is to the gravitational
deflection of light by a massive object. We may access this observable by computing the
change in momentum of a narrow (small $\phlperp$) beam of light passing with non-zero impact parameter $b$
past a massive point-like particle. At leading order, there is no radiation of momentum, so
the change in momentum of the wave is simply the negative of the
change in momentum of the massive point source: our starting point
is once again \eqn{LOImpulseClassical}.  

Before we discuss the details of the calculation, it is worth dwelling for a moment on our setup.
Eddington's famous observations demonstrated that starlight is deflected by the sun in accordance with
General Relativity. Near the
sun, light emitted by a distant star is essentially a spherical
wave, and so the incoming wave is extremely delocalized. In contrast, we have chosen
to study a collimated, narrow beam of light.
Nevertheless, the difference between our setup and Eddington's case is immaterial. We work in the situation
where the wavelength $\lambda$ of the light is very small compared to the impact parameter:
this is the domain of geometric optics, and also applies to Eddington's case. It is in the context
of geometric optics that the bending is well-defined; the geometric bending does not depend on the details of
the wave.

For our purposes the setup of a narrow beam in the far past is just a simpler place to start. The
reason is that we can then determine the bending of light by computing the impulse on the beam:
this impulse is directly the change in direction of the wave. By contrast the impulse on starlight
due to the sun involves integrating over the whole incoming 
spherical wavefront: this is not
related in a simple manner to the bending of light.

In the geometric-optics regime, we need the wavelength of the light $\lambda$ to be small. At
the same time we must suppress all quantum effects, so we choose $\lambda$ to be large compared to the Compton wavelength $\lcomp$ 
of our point source. To keep
our beam collimated, \eqn{PhotonConditionI} requires that $\phlperp \gg \lambda$. The requirement that our beam is  narrow is $\phlperp \ll b$. 
Thus there is a series of inequalities:
\[
\lcomp \ll \lambda \ll \phlperp \ll \lscatt \sim b\,.
\label{eq:classical+geometric}
\] 
Note that the scattering length $\lscatt$ is expected to be of 
order of the impact parameter in this case, as we are
considering a $t$ channel process. For simplicity, we consider a 
monochromatic beam with $\sigma_\parallel\rightarrow 0$.
The final length scale to consider is the size $\lpack$ of the 
point-particle's wave packet. As usual we require
$\lcomp \ll \lpack \ll \lscatt$. Once these conditions are met, 
there will be little overlap between the beam 
and the wave packet, so we do not anticipate that the values of the ratios $\lambda / \lpack$ or $\phlperp / \lpack$ will be
important.

The impulse given in \eqn{LOImpulseClassical} simplifies due to 
the constraints of \eqn{eq:classical+geometric}. 
Note that the quantity 
$|\qb \cdot \kb_2| \gg |\qb^2|$
in the second delta function, as $\kb_2 \sim 1/\lambda$ while $\qb \sim 1/\lscatt$.
The wavenumber $\qb$ is then dominantly in the plane of scattering. In this plane, the coherent waveshape
$\alphab_2$ is of width $1 / \phlperp$ so that we may approximate 
$\alphab_2^*(\kb_2-\qb) \simeq \alphab_2^*(\kb_2)$.
For the same reason, the explicit theta function in the impulse simplifies: 
$\Theta(\kb_2^\timecomponentlabel - \qb^\timecomponentlabel) = 1$.
Taking into account the sign demanded by momentum balance, the impulse on the wave is,
\[
\langle \Delta p^\mu_2 \rangle &= 
-g^2 \Lexp \int \df(\kb_2)\dd^4 \qb\;\del(2\qb\cdot p_1)\del(2\qb\cdot \kb_2) \, \;
|\alphab_2(\kb_2)|^2
\\[-3mm] &\hspace*{45mm} \times 
\, e^{-ib\cdot \qb} 
\,i \qb^\mu \,
\AmplB(p_1\,\hbar \kb_2^{\eta}\rightarrow p_1+\hbar \qb,\,\hbar (\kb_2-\qb)^{\eta})
\Rexp\,.
\]

\def\geom{\textrm{geom}}
The integral over $\kb_2$ is now in a great many respects analogous to the integral over the massive particle wave function which is
hidden in our double-angle brackets. In the geometric optics limit, $\alphab_2(\kb_2)$ is a steeply-peaked function of the wave number
peaked at $\kb_2 = \kbbeam$; in view of \eqn{eq:nphotons}, its normalization is related to the number of photons in the beam.
The amplitude, meanwhile, is a smooth function in this region. The $\kb_2$ integral then has the structure,
\[
\int \df(\kb_2) \, \del(2\qb\cdot \kb_2) \, |\alphab_2(\kb_2)|^2 \, f(\kb_2) \, \simeq f(\kbbeam) \int \df(\kb_2) \, \del(2\qb\cdot \kb_2) \, |\alphab_2(\kb_2)|^2 \,,
\]
where $f$ is a slowly-varying function. 
We thus encounter the convolution of a delta function and the sharply-peaked $|\alpha_2(k)|^2$. 
The result of the convolution is a broadened delta function centered at $\kb_2 = \kbbeam$. 
Neglecting the width (of order $\sigma_\perp$) of this function we have,
\[
\int \df(\kb_2) \, \del(2\qb\cdot \kb_2) \, |\alphab_2(\kb_2)|^2 \, f(\kb_2) \, \simeq f(\kbbeam) \, \Nphoton \hbar \, \del (2 \qb \cdot \kbbeam) \,.
\label{eq:moreOnNphoton}
\]
Notice the appearance of the number of photons $\Nphoton$ in the beam: this normalization constant emerges from the integral
over $|\alpha_2(k)|^2$. The classical geometric optics approximation does not have access to this number of photons, and
correspondingly it will cancel in our expression for the deflection angle below. Certain other physical quantities do involve this
number of photons: for example, the total momentum of the beam is,
\[
\totalBeamMomentum^\mu &= \int \df(\kb) |\alphab(\kb)|^2\,\kb^\mu \\
&\simeq \Nphoton \hbar \, \kbbeam^\mu \,.
\]
Returning to the impulse on the beam, use of \eqn{eq:moreOnNphoton} leads to the expression,
\[
\langle \Delta p^\mu_{\geom} \rangle &= 
-\Nphoton \hbar \, g^2 \Lexp \int \dd^4 \qb\;\del(2\qb\cdot p_1)\del(2\qb\cdot \kbbeam) \, \;
\\[-3mm] &\hspace*{35mm} \times 
\, e^{-ib\cdot \qb} 
\,i \qb^\mu \,
\AmplB(p_1\,\hbar \kbbeam^{\eta}\rightarrow p_1+\hbar \qb,\,\hbar (\kbbeam-\qb)^{\eta})
\Rexp\,.
\label{eq:generalImpulseGeometric}
\]
The subscript reminds us that the approximation is valid in the 
geometric-optics limit. 

At leading order, we only need the four-point tree-level amplitude.  As there are no contributions
singular in $\hbar$ at this order, we can simply retain only 
the terms that survive in the
classical limit:
\[
\AmplB(p_1\,k_2^{\eta}\rightarrow p_1',\,k_2'^{\eta})
&= \frac{p_1 \cdot k_2 \, p_1 \cdot k'_2}{q^2} \, \polvhconj{\eta}{(k_2^\vpp)} \cdot \polvh{\eta}{(k'_2)} + \cdots\,, \\
&= \frac{p_1 \cdot \kb_2 \, p_1 \cdot \kb'_2}{\qb^2} \, \polvhconj{\eta}{(\kb_2^\vpp)} \cdot \polvh{\eta}{(\kb'_2)} + \cdots\,,
\label{eq:GRComptonAmpClassical}
\]
where we have chosen the gauge $p_1 \cdot \polvh{\eta}{(k)} = 0$ for each polarization vector, and the ellipsis 
indicates terms which are suppressed by powers of $\hbar$.

This amplitude simplifies further in the geometric-optics limit. The inequalities \eqn{eq:classical+geometric} require
in particular that the wave number $\qb \sim 1/b \ll \kb_2$. We may therefore replace the scalar product $p \cdot \kb'_2$ with
$p \cdot \kb_2$ in \eqn{eq:GRComptonAmpClassical}, up to terms which are neglected in the geometric-optics limit. 
At the same time,
we may replace the polarization vector $\polvh{\eta}{(\kb'_2)}$ 
with $\polvh{\eta}{(\kb_2)}$ to the same order of approximation. 
The amplitude is then simply,
\[
\AmplB(p_1^\vpp\,k_2^{\eta}\rightarrow p_1',\,k_2'^{\eta})
= -\frac{(p_{1}^\vpp\cdot \kb_2^\vpp)^2}{\qb^2}+\cdots
\label{eq:GRComptonAmpGeometric}
\]
We note that the geometric-optics limit of the amplitude 
for the scattering of a photon off a massive scalar
is helicity-independent. Up to constant factors, it reduces
to the amplitude between one massless and one massive 
scalar\footnote{See the beautiful and pedagogical discussion in ref.~\cite{Lightbending} for more details.}. This is as expected
from the equivalence principle: if the classical limit weren't universal, 
the impulse and hence the scattering angle would have 
helicity-dependent contributions.

In order to the evaluate the impulse, we insert the 
geometric-optics amplitude~(\ref{eq:GRComptonAmpGeometric})
into the expression~(\ref{eq:generalImpulseGeometric}) for the impulse in the geometric-optics limit.  We obtain,
\[
\langle \Delta p^\mu_{\geom} \rangle &= i\kappa^2 \, \Nphoton \hbar \, (p_{1}\cdot \kbbeam)^2 \, \int \dd^4 \qb \;
 \del(2\qb\cdot p_1)\del(2\qb\cdot \kbbeam)\;
\, e^{-i b\cdot \qb}  \frac{\qb^\mu}{\bar{q}^2}\\
&= i\kappa^2 \, (p_{1}\cdot \totalBeamMomentum)^2 \, \int \dd^4 \qb \;
 \del(2\qb\cdot p_1)\del(2\qb\cdot \totalBeamMomentum)\;
\, e^{-i b\cdot \qb}  \frac{\qb^\mu}{\bar{q}^2}\,.
\label{eq:masslessimpulse}
\]
Here, we have replaced the general coupling $g$ by the appropriate 
gravitational coupling $\kappa$, and the wavenumber $\kbbeam$
by the total beam momentum $\totalBeamMomentum$. 
The second line of this equation is strikingly similar to the 
impulse in a scattering process between two \emph{massive} 
classical objects.
Indeed, the integral remaining in \eqn{eq:masslessimpulse} is essentially the same as the integral appearing in the LO impulse 
in ref.~\cite{KMOC}. It can easily
be performed by taking the light beam in the $z$ direction, $\totalBeamMomentum^\mu = (E,0,0,E)$.  The result is,
\begin{equation}
\langle \Delta p^\mu_{\geom} \rangle = 
-\kappa^2\, \frac{p_{1}\cdot \totalBeamMomentum}{8 \pi \, b^2} \, b^{\mu}\,.
\end{equation}
The impact parameter $b^\mu$ is directed from the massive particle 
towards the wave, so the sign
above indicates that the interaction is attractive. 

The scattering angle $\theta$ is then determined geometrically in terms of the impulse,
\begin{equation}
\sin\theta = \frac{|b \cdot \Delta p|}{|\v{b}| \, E}\,,
\end{equation}
once we have fixed a frame. We have taken the absolute value to drop the sign of the angle, understanding
that the bending is towards the scatterer.
Working in the rest frame of the massive scalar, and using
$\kappa^2=32 \pi G_{N}$, we reproduce the well-known value for the 
gravitational bending of light,
\begin{equation}
\theta=\frac{4G_{N}m}{|\v{b}|}+\cdots.
\end{equation}

As a final comment, it is satisfying that the impulse we have obtained in \eqn{eq:masslessimpulse} is essentially the 
same as the impulse on massive point particles as discussed in ref.~\cite{KMOC}. This occurred as the inequalities
\eqn{eq:classical+geometric} greatly simplified the impulse. These inequalities themselves are very similar to the Goldilocks
conditions \eqn{Goldilocks} for classical point-like particles.
The fact that the dynamics of massive particles is so similar to the behavior of waves in the geometric-optics regime was 
a celebrated aspect
of nineteenth and early twentieth century physics, known as the Hamiltonian analogy. This analogy was highlighted by
Schr\"odinger~\cite{schrodinger} and others as an important consideration in the early days of 
quantum mechanics.

\subsection{Higher Orders}

Although in sections~\ref{sec:thomsonImpulse} 
and~\ref{LightBendingSection} we focused on 
leading-order applications,
our formalism is completely general and \eqn{ImpulseMaster} holds 
to all perturbative orders. As we have seen, 
the leading-order contribution arises at $\Ord(g^2)$. The second 
term, $\WI2$, in the impulse of~\eqn{ImpulseMaster} involves 
one-loop amplitudes, and therefore contributes only starting
at $\Ord(g^4)$. 
Consequently, we can identify a further contribution, at $\Ord(g^3)$, which receives
no contribution from $\WI2$ but only from $\WI1$. It arises from 
the leading corrections to \eqn{AlternateTRepresentation},
\begin{equation}
\begin{aligned}
\delta T_3 \equiv \sum_{\tilde\hellabel, \tilde\hellabel',
\tilde\hellabel''} 
\int \df(\trn_1, &\trp_1,\tilde k_2^\vpp, \tilde k_2', \tilde k_3^\vpp) \; 
\\[-3mm]
\times \bigl[&
\langle \trp_1\tilde k_2^{\prime\tilde\hellabel'}| T 
| \trn_1 \tilde k_2^{\tilde\hellabel} 
\tilde k_3^{\tilde\hellabel''}\rangle \;
\creationh{\tilde\hellabel'}(\tilde k_2') \creation{}(\trp_1)\, 
\annihilation{}(\trn_1)\annihilationh{\tilde\hellabel}(\tilde k_2^\vpp)
\annihilationh{\tilde\hellabel''}(\tilde k_3^\vpp)
\\&+ \langle \trp_1 \tilde k_2^{\prime\tilde\hellabel'} 
\tilde k_3^{\tilde\hellabel''}| T 
| \trn_1 \tilde k_2^{\tilde\hellabel} \rangle \;
\creationh{\tilde\hellabel''}(\tilde k_3^\vpp)
\creationh{\tilde\hellabel'}(\tilde k_2') \creation{}(\trp_1)\, 
\annihilation{}(\trn_1)\annihilationh{\tilde\hellabel}(\tilde k_2^\vpp)
\bigr]\,,
\end{aligned}
\label{AlternateTRepresentation3}
\end{equation}
where the additional argument in the measure corresponds
to a factor of $\df(\tilde k_3^\vpp)$.

Inserting the integrand of $\delta T_3$ into the matrix element
in \eqn{ImpulseStep1}, we obtain (analogously 
to \eqn{IntegrandMatrixElement}),
\begin{equation}
\begin{aligned}
\langle p_1'\, \alpha_2^{\eta}|&\delta 
T_3|p_1\,\alpha_2^{\eta}\rangle =
\\ 
\bigl[&\langle \trp_1\tilde k_2^{\prime\tilde\hellabel'}| T 
| \trn_1 \tilde k_2^{\tilde\hellabel} 
\tilde k_3^{\tilde\hellabel''}\rangle \;
\langle p_1'\, \alpha_2^{\eta}|
\creationh{\tilde\hellabel'}(\tilde k_2') \creation{}(\trp_1)\, 
\annihilation{}(\trn_1)\annihilationh{\tilde\hellabel}(\tilde k_2)
\annihilationh{\tilde\hellabel''}(\tilde k_3)
|p_1\,\alpha_2^{\eta}\rangle
\\&
+ \langle \trp_1\tilde k_2^{\prime\tilde\hellabel'} 
\tilde k_3^{\tilde\hellabel''}| T 
| \trn_1 \tilde k_2^{\tilde\hellabel} \rangle \;
\langle p_1'\, \alpha_2^{\eta}|
\creationh{\tilde\hellabel''}(\tilde k_3)
\creationh{\tilde\hellabel'}(\tilde k_2') \creation{}(\trp_1)\, 
\annihilation{}(\trn_1)\annihilationh{\tilde\hellabel}(\tilde k_2)
|p_1\,\alpha_2^{\eta}\rangle
\bigr]
\\ =\,& \Del(\trn_1-p_1)\,\Del(\trp_1-p'_1)\,
      \delta_{\tilde\hellabel,\hellabel}
      \delta_{\tilde\hellabel',\hellabel}
      \alpha_2(\tilde k_2)\alpha_2^*(\tilde k_2')
\\&\times       
      \bigl[\delta_{\tilde\hellabel'',\hellabel}
      \,\alpha_2(k_3) 
         \langle \trp_1\tilde k_2^{\prime\tilde\hellabel'}| T 
              | \trn_1 \tilde k_2^{\tilde\hellabel} 
                \tilde k_3^{\tilde\hellabel''}\rangle
+\delta_{\tilde\hellabel'',\hellabel}\,\alpha_2^*(k_3)
      \langle \trp_1\tilde k_2^{\prime\tilde\hellabel'} 
\tilde k_3^{\tilde\hellabel''}| T 
| \trn_1 \tilde k_2^{\tilde\hellabel} \rangle
      \bigr]
\,.
\end{aligned}
\label{IntegrandMatrixElement3}
\end{equation}

The scattering matrix elements in this 
expression introduce five-point amplitudes,
\begin{equation}
\begin{aligned}
\langle \trp_1\,\tilde k_2^{\prime \tilde\hellabel'}| T
|\trn_1\,\tilde k_2^{\tilde\hellabel}
\,\tilde k_3^{\tilde\hellabel''}\rangle
&= \Ampl(\trn_1
\,\tilde k_2^{\tilde\hellabel}
\,\tilde k_3^{\tilde\hellabel''}\rightarrow \trp_1\,\tilde k_2^{\prime \tilde\hellabel'})
 \,\del^4(\trn_1+\tilde k_2+\tilde k_3-\trp_1-\tilde k_2')\,,
\\ \langle \trp_1\,\tilde k_2^{\prime \tilde\hellabel'}
\,\tilde k_3^{\tilde\hellabel''}| T|\trn_1\,\tilde k_2^{\tilde\hellabel}\rangle
&= \Ampl(\trn_1\,k_2^{\tilde\hellabel}\rightarrow 
\trp_1\,\tilde k_2^{\prime \tilde\hellabel'}
\,\tilde k_3^{\tilde\hellabel''})\,
 \del^4(\trn_1+\tilde k_2-\trp_1-\tilde k_2'-\tilde k_3)\,.
\end{aligned}
\label{ComptonMatrixElementWithEmission}
\end{equation}
By crossing, we could choose to identify,
\begin{equation}
\Ampl(\trn_1\,\tilde k_2^{\tilde\hellabel}
\,\tilde k_3^{\tilde\hellabel''}\rightarrow \trp_1\,\tilde k_2^{\prime \tilde\hellabel'})
=\Ampl(\trn_1\,\tilde k_2^{\tilde\hellabel}\rightarrow \trp_1,
\,\tilde k_2^{\prime \tilde\hellabel'},
\,(-\tilde k_3)^{-\tilde\hellabel''})\,.
\label{Crossing}
\end{equation}
Substituting these expressions into \eqn{ImpulseStep1}
and dropping tildes, we obtain,
\begin{equation}
\begin{aligned}
\WI1|_{g^3} &= 
\int \df(p_1^\vpp)\df(p_1')\df(k_2^\vpp)\df(k_2')\df(k_3^\vpp)\; \alpha_2^*(k_2')\alpha_2(k_2)
\\[-3mm] &\hspace*{20mm} \times 
e^{-ib\cdot(p_1'-p_1^\vpp)/\hbar} 
  \phi_1(p_1)\phi_1^*(p_1')
\,i (p_1'-p_1)^\mu 
\\ &\hspace*{20mm} \times \Bigl[\alpha_2(k_3)
\Ampl(p_1^\vpp\,k_2^{\eta}\,k_3^{\eta}\rightarrow p_1'\,k_2^{\prime \eta})
 \,\del^4(p_1^\vpp+k_2^\vpp+k_3^\vpp-p_1'-k_2')
\\ &\hspace*{20mm} \hphantom{\times \Bigl[}+\alpha_2^*(k_3)
\Ampl(p_1\,k_2^{\eta}\rightarrow p_1'\,k_2^{\prime \eta}\,k_3^{\eta})\,
 \del^4(p_1+k_2-p_1'-k_2'-k_3)\Bigr]
\,.
\end{aligned}
\label{ImpulseStep3g3}
\end{equation}

This $\Ord(g^3)$ term is interesting as it differs in structure
from contributions to the impulse for massive-particle
scattering studied in ref.~\cite{KMOC}. 
In that case, the first corrections arise at
$\Ord(g^4)$, from one-loop amplitudes in $\MI1$ and
cut one-loop amplitudes in $\MI2$.  We leave an
investigation of the new contributions~(\ref{ImpulseStep3g3})
to future work.

Another difference between purely massive scattering and particle-on-wave scattering relates to the radiation 
reaction. In the massive case~\cite{KMOC}, radiation reaction 
first occurs at next-to-next-to-leading order,
that is at $\Ord(g^6)$.
In contrast, radiation reaction arises at $\Ord(g^4)$ 
in wave--particle scattering. This radiation reaction must
contain contributions from the second term in the impulse, $\WI2$, 
which contributes at that order.

\section{Point-like Observables}
\label{ObserversSection}

In the previous section, we built on ref.~\cite{KMOC} to analyze
what we may call
\textit{global\/} observables, requiring
an array of detectors covering the celestial sphere at infinity in order to measure the
quantity.  This is most manifest for the total radiated momentum, defined by eq.~(3.33) 
of ref.~\cite{KMOC},
\begin{align}
\Rad^\mu \equiv \langle k^\mu \rangle &= {}_\textrm{in}{\langle} \psi | \, S^\dagger \operator K^\mu S \, | \psi \rangle_\textrm{in}
= {}_\textrm{in}\langle \psi | \, T^\dagger \operator K^\mu T \, | \psi \rangle_\textrm{in}\,.
\end{align}

\def\xz{x}
Even in electromagnetic scattering, achieving $4\pi$ coverage would make this
a challenging measurement.  In the gravitational context, where we would be looking
to detect emission from scattering of distant black holes, such a measurement would
be hopelessly impractical.  Instead, for the remainder of this article, we turn to what we may call
\textit{local\/} observables, which can be measured with a localized detector, albeit
still sitting somewhere on the celestial sphere, say at $\xz$.
The paradigm for such a measurement
is that of the waveform $W(t,\bhn;\xz)$ of radiation emitted during a scattering event in direction $\bhn$
from an event at the coordinate origin.  (That is, we adopt the convention that $-\bhn$
points back from the observer towards the scattering event.)
We will focus on electromagnetic
radiation here, but much of the formalism will carry over to 
the gravitational case.
Let us keep in mind that we will be interested in several 
detectors, all nearby
$\xz$, though with separations that are completely negligible compared to the distance
from the origin.

Local observables have a general structure which, as we will see, is determined by some source (the scattering event)
and the propagation of messengers over very large distances. In fact it is convenient to break up our
discussion of these observables along these lines. In the present section we will discuss this overall
structure in more detail, with a focus on the crucial aspect of propagation. In the following sections, we will extract
general expressions for local observables
from quantum field theory, and connect
to the Newman-Penrose formalism. Then we
will examine
global observables in cases where a classical wave scatters off
a massive
particle before turning to the physically important case where two massive particles scatter and radiate.

It will be easier to discuss and manipulate the Fourier transform of the waveform with respect to time. We will
refer to this as the spectral waveform $f(\omega,\bhn; \xz)$:
\begin{equation}
f(\omega,\bhn;\xz) = \int_{-\infty}^{+\infty} dt\; W(t,\bhn;\xz)\, e^{i\omega t}\,.
\label{SpectralFunction}
\end{equation}
Given a result for the spectral waveform, we can of course recover the time-dependent waveform
via an inverse Fourier transform.  Because we are interested in radiation produced
by long-range forces, the idealized waveforms for the scattering
processes we will consider stretch infinitely far back and forward in time.
The idealization is implicit in the infinite limits for the integral in \eqn{SpectralFunction}.
In an actual measurement, however, the waveform would be below the noise floor of
the detector for all times before a
 `signal start time' preceding the moment of closest approach, 
and likewise for all times after a `signal end time' following that moment.  We can then
take the theoretical waveforms to be approximations to actual ones cut off at the start
and end times.  Label the interval between the two by $\Delta t_s$.

\def\Gret{G_{\rm ret}}
Let us imagine that the point of closest approach during the
scattering event is 
at the coordinate origin,
$(t,{\bf x})=(0,{\bf 0})$. When a massless wave scatters off a 
point particle, the wave may overlap the particle; we take a 
suitable event of maximum overlap as the origin.
We can treat the scattering as occurring in a box
of temporal length $\Delta t_s$, and of spatial size $\Delta x_s$.  Radiation is emitted inside the box during the
scattering event, and then spreads out.  We will take an (idealized) measurement of the
radiation in some direction $\bhn$, at a much later time and at a point very far away in
that direction.  The details of the scattering --- the particles' interaction and spins ---
will determine the radiation emitted inside the box.  Modifying those details could radically
change the emission.  Those details, however, will have no effect on the propagation of
the radiation out to the distant measuring apparatus.  Only the spin of the radiated
field can have any effect.  We thus expect the form of the result to be a Green's
function convoluted with a source.  More precisely, given that we have only outgoing
radiation, we expect a retarded Green's function $\Gret$.  We can
then expand the Green's function in the large-distance limit to obtain the connection
between the observable and the emitted radiation inside the box.

\def\scalarJ{{\widetilde J}_s}
\def\scalarJx{J_s}
\def\Jtilde{{\widetilde J}}
\def\Jk{\Jtilde_{\vec\mu}}
\def\Jx{J_{\vec\mu}}
\def\Jsp{J_{\widevec{\alpha\beta}}}
The details of the scattering inside the box around $(0,{\bf 0})$ define a current for
our radiation.  In a real-world context, we are interested in electromagnetic or
gravitational radiation, but we can equally well treat the case of (massless) scalar
radiation as well.  The details of the scattering inside the box 
give rise to a wavenumber-space field-strength current,
$\Jk(\kb)$, where the notation $\vec\mu$ denotes a number of indices appropriate to
the radiated messenger: none for a scalar, two for a photon, 
and four for a graviton,
\begin{equation}
 \begin{array}{cl}
    \Jtilde(\kb):&\quad \textrm{scalar}\,,  \\
    \Jtilde_{\mu\nu}(\kb):&\quad \textrm{electromagnetism}\,,  \\
    \Jtilde_{\mu\nu\rho\sigma}(\kb):&\quad \textrm{gravity}\,.
 \end{array}
\end{equation}
In a slight abuse of language, we will refer to these
quantities simply as currents.
They will satisfy appropriate conservation conditions.  We will 
later obtain an expression for such a current in terms of 
scattering amplitudes.

Given this current, the usual position-space current can of course be obtained by taking a Fourier transform,
\begin{equation}
\Jx(x) = \int \dd^4 \kb\; \Jk(\kb)\,e^{-i\kb\cdot x}\,.
\label{SpatialCurrent}
\end{equation}
Clearly we can also write $\Jk(\kb)$ in terms of $\Jx(x)$ via an inverse transform,
\begin{equation}
\Jk(\kb) = \int d^4 x\;\Jx(x)\,e^{i \kb\cdot x}\,.
\label{MomentumCurrent}
\end{equation}
Both of these forms of the current will be helpful for us below.

\def\RadObs{R_{\vec\mu}}
\def\Gret{G_{\rm ret}}
\def\Gadv{G_{\rm adv}}
\def\vnh{{\v{\hat n}}}
As we will show in detail in the next section, we obtain
an $x$-dependent radiation observable in the general form,
\begin{equation}
\RadObs(x) = i\int \df(\kb)\;\bigl[\Jk(\kb)\, e^{-i\kb\cdot x}-\Jk^*(\kb)\, e^{+i\kb\cdot x}\bigr]\,,
\label{OriginalRadiation}
\end{equation}
that is, as an integral of the source $\Jk(\kb)$
over the on-shell massless phase space for the radiated messenger. 
Examples will include expectations of hermitian operators, 
such as the field-strength operator in electromagnetism, or the Riemann tensor in gravity.

The hermiticity properties of our radiation observables is manifest in \eqn{OriginalRadiation}. But notice that the observables
are defined as integrals over positive frequencies $\kb^\timecomponentlabel \geq 0$. Yet in writing the innocuous-seeming 
Fourier transform in \eqn{SpatialCurrent}, we have assumed knowledge of the current for both positive \emph{and} 
negative frequency. So we must fill a gap: what do we mean by the current for negative frequency? In fact, the 
reality condition provides the necessary information. Our currents are real in position space, and we may note that,
\[
\Jx(x) = \int \dd^4 \kb\; \theta(\kb^\timecomponentlabel) 
\left[ \Jk(\kb)\,e^{-i\kb\cdot x} + \Jk(-\kb) e^{i \kb \cdot x} \right] \,.
\]
The reality condition then leads to the relation,
\[
\Jk(-\kb) = \Jk^*(\kb) \,.
\label{eq:realityCondition}
\]
We use this relation to define the current for negative frequency.

A key simplification arises because the source event, occurring in our box, is sourced in a comparatively localized region compared to the very large propagation distance of the outgoing radiation. To access this simplification, we follow
a well-trodden path~\cite{schwingerBook} by rewriting
our radiation observables as integrals over the spatial
extent of the source. Thus, we express the observable of
\eqn{OriginalRadiation} in
terms of the spatial current $\Jx(x)$, yielding
\begin{equation}
\RadObs(x) = i\int \df(\kb)\, d^4y\; \Jx(y)\bigl[e^{-i\kb\cdot (x-y)}- e^{+i\kb\cdot (x-y)}\bigr]\,.
\label{RadiationII}
\end{equation}
Next, we interchange orders of integration. Judicious forethought reveals the combination
of phase space integrals to be a difference of retarded and advanced Green's functions, 
\begin{equation}
\RadObs(x) = \int d^4y\; \Jx(y)\bigl[\Gret(x-y)- \Gadv(x-y)\bigr]\,.
\label{RadiationIII}
\end{equation}
In the far future, where the observer measures the wavetrain emitted from the scattering
event, $\Gadv$ will vanish.  Put in an explicit form for $\Gret$, and switch back
to the wavenumber-space current in order to make the
complete dependence of the integrand on $x$ and $y$ manifest.
The result is,
\begin{equation}
\begin{aligned}
\RadObs(x) &= \int \dd\omega \dd^3 \v{\kb}\, d^4y\;  \Jk(\kb)\,e^{-i\kb\cdot y}\,
\frac{\delta(x^0-y^0-|\v{x}-\v{y}|)}{4\pi |\v{x}-\v{y}|}
\\ &= \int \dd\omega \dd^3 \v{\kb}\, d^3\v{y}\;  \Jk(\kb)\,
\frac{e^{-i\omega x^0}\,e^{+i\omega|\v{x}-\v{y}|}\,e^{+i\v{\kb}\cdot\v{y}}}{4\pi |\v{x}-\v{y}|} \,.
\end{aligned}
\label{RadiationIV}
\end{equation}
Notice that the integral is now over \textit{all\/} wavenumbers.
We have split the four-dimensional momentum integration into integrals over spatial and frequency components for later
convenience.

From the earlier discussion, 
we know that $\Jx(y)$ is concentrated around $y\simeq 0$, whereas
$x$ is far away ($x\gg y$).  Accordingly we can expand the integrand there, using,
\begin{equation}
\begin{aligned}
|\v{x}-\v{y}| &\sim \bigl[\v{x}^2-2\v{x}\cdot\v{y}\bigr]^{1/2}
\\&\sim |\v{x}| \Bigl(1-\frac{\vnh\cdot\v{y}}{|\v{x}|}\Bigr)\,.
\end{aligned}
\label{Expansion}
\end{equation}
We must be careful in performing this expansion: while it is sufficient to retain
the leading term in the denominator, we must retain formally subleading terms that
contribute to nontrivial phases.  Even in those exponents, 
we can of course still drop terms beyond
the subleading, as they give rise to no nontrivial phases.

Substituting the expansion~(\ref{Expansion}) into \eqn{RadiationIV}, we obtain,
\begin{equation}
\begin{aligned}
\RadObs(x) &= \int \dd\omega \dd^3 \v{\kb}\, d^3\v{y}\;  \Jk(\kb)\,
\frac{e^{-i\omega x^0}\,e^{+i\omega|\v{x}|}e^{-i\omega \bhn\cdot\v{y}}\,e^{+i\v{\kb}\cdot\v{y}}}
{4\pi |\v{x}|}\,;
\end{aligned}
\label{RadiationV}
\end{equation}
performing in turn the $\v{y}$ and $\v{k}$ integrals, we finally obtain,
\begin{equation}
\begin{aligned}
\RadObs(x) &= \frac{(2 \pi)^3}{4\pi |\v{x}|}\int \dd\omega \dd^3 \v{\kb}\;  \Jk(\kb)\,
e^{-i\omega x^0}\,e^{+i\omega|\v{x}|}\,\delta^3(\v{\kb}-\omega\bhn)
\\ &= \frac1{4\pi |\v{x}|}\int \dd\omega \;  \Jk(\omega, \omega\bhn)\,
e^{-i\omega (x^0-|\v{x}|)} \,.
\end{aligned}
\label{RadiationVI}
\end{equation}

We can thus identify the waveform with the coefficient of the leading-power
term $|\v{x}|^{-1}$, 
\begin{equation}
\begin{aligned}
W_{\vec{\mu}} (t,\bhn;\xz) &= \frac1{4\pi}\int \dd\omega \;  \Jk(\omega, \omega\bhn)\,
e^{-i\omega (\xz^0-|\v{x}|)}
\end{aligned}\,.
\label{WaveformI}
\end{equation}
In this equation, $t$ represents the observer's clock time.  We could take it to be
$\xz^0$, or $\xz^0-|\v{\xz}|$, or some other convenient time.  We must nonetheless retain
the separate dependence on $\xz^0$ and $|\v{x}|$, because these quantities will differ
between the cluster of nearby observers in which we are interested.  That is, the absolute
phase of the waveform at any given observer's location is not measurable and is therefore
irrelevant, but the relative phases between nearby observers are measurable.

Choosing $t = \xz^0-|\v{\xz}|$, the corresponding spectral waveform is then simply,
\[
f_{\vec{\mu}}(\omega, \bhn) = \frac1{4\pi} \Jk(\omega, \omega\bhn) \,.
\label{eq:spectralWaveform}
\]
More precisely, \eqn{eq:spectralWaveform} is the waveform for positive frequencies. 
For negative frequencies, the waveform follows from \eqn{eq:realityCondition},
\[
f_{\vec{\mu}}(\omega, \bhn) = \frac1{4\pi} \Jk^*(-\omega, -\omega\bhn) \,;
\label{eq:spectralWaveformNegativeFrequency}
\]
notice that $-\omega$ is now positive.
In both cases, once we know the current $\Jk(\kb)$, we can 
immediately write down the spectral waveform.

\def\Fop{\mathbb{F}^{\mu\nu}}

\def\vsuperp{\vphantom{()}}

\section{Spectral Waveforms}
\label{SpectralFunctionSection}

As we have seen, the waveform is directly related to the current $\Jk(\kb)$ generated by the scattering event. We must choose a 
specific local radiation observable to determine this current using its definition, \eqn{OriginalRadiation}. In this section we will study examples in both electrodynamics and gravity. Let us begin with a simple case: the field-strength tensor~\eqref{eqn:fieldstrengthdn} in electrodynamics.

We choose an observer at $x$, in the far future of the event, equipped to measure the expectation value of the electric and magnetic field at
the point $x$. The observable is therefore
\[
\Foutdn \equiv {}_{\textrm{out}}\langle\psi|\Fopdn(x)|\psi\rangle_{\textrm{out}}  \,.
\]
We can rewrite the outgoing state in terms of the incoming state using
the time-evolution operator or $S$-matrix,
\[
\Foutdn = {}_{\textrm{in}}\langle \psi| S^\dagger \Fopdn(x) S | \psi \rangle_{\textrm{in}} \,,
\]
where (as usual) $|\psi\rangle_\textrm{in}$ is the incoming state in the far past. This state could contain, for example, two isolated massive point-like particles,
or a single isolated massive particle and a coherent state describing incoming radiation. A state of the former type would be appropriate to
study radiation emitted as two particles scatter, while a state of the latter type can be used to study the scattered radiation field in a Thomson
scattering process. We will study both of these examples in detail later in this article.

Inserting the expression for the field-strength 
tensor~(\ref{eqn:fieldstrengthdn}) into this expectation value, 
and converting to integrals
over wavenumbers, we learn that,
\begin{equation}
\begin{aligned}
\Foutdn= -2i \hbar^{3/2} \sum_\hellabel \int \df(\kb) \, 
\bigl[ &\langle \psi | S^\dagger \annihilationh{\hellabel}(k) S | \psi \rangle \,\kb_{[\mu}^\vmu \polvhconj{\hellabel}_{\nu]}(\kb)\, e^{-i \kb\cdot x} \\
& - \langle \psi | S^\dagger \creationh{\hellabel}(k) S | \psi \rangle \,\kb_{[\mu}^\vmu \polvh{\hellabel}_{\nu]}(\kb)\, e^{+i\kb \cdot x} \bigr] \,,
\end{aligned}
\end{equation}
where we have again dropped the `in' subscript, leaving it implicit 
in the rest of our discussion.  (Recall that $k$ is just a label
for the creation and annihilation operators, and we can use
$\kb$ interchangeably for this purpose.)

We now see the virtue of our definition of the general class of radiation observables in \eqn{OriginalRadiation}. Evidently the expectation 
value $\Foutdn$ is of precisely this form, and we can read off the current $\Jk(\kb)$ as
\[
\Jtilde_{\mu\nu}(\kb) = -2 \hbar^{3/2} \sum_\hellabel \langle \psi | S^\dagger \annihilationh{\hellabel}(k) S | \psi \rangle &\,\kb_{[\mu}^\vmu \polvhconj{\hellabel}_{\nu]}(\kb) \,.
\]
The discussion of the previous section therefore applies, and we see from \eqn{eq:spectralWaveform} that the corresponding spectral waveform 
is,
\[
f_{\mu\nu}(\omega,\bhn) &= -\frac1{2\pi} \hbar^{3/2} \sum_\hellabel \langle \psi | S^\dagger \annihilationh{\hellabel}(k) S | \psi \rangle\,\kb_{[\mu}^\vmu \polvhconj{\hellabel}_{\nu]}(\kb) \Big|_{\kb = (\omega, \omega \bhn)} \,,
\label{eq:allOrdersSpectralEM}
\]
for positive frequency ($\omega >0$).
For negative frequency ($\omega < 0$) the waveform is,
\[
f_{\mu\nu}(\omega,\bhn) &= -\frac1{2\pi} \hbar^{3/2} \sum_\hellabel \langle \psi | S^\dagger \creationh{\hellabel}(k) S | \psi \rangle\,\kb_{[\mu}^\vmu \polvh{\hellabel}_{\nu]}(\kb) \Big|_{\kb = -(\omega, \omega \bhn)} \,.
\]
This result holds to all orders in perturbation theory.

It is straightforward to extend this result to gravity. We work in Einstein gravity, and assume that the spacetime is asymptotically Minkowskian.
In this case our observer at $\xz$ is very far from the source of gravitational waves, and is equipped to measure the expectation value of
the local spacetime curvature $\langle R^\textrm{out}_{\mu\nu\rho\sigma}(x) \rangle$. The corresponding spectral waveform is nothing 
but the double copy of~\eqn{eq:allOrdersSpectralEM},
\[
f_{\mu\nu\rho\sigma}(\omega,\bhn) &= \frac{i \kappa}{2\pi} \hbar^{3/2} \sum_\hellabel \langle \psi | S^\dagger \annihilationh{\hellabel}(k) S | \psi \rangle\,
\kb_{[\mu}^\vmu \polvhconj{\hellabel}_{\nu]}(\kb) 
\;\kb_{[\rho}^\vmu \polvhconj{\hellabel}_{\sigma]}(\kb) 
\Big|_{\kb = (\omega, \omega \bhn)} \,,
\label{eq:fRiemann}
\]
for $\omega > 0$. In this equation, the operator $\annihilationh{\hellabel}(k)$ 
annihilates perturbative gravitational states. We have included a 
factor $\kappa / 2$ so that the Riemann tensor has the
conventional normalization. Noting that
the metric perturbation falls off as inverse distance, it follows that non-linear terms in the Riemann tensor produce
corrections which fall off faster than inverse distance. Consequently, we have neglected them.
Notice that all possible traces
of \eqn{eq:fRiemann} vanish, consistent with the fact that the Riemann tensor in vacuum
equals the Weyl tensor.
The waveform for negative frequency is,
\[
f_{\mu\nu\rho\sigma}(\omega,\bhn) &= -\frac{i \kappa}{2\pi} \hbar^{3/2} \sum_\hellabel \langle \psi | S^\dagger \creationh{\hellabel}(k) S | \psi \rangle\,
\kb_{[\mu}^\vmu \polvh{\hellabel}_{\nu]}(\kb) 
\;\kb_{[\rho}^\vmu \polvh{\hellabel}_{\sigma]}(\kb) 
\Big|_{\kb = -(\omega, \omega \bhn)} \,.
\]

The Lorentz indices on these observables reflects the tensor structure of electrodynamics and gravity. 
In both cases, however, there are only two possible polarizations of the outgoing radiation. 
It is helpful to project the waveform onto one of these polarizations. Classically, a convenient way
to do so is to use the Newman--Penrose (NP)~\cite{Newman:1961qr} 
formalism, which is intimately connected to the spinor-helicity 
method of scattering 
amplitudes~\cite{Emond:2020lwi,Splitsignature,BGKV}. We can adopt the same idea in the present context. For us, a simple route to the NP formalism is to 
pick a complex basis of vectors which is aligned with our setup. We choose the vectors\footnote{We use capital letters to denote
the elements of our NP basis rather than the more traditional lower case symbols in order to distinguish the vectors
from loop momenta, masses, et cetera.}
\[
L^\mu = \kb^\mu / \omega = (1, \bhn)^\mu, \quad N^\mu = \reference^\mu, 
\quad M^\mu = \polvh{+}{}^\mu, \quad M^{*\mu} = \polvh{-}{}^\mu \,.
\]
The null vector $\reference$ is simply a gauge choice, satisfying 
$\reference \cdot \polvh{\pm} = 0$ and 
$L \cdot N = L \cdot \reference = 1$. Furthermore note that $M\cdot M^* = -1$.
The scaling of the NP vector $L$ ensures that it does not depend on frequency $\omega$, and is dimensionless. Indeed
the polarization vectors $\polvh{\pm}$ do not depend on the scaling of $\kb$ so they are also independent of frequency.
These vectors therefore make sense as a spacetime basis, not merely as a basis in Fourier space.

\def\npPhi{\Phi}
\def\npPsi{\Psi}
It is easy to check that the only non-zero components of 
$f_{\mu\nu}$ in the NP basis are $f_{\mu\nu} M^{*\mu} N^\nu$ and 
$f_{\mu\nu} M^{\mu} N^\nu$. These are the leading radiative NP scalar, traditionally~\cite{Penrose:1986ca} denoted $\npPhi_2^0$, and its conjugate. 
We can write these NP scalars as Fourier transforms:
\[
\npPhi_2^0(t, \bhn) 
&= \int \dd \omega \, e^{-i \omega t} \, \tilde\npPhi_2^0(\omega, \bhn) \,.
\]
Notice that we commuted the NP basis vectors through the frequency integration sign. This is permissible as the basis vectors are independent of frequency.
For positive frequency $\omega$, we find,
\[
\tilde \npPhi_2^0(\omega, \bhn) = -\frac\omega{4\pi} \hbar^{3/2} \langle \psi | S^\dagger \annihilationh{-}(k) S | \psi \rangle\Big|_{\kb = (\omega, \omega \bhn)} \,,
\]
while 
for negative frequency, the corresponding expression reads,
\[
\tilde \npPhi_2^0(\omega, \bhn) = +\frac\omega{4\pi} \hbar^{3/2} \langle \psi | S^\dagger \creationh{+}(k) S | \psi \rangle\Big|_{\kb = -(\omega, \omega \bhn)} \,.
\]
Combining these results, we find that the time-domain NP scalar is,
\[
\npPhi_2^0(t, \bhn) 
= -\frac{\hbar^{3/2}}{4\pi} \int \dd \omega \, \Theta(\omega) \, \omega \bigl[ &e^{-i \omega t} \langle \psi | S^\dagger \annihilationh{-}(k) S | \psi \rangle 
\\ &
+  e^{+i \omega t} \langle \psi | S^\dagger \creationh{+}(-k) S | \psi \rangle \bigr] \Big|_{\kb = (\omega, \omega \bhn)} \,.
\]

In gravity, the corresponding radiative NP scalar is defined by
\[
\npPsi_4(x) = - N_\mu M^*_\nu N_\rho M^*_\sigma \langle W^{\mu\nu\rho\sigma}(x) \rangle \,,
\]
where $W^{\mu\nu\rho\sigma}(x)$ is the Weyl tensor, equal to the Riemann tensor in our case. Expanded at large distances, the leading
term in the NP scalar is $\npPsi_4^0$:
\[
\npPsi_4(x) = \frac{1}{|\v x|} \npPsi_4^0  + \cdots \,.
\]
This object is directly relevant
to gravitational waveforms~\cite{NumericalRelativity,Boyle:2019kee}. We find that the spectral version of the NP scalar is,
\[
\tilde \npPsi_4^0(\omega, \bhn) = -i \frac{\kappa \, \omega^2}{8\pi} \hbar^{3/2} \langle \psi | S^\dagger \annihilationh{--}(k) S | \psi \rangle\Big|_{\kb = (\omega, \omega \bhn)} \,,
\]
for positive $\omega$.
Let us emphasize once again that these results hold to all orders of perturbation theory.

NP scalars are particularly well-suited for comparison with helicity amplitudes in quantum field theory. However, they may be slightly less familiar than the more elementary field strengths; field strengths also have the virtue of being hermitian quantities. Therefore, 
in the remainder of this article, we will also study the expectation of the radiative field-strength tensor in perturbation theory. 
This entails rewriting the scattering matrix in terms of the transition matrix
$T$, $S=1+i T$,
\def\vtskip{\mskip2mu}
\begin{equation}
\begin{aligned}
 \Foutdn &= 
 \langle\psi|(1-iT^\dagger)\Fopdn(x)\vtskip
(1+i T)|\psi\rangle
\\ &= \langle\psi|\Fopdn(x)|\psi\rangle
+ 2 \Re i\langle\psi|\Fopdn(x) T|\psi\rangle
+\langle\psi|T^\dagger\Fopdn(x) T|\psi\rangle\,.
\end{aligned}
\label{FexpectationII}
\end{equation}
The first term in \eqn{FexpectationII} is the expectation value of the field strength due to any incoming radiation
which may be present in $|\psi\rangle_\textrm{in}$; the following
term is linear 
in amplitudes, and thus of $\Ord(g^3)$ (or higher); the last term is quadratic
in amplitudes (or equivalently, linear in a cut amplitude), and contains
terms of $\Ord(g^5)$ and higher.  

Using unitarity, we can rewrite \eqn{FexpectationII},
\begin{equation}
\Foutdn(x) = \langle\psi|\Fopdn(x)|\psi\rangle + 
i\langle\psi|[\Fopdn(x),T]|\psi\rangle
+\langle\psi|T^\dagger[\Fopdn(x), T]|\psi\rangle\,.
\label{FMaster}
\end{equation}
The commutator in the second term of this expression is reminiscent of the form of the impulse
$\Delta p$ (although in case of the field strength, the first term above need not vanish). This second form
of the field strength can be both instructive and useful, but it has a slight disadvantage that
reality properties are somewhat obscured compared to~\eqn{FexpectationII}.
When taking the classical limit, we are interested in the
leading term in the large-distance expansion as well; 
for such radiation observables,
we will understand the $\LexpT\cdots\RexpT$ notation to impose
that expansion as well.

We will use this observable to analyze emitted radiation in the
scattering of two charged particles in
\sect{EmissionSection}.  We first continue our analysis of 
Thomson scattering in the next section.

\section{From Compton Scattering to Thomson Scattering}
\label{ThomsonScatteringSection}
In \sect{sec:thomsonImpulse}, we considered the
Thomson scattering process: electromagnetic
scattering of a classical beam off of
a massive point charge. In our earlier
discussion we studied
the impulse suffered by the massive particle 
during the process.
We are now equipped to deepen our analysis
by determining the scattered
light generated during Thomson scattering. We will
do so by using the results of the previous
section to compute the NP scalar $\npPhi_2^0$ which describes
that scattered light at very large distances.

In this situation, our initial state~\eqn{WaveInitialState} describes
an isolated massive particle, and a localized
beam of
incoming classical radiation described as in
\sect{LocalizedBeams} by a coherent state with
an appropriate waveshape function.
Correspondingly, the incoming state generates a
non-vanishing expectation value for the
electromagnetic field strength tensor. This
is the incoming classical radiation 
$\langle F_{\mu\nu}^\textrm{in}(x)\rangle$:
\[
\langle F_{\mu\nu}^\textrm{in}(x) \rangle =
\langle \psi_w | \Fopdn |\psi_w \rangle \,.
\]
In particular, there is a non-vanishing NP scalar $\npPhi_2^0$ in the far past. 

To focus attention on the scattered light, it is convenient to study the overall change in the NP scalar during the process,
\[
\Delta \npPhi_2^0(\omega, \bhn) = -\frac\omega{4\pi} \hbar^{3/2} \left[ \langle \psi_w | S^\dagger \annihilationh{-}(k) S | \psi_w \rangle
-\langle \psi_w | \annihilationh{-}(k) | \psi_w \rangle \right]\Big|_{\kb = (\omega, \omega \bhn)} \,.
\label{eq:npDeltaPhiDefAllOrder}
\]
This simply subtracts the contribution of the incoming beam to the radiation field in the future. We will compute this quantity at leading order, focusing on the positive-frequency
part throughout.

\graphicspath{ {images/} }
\begin{figure}[tb]
\centering
\includegraphics[draft=false,width=0.7\textwidth]{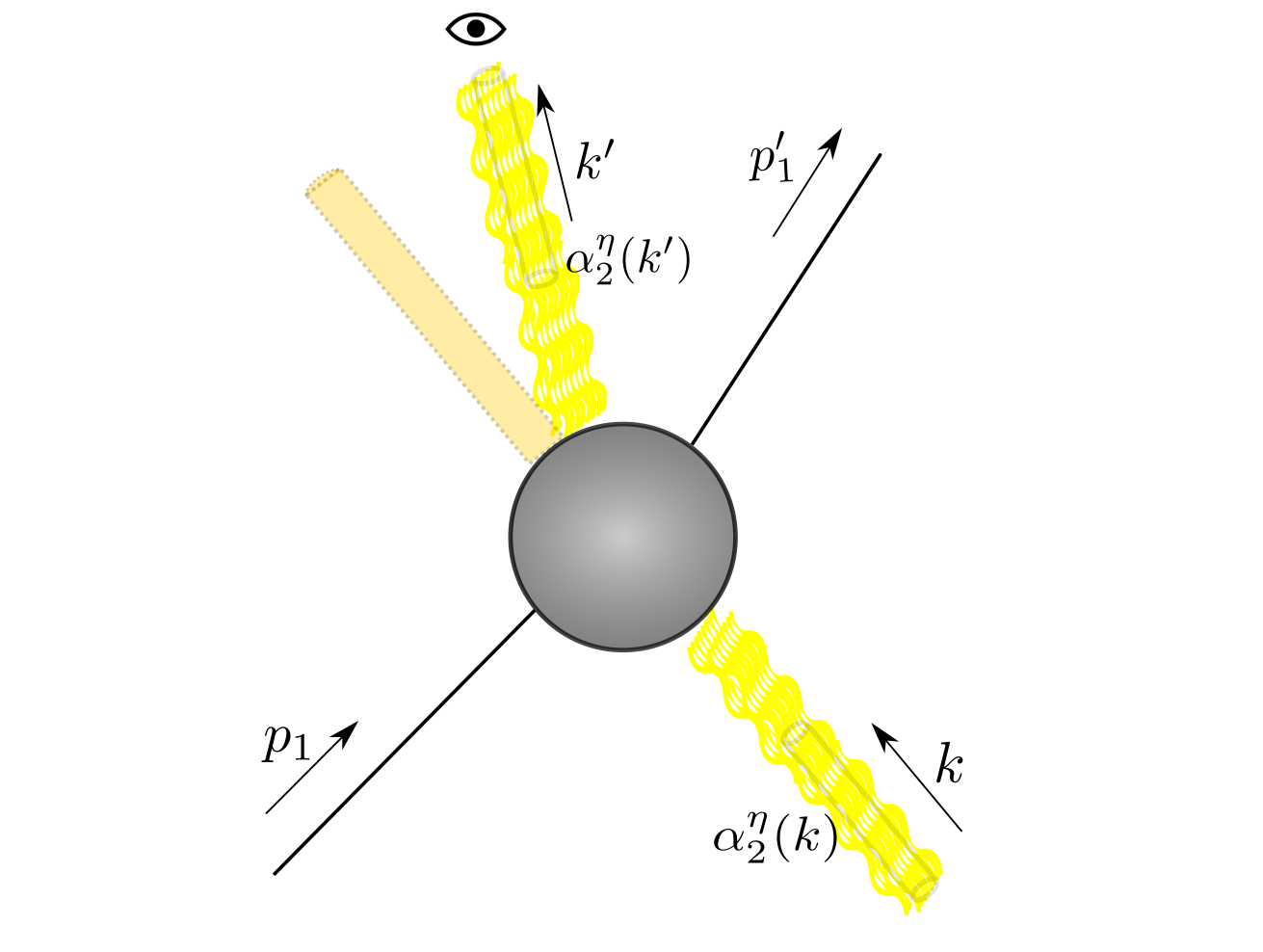}
\caption{The observer measures the field strength of the outgoing wave.}
\label{fig:waveform_Thomson}
\end{figure}

Using unitarity of the $S$ matrix, we may write $\Delta \npPhi_2^0$ in terms of a commutator,
\[
\Delta \npPhi_2^0(\omega, \bhn) = -\frac{i}{4\pi} \omega\hbar^{3/2} \langle \psi_w| [\annihilationh{-}(k_2') , T] | \psi_w \rangle\Big|_{\kb_2' = (\omega, \omega \bhn)} \,.
\label{eq:npDeltaPhiCommutator}
\]
We relabeled the quantity $\kb$ appearing in \eqn{eq:npDeltaPhiDefAllOrder} as $\kb_2'$ because, as we will see below, it has the interpretation of the wavevector
associated with the outgoing wave which was 
denoted $\kb_2'$ in \Sect{ScatteringLightSection}.

To compute the commutator $[\annihilationh{-}(k_2') , T]$, we make
use of \eqn{AlternateTRepresentation} to
expand the $T$ matrix in terms of creation and annihilation operators.
Dropping the
terms in the ellipsis of \eqn{AlternateTRepresentation}, the commutator is easily computed to be,
\[
[\annihilationh{-}(k_2') , T] = \sum_{\tilde\hellabel} \int \df(\trn_1, \trp_1, \tilde k_2^\vpp) \; \langle \trp_1 k_2'^{-} | T 
| \trn_1 \tilde k_2^{\tilde\hellabel}  \rangle \; 
\creation{}(\trp_1) \annihilation{}(\trn_1)\, 
\annihilationh{\tilde\hellabel}(\tilde k_2^\vpp) \,.
\]
Inserting this result in \eqn{eq:npDeltaPhiCommutator}, and expanding the state $|\psi_w\rangle$ using its definition~\eqn{WaveInitialState} specialized to the case $b = 0$
we easily find that,
\[
\Delta &\npPhi_2^0(\omega, \bhn) =
\\&\hphantom{=}\, -\frac{i}{4\pi} \omega\hbar^{3/2} \sum_\hellabel \int \df(p_1, p_1', k_2) \, \phi^*(p_1') \phi(p_1) \, \langle p_1' k_2'^{-} | T |  p_1 k_2^{\hellabel} \rangle \, \langle \alpha^+ | \annihilationh{\hellabel}(k_2) | \alpha^+ \rangle 
\\
&= -\frac{i}{4\pi} \omega\hbar^{3/2} \int \df(p_1, p_1', k_2) \, \phi^*(p_1') \phi(p_1) \, \langle p_1' k_2'^{-} | T |  p_1 k_2^{+}  \rangle \, \alpha(k_2) \,.
\]
The matrix element of the transition operator yields the Compton amplitude, as well as the usual delta function enforcing overall momentum conservation. We may
perform the $p_1'$ integral using this delta function to find that,
\[
\Delta &\npPhi_2^0(\omega, \bhn) =
\\&\hphantom{=}\,
-\frac{i}{4\pi} \omega\hbar^{3/2} \int \df(p_1, k_2) \, \del (2p_1 \cdot (k_2' - k_2))\, | \phi(p_1) |^2\, \alpha(k_2) \, \mathcal{A}(p_1 k_2^+ \rightarrow  p_1' k_2'^-) \,.
\]
We replaced the (conjugated) wavefunction $\phi^*(p_1' + k_2^\vpp - k_2')$ by $\phi^*(p_1')$ because the difference $(k_2^\vpp - k_2')/\hbar = \kb_2^\vpp - \kb_2'$ is small
(of order $1/\lambda$) compared
to the width of the wavefunction (which is of order $1/\ell_w$). The integral over the wavefunction is now precisely of the form required for the double-angle-bracket notation of ref.~\cite{KMOC} so that we arrive at,
\[
\Delta \npPhi_2^0(\omega, \bhn) = -\frac{i}{4\pi} \Lexp  \omega \hbar^{3/2} \int \df(k_2)  \, \del (2p_1 \cdot (k_2' - k_2)) \, \alpha(k_2) \, \mathcal{A}(p_1 k_2^+ \rightarrow p_1' k_2'^-) \Rexp \,.
\]
Finally, we insert the explicit Compton amplitude of \eqn{eq:comptonGeneral}, and replace the remaining integral over $k_2$ with an integral over the associated wavenumber $\kb_2$ to learn
that the LO NP scalar due to the scattering process is,
\[
\Delta \npPhi_2^0(\omega, \bhn) =i \frac{Q^2 e^2}{16 \pi} \Lexp \omega \int \df(\kb_2) \, \del (2p_1 \cdot (\kb_2' - \kb_2)) \, \alphab(k_2) \, m^2 \frac{\langle \kb_2 \kb_2'\rangle}{[ \kb_2 \kb_2'] \, \kb_2 \cdot p_1} \Rexp \,.
\label{eq:deltaNPPhiFinal}
\]
The same result would also be obtained from a classical analysis of the leading order radiation field of a point charge moving under the influence of 
an incoming classical wave.

Alternatively, it is possible to compute the expectation value of the field strength in the very far future. Focusing again on the change in the field strength,
\[
\langle \Delta F_{\mu\nu}(x)\rangle \equiv
\langle F_{\mu\nu}^\textrm{out}(x) \rangle -
\langle F_{\mu\nu}^\textrm{in}(x) \rangle \,,
\]
it is straightforward to use \eqn{FexpectationII} and find that,
\[
\langle \Delta F_{\mu\nu}(x)\rangle &=
 i\langle\psi_w|[\Fopdn(x), T]|\psi_w\rangle
+ \cdots.
\label{eq:DeltaFMaster}
\]
We have indicated higher order terms are present in the ellipsis. 
It may be worth emphasizing once again that this result is the same as one would find
be direct computation using background field methods:
\begin{equation}
\begin{aligned}
\langle \Delta F^{\mu\nu}(x)\rangle &= 
\langle \psi_w| S^{\dagger}  \Fop(x) S |  \psi_w \rangle 
   - \langle \psi_w| \Fop(x) |  \psi_w \rangle 
\\   &= \int  \df(p_1) \df(p_1) \phi(p_1) \phi^*(p'_1) e^{-i b\cdot (p_1'-p_1)/\hbar}  \\ 
&\hspace*{12mm}\times \Bigl\{ i \langle p_1'|  [\Fop(x), T (\Acl^{(\hellabel)}) ]| p_1 \rangle 
+ \langle p_1'| T^{\dagger}(\Acl^{(\hellabel)})
 \Fop(x) T(\Acl^{(\hellabel)}) | p_1 \rangle \Bigr\}\,,
\end{aligned}
\end{equation}
where $\Acl^{(\eta)}(x)$ denotes the
classical background field corresponding to
our coherent state, and we once again used the relation 
$\coherent{\alpha,(\eta)}^{\dagger} \coherent{\alpha,(\eta)} = \mathbbm{1}$. 

Returning to the LO computation of the scattered field strength, by inserting the definition of the field strength
operator, we now encounter two commutators:
\[
\langle \Delta F_{\mu\nu}(x)\rangle =
 \frac{2}{\hbar^{3/2}} \sum_{\hellabel'} \int \df(k') & \left[ \langle\psi_w|[ \annihilationh{\eta'}(\kb'), T]|\psi_w\rangle \, \kb'_{[\mu} \polvhconj{\eta'}_{\nu]}(k') \, e^{-i k'\cdot x / \hbar} 
\right.  \\ & \left.  
\quad -\langle\psi_w|[ \creationh{\eta'}(\kb'), T]|\psi_w\rangle \, \kb'_{[\mu} \polvh{\eta'}_{\nu]}(k') \, e^{+i k'\cdot x / \hbar} \right] \,.
\]
The first of these was computed explicitly above; the second is very similar. After a short computation, the field strength can be expressed as,
\[
\langle \Delta F_{\mu\nu}(x)\rangle = \Re\, \Lexp 4 g^2 \sum_{\hellabel'} \int \df(\kb_2, \kb_2') \,& \del(2p_1 \cdot (\kb_2'-\kb_2)) \, \alphab(\kb_2) \\
& \times\AmplB(p_1 k_2^+ \rightarrow p_1' k'{}^{\hellabel'}) 
\, \kb'_{2[\mu} \polvhconj{\eta'}_{\nu]}(\kb_2') \, e^{-i \kb_2'\cdot x} \Rexp \,.
\label{eq:DeltaFSintegrand}
\]

Comparison with the NP scalar is facilitated by performing the $\kb_2'$ integral using the methods of \sect{ObserversSection}. Indeed, the field strength change of \eqn{eq:DeltaFSintegrand} is of the general form of the radiation observable~\eqn{OriginalRadiation}. The corresponding current is,
\[
\tilde J_{\mu\nu}&(\kb_2) = 
\\ &-4 i \Lexp \sum_{\hellabel'} \int \df(\kb_2') \, \del (2p_1 \cdot (\kb_2'-\kb_2)) \, \alphab(\kb_2) \AmplB(p_1 k_2^+ \rightarrow p_1' k_2'{}^{\hellabel'})\, \kb'_{2[\mu} \polvhconj{\eta'}_{\nu]}(\kb_2') \Rexp \,.
\]
The NP scalar can be obtained directly from this current as,
\[
\Delta \npPhi_2^0(\omega, \bhn) = \frac{1}{4\pi} \tilde J_{\mu\nu}(\kb) M^{*\mu} N^\nu \,.
\]
Performing the dot products, we recover our earlier result, \eqn{eq:deltaNPPhiFinal}.

Earlier, we identified incoming classical radiation with coherent states. The reader may
wonder then about the nature of outgoing radiation. A necessary condition for the outgoing radiation to be represented by a coherent state is that expectation values of observables, 
such as the field strength, should factorize. We have proved this explicitly earlier, see \eqn{eq:towardsClassicalFactorization}. 
Perhaps surprisingly, it turns out that this is also a sufficient 
condition. Indeed, one can work out the constraints on the probability 
density of the outgoing (pure) radiation: in the coherent state space 
(also called the Glauber--Sudarshan representation), the classical 
factorization of observables implies that the distribution has zero 
variance. In turn, this makes the distribution degenerate, 
\textit{i.e.\/}~supported on isolated points. But as shown by 
Hillery~\cite{Hillery:1985}, the normalization condition together with 
the purity constraint suffices to reduce the sum of delta functions 
in the 
coherent state space to just a single delta function.
That is, we have only a 
single outgoing coherent state in the classical limit. In 
appendix~\ref{app:opFactorisation}, we 
prove that the factorization condition holds at the lowest order 
in the coupling constant, which makes the outgoing radiation state of the Thomson 
scattering coherent up to order $g^2$. 
A more detailed discussion on this 
point will appear in forthcoming work~\cite{CGMORSW}.

\section{Emission Waveform}
\label{EmissionSection}

We turn now to photon emission in the scattering of two charged point particles.
At leading order in perturbation theory, only the second term 
in \eqn{FexpectationII} (or similarly, in \eqn{FMaster})
contributes.  
It will be of order $\Ord(g^3)$, whereas the second term will be of $\Ord(g^5)$.

If we now substitute the expression~(\ref{eqn:fieldstrengthdn}), 
along with that~(\ref{InitialState}) for
the initial-state wavefunction for the scattering particles 
into the first term
of \eqn{FexpectationII}, we obtain,
\begin{equation}
\begin{aligned}
\langle F^{\mu\nu}(x)\rangle_1 &= 
  \frac{4}{\hbar^{\expfrac32}}
    \Re\sum_{\hellabel}
     \int \df(p_1) \df(p_2) \df(p'_1) \df(p'_2) \df(k)\; 
\\[-3mm] &\hspace*{25mm}\times
 e^{-i b\cdot (p_1'-p_1)/\hbar} \phi(p_1)\phi^{*}(p'_1) \phi(p_2)\phi^{*}(p'_2)
\\ &\hspace*{25mm} \times
   k^{[\mu\vsuperp}\pol^{(\hellabel)\nu]*} e^{-ik\cdot x/\hbar}
      \langle p'_1\, p'_2| \annihilationh{\hellabel}(k) \,T|p_1\,p_2\rangle
\\ &= 
   \frac{4}{\hbar^{\expfrac32}}
 \Re\sum_{\hellabel}
  \int \df(p_1) \df(p_2) \df(p'_1) \df(p'_2) \df(k)\;
\\[-3mm] &\hspace*{25mm}\times
 e^{-i b\cdot (p_1'-p_1)/\hbar} \phi(p_1)\phi^{*}(p'_1) \phi(p_2)\phi^{*}(p'_2)
\\ &\hspace*{25mm} \times
   k^{[\mu\vsuperp}\pol^{(\hellabel)\nu]*} e^{-ik\cdot x/\hbar}
      \langle p'_1\, p'_2\,k^{\hellabel}| T|p_1\,p_2\rangle
\,.
\end{aligned}
\label{FFirstTerm}
\end{equation}
We can identify the matrix element as a five-point amplitude,
\begin{equation}
\begin{aligned}
\langle p'_1\, p'_2\,k^{\hellabel}| T|p_1\,p_2\rangle &= \Ampl(p_1,p_2\rightarrow p'_1,p'_2,k^{\hellabel})
    \del^4(p_1+p_2-p'_1-p'_2-k)\,.
\end{aligned}
\label{FirstTermAmplitudes}
\end{equation}
At leading order, we replace the amplitude by its
LO contribution, given
by a tree-level expression.
To compute the required waveform, we must identify the expectation of $F^{\mu\nu}(x)$ as
the spatial current $\Jx(x)$ in \eqns{SpatialCurrent}{MomentumCurrent}, and
via \eqn{MomentumCurrent}, in \eqn{WaveformI}.

Beyond leading order, the expectation of $F^{\mu\nu}(x)$ 
will receive higher-order contributions
to the amplitudes in \eqn{FirstTermAmplitudes}, alongside contributions from the last term in \eqn{FMaster},
\begin{equation}
\begin{aligned}
\langle F^{\mu\nu}(x)\rangle_2 &= 
   -\frac{2i}{\hbar^{\expfrac32}}
   \sum_{\hellabel}
 \int \df(p_1^\vpp) \df(p_2^\vpp) \df(p'_1) \df(p'_2) \df(k)\; 
\\[-3mm] &\hspace*{20mm}\times
  e^{-i b\cdot (p_1'-p_1)/\hbar} \phi(p_1)\phi^{*}(p'_1) \phi(p_2)\phi^{*}(p'_2)
\\ &\hspace*{20mm} \times
  \bigl[ k^{[\mu\vsuperp}\pol^{(\hellabel)\nu]*} e^{-i k\cdot x/\hbar}
      \langle p'_1\, p'_2| T^\dagger \annihilationh{\hellabel}(k) \,T|p_1\,p_2\rangle
\\ &\hspace*{20mm} \hphantom{\times\bigl[]}
      -  k^{[\mu\vsuperp}\pol^{(\hellabel)\nu]} e^{+i k\cdot x/\hbar}
                 \langle p'_1\, p'_2| T^\dagger \creationh{\hellabel}(k) T|p_1\,p_2\rangle\bigr]
\end{aligned}
\end{equation}
Insert a complete set of states to the right of each $T^\dagger$,
\begin{equation}
\langle\psi|T^\dagger \mathbb{\Fop} T|\psi\rangle
=\sum_X \int \df(r_1)\df(r_2)\;\langle\psi|T^\dagger|r_1\,r_2\,X\rangle
\langle r_1\,r_2\,X|\mathbb{\Fop} T|\psi\rangle\,,
\end{equation}
where the sum over $X$ is over all states, including no additional particles, and includes
an implicit integral over momenta of any particles in $X$ and a sum over any other quantum numbers.
As in ref.~\cite{KMOC}, we assume that each of the incoming massive particles carries a
separately conserved global charge, so that each intermediate state has one net particle of
each type.  We can ignore additional particle-antiparticle pairs of the massive particles,
as these contributions will disappear in the classical limit.
As there are no messengers in the initial state, and hence
no coherent states, there is no need to sum over arbitrary
numbers of messengers.  Accordingly, we do not need
to switch to a
coherent-friendly representation~(\ref{AlternateTRepresentation})
of the $T$ matrix.
We obtain,
\begin{equation}
\begin{aligned}
\langle F^{\mu\nu}(x)\rangle_2 &= 
  -\frac{2i}{\hbar^{\expfrac32}}
  \sum_X\sum_{\hellabel}
    \int \df(r_1)\df(r_2)\df(p_1) \df(p_2) \df(p'_1) 
    \df(p'_2)\df(k) \, 
\\[-3mm] &\hspace*{20mm}\times
   e^{-i b\cdot (p_1'-p_1)/\hbar} \phi(p_1)\phi^{*}(p'_1) \phi(p_2)\phi^{*}(p'_2)
\\ &\hspace*{20mm} \times
  \bigl[ k^{[\mu\vsuperp}\pol^{(\hellabel)\nu]*} e^{-ik\cdot x/\hbar}
      \langle p'_1, p'_2| T^\dagger |r_1\,r_2\,X\rangle
                   \langle r_1\,r_2\,X|\annihilationh{\hellabel}(k) \,T|p_1\,p_2\rangle
\\ &\hspace*{20mm} \hphantom{\times\bigl[]}
      -  k^{[\mu\vsuperp}\pol^{(\hellabel)\nu]} e^{+i k\cdot x/\hbar}
                 \langle p'_1, p'_2| T^\dagger|r_1\,r_2\,X\rangle
                   \langle r_1\,r_2\,X| \creationh{\hellabel}(k) T|p_1\,p_2\rangle\bigr]
\\&=    -\frac{2i}{\hbar^{\expfrac32}}
  \sum_X\sum_{\hellabel}\int \df(r_1)\df(r_2)\df(p_1) 
          \df(p_2) \df(p'_1) \df(p'_2)\df(k)\, 
\\[-3mm] &\hspace*{20mm}\times
   e^{-i b\cdot (p_1'-p_1)/\hbar} \phi(p_1)\phi^{*}(p'_1) \phi(p_2)\phi^{*}(p'_2)
\\ &\hspace*{15mm} \times
   \bigl[ k^{[\mu\vsuperp}\pol^{(\hellabel)\nu]*} 
          e^{-i k\cdot x/\hbar}
      \langle p'_1, p'_2| T^\dagger |r_1\,r_2\,X\rangle
                   \langle r_1\,r_2\,k^{\hellabel}\,X|T|p_1\,p_2\rangle
\\ &\hspace*{20mm} \hphantom{\times\bigl[]}
      -  k^{[\mu\vsuperp}\pol^{(\hellabel)\nu]} 
         e^{+i k\cdot x/\hbar}
                 \langle p'_1\, p'_2| T^\dagger|r_1\,r_2\,k^{\hellabel}\,X\rangle
                   \langle r_1\,r_2\,X| T|p_1\,p_2\rangle\bigr]\,.
\end{aligned}
\label{FSecondTerm}
\end{equation}
In the second term within brackets, 
the creation operator requires a photon in the intermediate state,
and eliminates it from the bra.  We then relabeled $X$ to exclude it.  Note as well
that at next-to-next-leading order and beyond, we necessarily require amplitudes with
three incoming particles.  These can just as easily be obtained by crossing.
The term~(\ref{FSecondTerm})
has the interpretation of a cut of an amplitude, just as for the 
second term in the impulse in ref.~\cite{KMOC}, as seen in 
eqs.~(3.26--3.31) therein.

This contribution first appears at next-to-leading order.  At this 
order, we are interested
in contributions with $X=\emptyset$, and we can identify the 
required matrix elements
as a combination of four- and five-point amplitudes,
\begin{equation}
\begin{aligned}
\langle r_1\,r_2| T|p_1\,p_2\rangle &= 
   \Ampl(p_1\, p_2 \rightarrow r_1\, r_2) \del^4(p_1+p_2-r_1-r_2)\,,
\\
\langle p'_1, p'_2| T^\dagger |r_1\,r_2\rangle &= 
   \Ampl^*(p'_1, p'_2 \rightarrow r_1, r_2) \del^4(p'_1+p'_2-r_1-r_2)\,,
\\
\langle r_1\,r_2\,k^{\hellabel}|T|p_1\,p_2\rangle &= 
   \Ampl(p_1, p_2 \rightarrow r_1, r_2, k^{\hellabel})  \del^4(p_1+p_2-r_1-r_2-k)\,,
\\
\langle p'_1\, p'_2| T^\dagger|r_1\,r_2\,k^{\hellabel}\rangle &= 
   \Ampl^*(p'_1, p'_2 \rightarrow r_1, r_2, k^{\hellabel}) \del^4(p'_1+p'_2-r_1-r_2-k)\,.
\end{aligned}
\label{SecondTermAmplitudes}
\end{equation}
For the next-to-leading order contribution to 
$\langle F^{\mu\nu}(x)\rangle$, we use
tree-level amplitudes in \eqn{SecondTermAmplitudes}.

\section{The Detected Wave at Leading Order}
\label{ExplicitWaveformSection}

The leading-order contribution to the waveform will
arise at $\Ord(g^3)$, as described in the previous section.
We apply the approach of ref.~\cite{KMOC} to 
\eqn{FFirstTerm}.  Similarly to that reference,
and to~\sect{ScatteringLightSection}, we define the 
momentum mismatches,
\begin{equation}
\begin{aligned}
q_1 &= p'_1-p_1\,, \\
q_2 &= p'_2-p_2\,;
\end{aligned}
\label{MomentumMismatchesWaveform}
\end{equation}
and trade the integrals over the $p_i'$ for integrals over
the $q_i$,
\begin{equation}
\begin{aligned}
\langle F^{\mu\nu}&(x)\rangle_1 = 
\\&\hspace*{-6mm} \frac{4}{\hbar^{\expfrac32}}
  \Re\sum_{\hellabel}\int \df(p_1) \df(p_2) 
       \dd^4 q_1 \dd^4 q_2 \df(k)\;
  \del(2p_1\cdot q_1+q_1^2)\del(2p_2\cdot q_2+q_2^2)
\\[-2mm] &\hspace*{12mm}\times
 e^{-i b\cdot q_1/\hbar}
 \Theta(p_1^\timecomponentlabel+q_1^\timecomponentlabel)
 \Theta(p_2^\timecomponentlabel+q_2^\timecomponentlabel)
  \phi(p_1)\phi^{*}(p_1+q_1) \phi(p_2)\phi^{*}(p_2+q_2)
\\ &\hspace*{12mm} \times
   k^{[\mu\vsuperp}\pol^{(\hellabel)\nu]*} e^{-ik\cdot x/\hbar}
       \Ampl(p_1,p_2\rightarrow p_1+q_1,p_2+q_2,k^{\hellabel})
    \del^4(q_1+q_2+k)
\,.
\end{aligned}
\label{FatLO}
\end{equation}
We can take the classical limit, and change to the required
wavenumber variables for the $q_i$ and $k$, 
\begin{equation}
\begin{aligned}
\langle F^{\mu\nu}&(x)\rangle_{1,\class} = 
\\&\hspace*{-8mm} {g^3}
  \Lexp \hbar^{2}\!\Re\sum_\hellabel\int \df(\kb)
  \kb^{[\mu\vsuperp}\pol^{(\hellabel)\nu]*} e^{-i\kb\cdot x}
\\[-2mm] &\hspace*{30mm}\times
   \prod_{i=1,2}\int \dd^4 \qb_i\;\del(p_i\cdot \qb_i)\; 
        e^{-i b\cdot \qb_1} 
    \del^4(\qb_1+\qb_2+\kb)
\\&\hspace*{45mm}\times
       \AmplB(p_1,p_2\rightarrow 
             p_1+\hbar \qb_1,p_2+\hbar \qb_2,\hbar \kb^{\hellabel})
\Rexp\,.
\hspace*{-9mm}
\end{aligned}
\label{FatLOcl}
\end{equation}
\def\wb{\bar w}
We have also extracted powers of $\hbar$ from the coupling,
and dropped the $\hbar$-suppressed terms inside the
on-shell delta functions as well as the positive-energy
theta functions.  We recognize the inner integral in the
second term as the radiation kernel defined in eq.~(4.42)
of ref.~\cite{KMOC} (after changing variables there
$p_i\rightarrow p_i-\hbar \wb_i$ and $\wb_i\rightarrow -\qb_i$),
\begin{equation}
\begin{aligned}
\RadKerCl(\kb^\hellabel;b) 
& \equiv \hbar^2 
   \prod_{i=1,2}\int \dd^4 \qb_i\;\del(p_i\cdot \qb_i)\; 
        e^{-i b\cdot \qb_1} 
    \del^4(\qb_1+\qb_2+\kb)
\\&\hspace*{45mm}\times
       \AmplB(p_1,p_2\rightarrow 
             p_1+\hbar \qb_1,p_2+\hbar \qb_2,\hbar \kb^{\hellabel})
\,.
\end{aligned}
\label{eq:defOfRLO}
\end{equation}
We have made the impact parameter an explicit argument here.
At LO, we can then write,
\begin{equation}
\begin{aligned}
\langle F^{\mu\nu}(x)\rangle_{1,\class} &= 
  g^3 \Lexp \Re\sum_{\hellabel}\int \df(\kb)
  \kb^{[\mu\vsuperp}\pol^{(\hellabel)\nu]*} e^{-i\kb\cdot x}
      \RadKerCl(\kb^{\hellabel};b)
\Rexp\,.
\hspace*{-9mm}
\end{aligned}
\label{FatLOcl2}
\end{equation}
The integrand has the form of the radiation observables
introduced in \sect{SpectralFunctionSection}.
The spectral waveform is then,
\begin{equation}
\begin{aligned}
f_{\mu\nu}(\omega,\bhn) = 
 -\frac{i g^3}{8 \pi}\sum_{\hellabel} \bigl[&\Theta(\omega)    
  \kb^{[\mu\vsuperp}\pol^{(\hellabel)\nu]*}
     \RadKerCl(\kb^{\hellabel};b)\big|_{\kb=\omega(1,\bhn)}
\\ &-\Theta(-\omega)    
  \kb^{[\mu\vsuperp}\pol^{(\hellabel)\nu]}
     \RadKerCl{}^*(\kb^{\hellabel};b)\big|_{\kb=-\omega(1,\bhn)}     
     \bigr]
\end{aligned}
\label{SpectralFunctionResult}
\end{equation}
The corresponding result for the Fourier-space NP scalar is,
\begin{equation}
\tilde\npPhi_2^0(\omega,\bhn) = -\frac{i g^3 \omega}{16\pi}
      \Lexp \Theta(\omega)\RadKerCl(\omega(1,\bhn)^{-};b)
      +\Theta(-\omega)\RadKerCl{}^*(-\omega(1,\bhn)^{+};b)\Rexp\,.
\end{equation}
Equivalently, we may write,
\begin{equation}
\begin{aligned}
\npPhi_2^0(t,\bhn) =
  -\frac{i g^3}{16\pi} \Lexp \int \hat{d}\omega \, \Theta(\omega) 
  \;\omega \bigl[
  &\,e^{-i\omega\cdot t}
      \RadKerCl(\omega(1,\bhn)^{-};b)
\\&
  -\,e^{+i\omega\cdot t}
      \RadKerCl{}^*(\omega(1,\bhn)^{+};b)\bigr]
\Rexp\,.
\hspace*{-9mm}
\end{aligned}
\end{equation}
As the LO radiation kernel $\RadKerCl$ 
is given by a five-point amplitude,
the waveform as a function of
frequency $\omega$, is simply the five-point amplitude
up to the additional factor of $\omega$.

The explicit form of \eqn{eq:defOfRLO} for electromagnetic
scattering is given in eq.~(5.46) of ref.~\cite{KMOC},
and reproduced as \eqn{RadiationKernel}.  We evaluate it
in appendix~\ref{IntegralsAppendix}, to obtain,
\begin{equation}
\begin{aligned}
\RadKerCl(\kb;b) &= \frac{Q_1^2 Q_2}{m_1\, u_1\cdot \kb} 
                   \bigl[ u_2\cdot\kb\, u_1\cdot\pol - u_1\cdot\kb\,u_2\cdot\pol\bigr]\,
                   I_3
\\ &\phantom{=\,}
     - \frac{Q_1^2 Q_2\gamma}{m_1\, u_1\cdot \kb (\gamma^2-1)} 
                   \bigl[ u_1\cdot\kb\, (u_1-\gamma u_2)\cdot\pol 
                        - (u_1-\gamma u_2)\cdot\kb\,u_1\cdot\pol\bigr]\,
                   I_{3}
\\ &\phantom{=\,}                    
     +\frac{Q_1^2 Q_2\,\gamma\,e^{i b\cdot \kb}}{m_1\, u_1\cdot \kb} 
            \bigl[u_1\cdot \kb\,\tb\cdot\pol-\tb\cdot\kb\,u_1\cdot\pol\bigr]
\\ &\phantom{=\,+\,}\times 
         \frac{i}{2\pi\,({\gamma^2-1})} 
            K_1\bigl(\sqrt{-b^2}\,u_1\cdot \kb/\sqrt{\gamma^2-1}\bigr)
\\ &\phantom{=\,}+\bigl(1\leftrightarrow 2\textrm{\ modulo phases}\bigr)
\\ &= \frac{Q_1^2 Q_2\,e^{i b\cdot \kb}}{m_1\, u_1\cdot \kb} 
                   \bigl[ u_2\cdot\kb\, u_1\cdot\pol - u_1\cdot\kb\,u_2\cdot\pol\bigr]\,
\\ &\phantom{=\,+\,}\times 
                   \frac{1}{2\pi\,\sqrt{\gamma^2-1}} 
  K_0\bigl(\sqrt{-b^2}\,u_1\cdot \kb/\sqrt{\gamma^2-1}\bigr)
\\ &\phantom{=\,}                    
     +\frac{Q_1^2 Q_2\gamma\,e^{i b\cdot \kb}}{m_1\, u_1\cdot \kb} 
            \bigl[u_1\cdot \kb\,\tb\cdot\pol-\tb\cdot\kb\,u_1\cdot\pol\bigr]
\\ &\phantom{=\,+\,}\times 
         \frac{i}{2\pi\,({\gamma^2-1})} 
            K_1\bigl(\sqrt{-b^2}\,u_1\cdot \kb/\sqrt{\gamma^2-1}\bigr)
\\ &\phantom{=\,}+\bigl(1\leftrightarrow 2\textrm{\ modulo phases}\bigr)\,.
\end{aligned}
\end{equation}
A side calculation shows that (with $\reference$ a null reference momentum),
\begin{equation}
\begin{aligned}
u_2\cdot\kb\, u_1\cdot\pol - &u_1\cdot\kb\,u_2\cdot\pol =
\\ &\phantom{=} 
\frac1{\sqrt2 \spa{\reference}.{\kb}}\bigl[
  \nsand{\kb}.{u_2}.{\kb}\nsand{\reference}.{u_1}.{\kb}
  -  \nsand{\kb}.{u_1}.{\kb}\nsand{\reference}.{u_2}.{\kb}\bigr]
\\ &= 
\frac1{\sqrt2}  \nsandbb{\kb}.{u_2\, u_1}.{\kb}
\end{aligned}
\end{equation}
for positive-helicity emission, and
\begin{equation}
\frac1{\sqrt2}  \nsandaa{\kb}.{u_2\, u_1}.{\kb}
\end{equation}
for negative-helicity emission.

Then,
\begin{equation}
\begin{aligned}
\RadKerCl(\kb^{+};b) 
 &= \frac{Q_1^2 Q_2\,e^{i b\cdot \kb}}{2\sqrt2\pi m_1\, u_1\cdot \kb\,\sqrt{\gamma^2-1}} 
\\ &\phantom{=\,}\hspace*{8mm}\times
\biggl\{ \nsandbb{\kb}.{u_2\, u_1}.{\kb}\,
  K_0\bigl(\sqrt{-b^2}\,u_1\cdot \kb/\sqrt{\gamma^2-1}\bigr)
\\ &\phantom{=\,}\hspace*{8mm}\phantom{\times\,\biggl\{}
     +\frac{i\nsandbb{\kb}.{b\, u_1}.{\kb}}{\sqrt{\gamma^2-1}\,\sqrt{-b^2}} 
            K_1\bigl(\sqrt{-b^2}\,u_1\cdot \kb/\sqrt{\gamma^2-1}\bigr)\biggr\}
\\ &\phantom{=\,}+
\frac{Q_1 Q_2^2}{2\sqrt2\pi m_2\, u_2\cdot \kb\,\sqrt{\gamma^2-1}} 
\\ &\phantom{=\,}\hspace*{8mm}\times
\biggl\{ \nsandbb{\kb}.{u_1\, u_2}.{\kb}\,
  K_0\bigl(\sqrt{-b^2}\,u_2\cdot \kb/\sqrt{\gamma^2-1}\bigr)
\\ &\phantom{=\,}\hspace*{8mm}\phantom{\times\,\biggl\{}
     +\frac{i\nsandbb{\kb}.{b\, u_2}.{\kb}}{\sqrt{\gamma^2-1}\,\sqrt{-b^2}} 
            K_1\bigl(\sqrt{-b^2}\,u_2\cdot \kb/\sqrt{\gamma^2-1}\bigr)\biggr\}\,.
\end{aligned}
\end{equation}
There is a similar result for the other photon helicity.

Using the integrals,
\begin{equation}
\begin{aligned}
\int_0^\infty &d\omega\;\omega e^{-i\omega(t+a_0)}
             K_0(\omega a_1) =
\\& \frac1{a_1^2 + (a_0 + t)^2} - 
\frac{(t+a_0)}{[a_1^2 + (a_0 + t)^2]^{3/2}}
\arcsinh\Bigl(\frac1{a_1}(t+a_0)\Bigr)
\\&-\frac{i\pi}2 \frac{(t+a_0)}{[a_1^2 + (a_0 + t)^2]^{3/2}}
\,,
\\
\int_0^\infty &d\omega\;\omega e^{-i\omega(t+a_0)}
             K_1(\omega a_1) =
\\& \frac{\pi a_1}{2[a_1^2 + (a_0 + t)^2]^{3/2}}
-i\frac{(a_0+t)}{a_1 [a_1^2 + (a_0 + t)^2]}
\\&-i \frac{a_1}{[a_1^2 + (a_0 + t)^2]^{3/2}}
\arcsinh\Bigl(\frac1{a_1}(t+a_0)\Bigr)\,;
\end{aligned}    
\end{equation}
and defining,
\def\adenom{\rho}
\def\bn{{\bf b}\cdot \bhn}
\def\nh{{\hat n}}
\begin{equation}
\begin{aligned}
u_{i,\bhn} &\equiv u_i\cdot \kb/\omega = u_i\cdot(1,\bhn)\,,
\\ \adenom_1(t) &\equiv -b^2 u_{1,\bhn}^2+(\gamma^2-1)(t+\bn)^2\,,
\\ \adenom_2(t) &\equiv -b^2 u_{2,\bhn}^2+(\gamma^2-1)t^2\,,
\end{aligned}    
\end{equation}
along with,
\def\Util{\Xi}
\def\bv{{\bf v}}
\begin{equation}
\begin{aligned}
\Util_{ia}^{\zeta}(t,\bhn;\bv) &=
\frac{\sqrt{\gamma^2-1}}
      {\adenom_1(t)} 
      -\zeta\frac{(\gamma^2-1)(t+\bv\cdot\bhn)}
            {\adenom_1^{3/2}(t)}
             \arcsinh\biggl(
                \frac{\sqrt{\gamma^2-1}}{\sqrt{-b^2} u_{1,\bhn}}(t+\bv\cdot\bhn)\biggr)
\\&\phantom{=\,}
-\frac{i\pi}2 \frac{(\gamma^2-1)(t+\bv\cdot\bhn)}
        {\adenom_1^{3/2}(t)}
\,,
\\
\Util_{ib}(t,\bhn;\bv) &=
         \frac{\pi u_{1,\bhn}}
                {\adenom_1^{3/2}(t)}
                +i\frac{\sqrt{\gamma^2-1} (t+\bv\cdot\bhn)}
                    {b^2 u_{1,\bhn}
                     \adenom_1(t)}
\\&\phantom{=\,}
            -i\frac{u_{1,\bhn}}
           {\adenom_1^{3/2}(t)}
           \arcsinh\biggl(\frac{\sqrt{\gamma^2-1}}
                              {\sqrt{-b^2} u_{1,\bhn}}
                              (t+\bv\cdot\bhn)\biggr)
\,,
\end{aligned}
\end{equation}
we can write,
\begin{equation}
\begin{aligned}
\npPhi_2^0&(t,\bhn) =
\\ &-\frac{i g^3 Q_1^2 Q_2}
  {(4 \pi)^3\sqrt2 m_1\, u_{1,\bhn}} 
\Bigl[ \nsandaa{\nh}.{u_2\, u_1}.{\nh} 
             \,\Util_{1a}^+(t,\bhn;{\bf b})
             -\nsandbb{\nh}.{u_2\, u_1}.{\nh}
             \,\Util_{1a}^-(t,\bhn;{\bf b})
\\&
   \hspace*{30mm}
+i \bigl(\nsandaa{\nh}.{b\, u_1}.{\nh} 
                -{\nsandbb{\nh}.{b\, u_1}.{\nh}}\bigr)
\,\Util_{1b}(t,\bhn;{\bf b})\Bigr]
\\&
-\frac{i g^3 Q_1 Q_2^2}
 {(4 \pi)^3\sqrt2 m_2\, u_{2,\bhn}} 
\Bigl[ \nsandaa{\nh}.{u_1\, u_2}.{\nh} 
             \,\Util_{2a}^+(t,\bhn;{\bf 0})
             -\nsandbb{\nh}.{u_1\, u_2}.{\nh}
               \,\Util_{2a}^-(t,\bhn;{\bf 0})
\\&
   \hspace*{30mm}
+i \bigl(\nsandaa{\nh}.{b\, u_2}.{\nh} 
         -{\nsandbb{\nh}.{b\, u_2}.{\nh}}\bigr)
\,\Util_{2b}(t,\bhn;{\bf 0})
           \Bigr]
\,.
\end{aligned}    
\end{equation}
Here, $\keta{\hat n}$ and $\ketb{\hat n}$ are spinors
built out of the null vector $(1,\bhn)$.

\section{Connection to Radiated Momentum}
\label{RadiatedMomentumSection}

In \sect{SpectralFunctionSection}, we presented
the general form for the waveform observable.  We worked
out the leading-order form in two-particle scattering
in \sect{EmissionSection}, and computed the explicit
form for electromagnetic scattering in the previous
section.  The appearance of the radiation kernel suggests
a connection to the radiated momentum previously
computed in ref.~\cite{KMOC}.  Let us elucidate that connection 
in this section.

In eq.~(3.33) of ref.~\cite{KMOC}, we find an expression for time-averaged radiated
momentum,
\begin{equation}
\Rad^\mu \equiv \langle k^\mu \rangle = {}_\textrm{in}{\langle} \psi | \, S^\dagger \operator K^\mu S \, | \psi \rangle_\textrm{in}
= {}_\textrm{in}\langle \psi | \, T^\dagger \operator K^\mu T \, | \psi \rangle_\textrm{in}\,.
\end{equation}
This quantity is also integrated over the entire celestial sphere; we need a more differential
observable.  Furthermore, this expression is related to the energy
emitted, rather than the amplitude of the emitted wave.  

\def\pow{{\textrm{pow}}}
\def\fpow{f_\epsilon}
\def\fpowcl{f_{\epsilon,\class}}
We can use Mellin transforms to extract a more restricted observable, passing
through the spectral waveform to relate the emitted power to the amplitude.
Write the expectation of the
observable $\langle (k^\timecomponentlabel)^{z-1}\rangle$,
\begin{equation}
\Rad(z) \equiv \langle (k^\timecomponentlabel)^{z-1} \rangle = 
{}_\textrm{in}\langle \psi | \, T^\dagger (\operator K^\timecomponentlabel)^{z-1} T 
\, |\psi \rangle_\textrm{in}\,.
\end{equation}
The inverse Mellin transform is related to the 
unpolarized energy density function,
\begin{equation}
\fpow(E) =  -i E\int_{c-i\infty}^{c+i\infty} dz\; E^{-z} \Rad(z)\,,
\end{equation}
where the integral is taken along a line parallel to the imaginary axis,
with $c\in (0,1)$ (or a
deformation of that contour that doesn't cross any poles or branch points)\footnote{With our conventions, the expected power of $(2\pi)^{-1}$ is in the forward rather than the inverse Mellin 
transform.}.  The total energy is given by the integral,
\begin{equation}
E_\textrm{tot} = \int_0^\infty dE\,\fpow(E)\,.
\label{TotalEnergy}
\end{equation}

\def\QRadKer{\mathcal{\hat R}}
Using the form in eq.~(3.38) of ref.~\cite{KMOC}, we can write,
\begin{equation}
\Rad(z) = \sum_X \int
\df(k)\df(\finalk_1)\df(\finalk_2)\;
  (k_X^\timecomponentlabel)^{z-1} \,
  \sum_\hellabel|\QRadKer(k^\hellabel,\finalk_X)|^2\,,
\label{RadiationMellin}
\end{equation}
for the expression in the quantum theory.  In this equation, $\QRadKer$
represents the quantum radiation kernel, given by the integral over
wavefunctions inside the absolute square in eq.~(3.38).  The quantum radiation
kernel is expressed directly in terms of a scattering amplitude.

In the classical limit, the density function is more naturally
a function of frequency rather than of energy,
\begin{equation}
\fpowcl(\omega) =  -i\, \omega \int_{c-i\infty}^{c+i\infty} dz\; 
  \omega^{-z} \Rad_{\class}(z)\,,
\end{equation}
so that $R_{\class}(z) = \hbar^{-z-1} R(z)$.
We can use eqs.~(4.40--4.41) of ref.~\cite{KMOC} to write,
\begin{equation}
\Rad_{\class}(z) = \sum_X 
\hbar^{-z-1}\Lexp \int \! \df(k) \, (k_X^\timecomponentlabel)^{z-1}
\sum_\hellabel\left |\RadKer(k^\hellabel, \finalk_X) \right|^2 \Rexp\,.
\label{RadiationMellinClassical}
\end{equation}
The radiation kernel here is expressed in terms of 
the appropriate limit of
a quantum scattering amplitude.

We next need to restrict the measured radiation from the entire celestial sphere to 
a narrow cone in a given direction.  We take the limit of the cone, and measure
only the radiation in a given direction from the scattering event.  We implicitly
assume that the measurement distance is much larger than the impact parameter, so
that there is a unique and well-defined direction.
It's not clear exactly what a formal expression for the operator would be,
but what we want is,
\begin{equation}
\operator {K}^\mu \,\delta^{(2)}(\hat{\operator K}-\bhn)\,,
\end{equation}
for radiation in the $\bf\hat n$ direction.    This operator is to be understood as inserting,
\begin{equation}
\sum_{i\in \textrm{messengers}} k_i^\mu \,\delta^{(2)}({\bf\hat k}_i-\bhn)\,,
\label{EffectiveInsertion}
\end{equation}
into a sum over states or equivalently the phase-space integral. 
Focusing on the energy component, this can be understood as a light ray 
operator~\cite{Ore:1979ry,Sveshnikov:1995vi,Korchemsky:1997sy,Korchemsky:1999kt,Gonzo:2020xza} given by,
\begin{equation}
\Eop(\hat{\mathbf{n}})=\int_{-\infty}^{+\infty} d u \lim _{r \rightarrow \infty} r^{2} \Top_{u u}\left(u, r, \mathbf{\hat{n}}\right)
\end{equation}
where $u$ denotes the light-cone time $u=t-r$ and 
$\Top_{u u}\left(u, r, \mathbf{\hat{n}}\right)$ is the 
(light-cone) time-time component of the stress-energy tensor (in 
gravity, this will be replaced by the Bondi news squared operator~\cite{Gonzo:2020xza}).
By applying the saddle point approximation for the fields in the energy momentum tensor, the plane wave expansion will localize to the point on the sphere in the direction of propagation. Schematically we will have (see refs.~\cite{He:2014laa,Strominger:2017zoo} for further details)
\begin{equation}
e^{i {x \cdot k}/{\hbar}}=e^{i \omega u+i \omega r\left(1-\bf \hat{n} \cdot \bf \hat{k} \right)} \stackrel{r \rightarrow \infty}{\sim} \frac{1}{i \omega r} e^{i \omega u} \delta^{(2)}\left(\bf \hat{n}-\bf \hat{k}\right)
\end{equation}
where $\omega = \bar{k}^t$. Then one finds,
\begin{equation}
\Eop(\hat{\mathbf{n}})=
\sum_{\hellabel}\int \df(k) \, k^{t}\:  
\delta^{(2)}\left(\bf \hat{n}-\bf \hat{k}\right) \left[\creationh{\hellabel}(k)\annihilationh{\hellabel}(k)\right]
\end{equation}
where the action on on-shell particle states is equivalent to the time 
component of \eqn{EffectiveInsertion}. 
\def\twotextlines#1#2{\genfrac{}{}{0pt}{}{\textrm{#1}}{\textrm{#2}}}
The analogous Mellin kernel for 
$(\operator K^\timecomponentlabel)^{z-1}$ is
presumably,
\begin{equation}
(\operator {K}^\timecomponentlabel)^{z-1}
\,\delta^{(2)}(\hat{\operator K}-\bhn)\,,
\end{equation}
which is to be understood as inserting,
\begin{equation}
\sum_{i\in \twotextlines{distinct}{messengers}} 
\hspace*{2mm}\Bigl(\hspace*{-4mm}\sum_{\genfrac{}{}{0pt}{}{j\parallel i}{j\in\textrm{messengers}}}
  \hspace*{-4mm} k_j^\timecomponentlabel\Bigr)%
  \mathop{\vphantom{a}}^{z-1}
 \,\delta^{(2)}({\bf\hat k}_i-\bhn)\,,
\label{RadiationMellinInsertion}
\end{equation}
into a sum over states or the phase-space integral.  The sum 
over distinct messengers
is a sum over messengers which are not collinear; the sum over 
the collinear messengers
is taken in the inner sum.  The inner sum includes $i$ itself.

This form is motivated by a subtlety about overlapping directions: 
if ${\bf \hat k}_j = {\bf\hat k}_l$
with the remaining directions distinct
we want,
\begin{equation}
\sum_{\genfrac{}{}{0pt}{}{i\in\textrm{messengers}}{i\neq j,l}}
(k_i^\timecomponentlabel)^{z-1} \,\delta^{(2)}({\bf\hat k}_i-\bhn)
+(k^\timecomponentlabel_j+k^\timecomponentlabel_l)^{z-1}
\,\delta^{(2)}({\bf\hat k}_j-\bhn)\,,
\end{equation}
which is what \eqn{RadiationMellinInsertion} is designed to give.  At leading order
this subtlety is irrelevant.

The analog to \eqn{RadiationMellin} is,
\begin{equation}
\begin{aligned}
\Rad(z,\bhn) &= 
\hspace*{-4mm}\sum_{i\in \twotextlines{distinct}{messengers}} 
\sum_X \int
\df(k_i)\df(\finalk_1)\df(\finalk_2)\;
\hspace*{2mm}\Bigl(\hspace*{-4mm}\sum_{\genfrac{}{}{0pt}{}{j\parallel i}{j\in\textrm{messengers}}}
  \hspace*{-3mm} k_j^\timecomponentlabel\Bigr)%
   \mathop{\vphantom{a}}^{z-1}
\\&\hphantom{=}\,\hspace*{40mm}\times
 \,\delta^{(2)}({\bf\hat k}_i-\bhn)\sum_\hellabel
  \,|\QRadKer(k_i^\hellabel,\finalk_X)|^2\,,
\end{aligned}
\label{RadiationMellinDirected}
\end{equation}
and to \eqn{RadiationMellinClassical},
\begin{equation}
\Rad_{\class}(z,\bhn) = 
\hspace*{-3mm}\sum_{i\in \twotextlines{distinct}{messengers}} 
\hspace*{-2mm}
\hbar^{-z-1}\Lexp \int \! \df(k_i) \, 
\Bigl(\hspace*{-4mm}\sum_{\genfrac{}{}{0pt}{}{j\parallel i}{j\in\textrm{messengers}}}
  \hspace*{-4mm} k_j^\timecomponentlabel\Bigr)%
   \mathop{\vphantom{a}}^{z-1}
 \,\delta^{(2)}({\bf\hat k}_i-\bhn)
\,\sum_\hellabel\left |\RadKer(k_i^\hellabel, \finalk_X) \right|^2 \Rexp\,.
\label{RadiationMellinDirectedClassical}
\end{equation}

\def\lo{(0)}
At LO, \eqn{RadiationMellinDirectedClassical} simplifies to just,
\begin{equation}
\Rad_{\class}^{\lo}(z,\bhn) = g^{6} 
\Lexp \int \! \df(\kb) \, (\kb^\timecomponentlabel)^{z-1}\,\delta^{(2)}({\bf\hat k}-\bhn)\,
\sum_\hellabel\left |\RadKer^{\lo}(\kb^\hellabel;b) \right|^2 \Rexp\,.
\label{RadiationMellinDirectedClassicalLO}
\end{equation}
The corresponding result for the spectral density
in the $\bhn$ direction is,
\begin{equation}
\begin{aligned}
\fpowcl(\omega,\bhn) &= 
 g^{6} \omega
\Lexp \int \! \df(\kb) \,
 \frac{\del(\ln \kb^\timecomponentlabel-\ln \omega)}{\kb^\timecomponentlabel} 
 \,\delta^{(2)}({\bf\hat k}-\bhn)
  \,\sum_\hellabel\left |\RadKer^{\lo}(\kb^\hellabel;b) \right|^2
  \Rexp\,.
\end{aligned}
\label{SpectralLO1}
\end{equation}
Writing out,
\begin{equation}
\begin{aligned}
\df(\kb) &= \frac{d^3{\bf \kb}}{2 (2\pi)^3\,|\bf \kb|} 
\\ &= \frac{|{\bf \kb}| d|{\bf \kb}|\,d\Omega_{\bf \kb}}{2 (2\pi)^3}\,,
\end{aligned}
\end{equation}
we can perform the integrals in \eqn{SpectralLO1} to obtain,
\begin{equation}
\begin{aligned}
\fpowcl(\omega,\bhn) &= 
 \frac{g^{6} \omega^2}{8 \pi^2}\,\sum_{\hellabel}
\Lexp \, \left |\RadKer^{\lo}(\omega (1,\bhn)^{\eta};b) \right|^2\,
 \Rexp\,.
\end{aligned}
\label{SpectralLO2}
\end{equation}

We can now compare this with the amplitude of each component of the waveform, expanded at the leading order order in the coupling: for $|f_{\mu\nu} M^{*\mu} N^\nu|$ and $|f_{\mu\nu} M^{\mu} N^\nu|$  we have, respectively
\begin{equation}
\begin{aligned}
|f_{\mu\nu}(\omega(1,\bhn)) M^{*\mu} N^\nu| &= \frac{\omega}{16 \pi}  g^{3} \left | \Lexp \, \RadKerCl(\omega(1,\bhn)^{-};b) \Rexp\, \right| \\
|f_{\mu\nu}(\omega(1,\bhn)) M^{\mu} N^\nu| &= \frac{\omega}{16 \pi} g^{3}\left | \Lexp \, \RadKerCl(\omega(1,\bhn)^{+};b) \Rexp\, \right| 
\end{aligned}
\label{SpectralLOmodulus}
\end{equation}
At LO, we can also write 
\begin{equation}
\begin{aligned}
\Lexp \, \left |\RadKer^{\lo}(\omega (1,\bhn)^{\eta};b) \right|^2 \Rexp\, = \left |  \Lexp \, \RadKer^{\lo}(\omega (1,\bhn)^{\eta};b) \Rexp\,  \right|^2 
\end{aligned}
\end{equation}
and therefore we can express the spectral density of
emission from \eqn{SpectralLO2} in terms of the amplitudes of the two 
helicity components of the waveform,
\begin{equation}
\begin{aligned}
\fpowcl(\omega,\bhn) &= 32 \big[ |f_{\mu\nu}(\omega(1,\bhn)) M^{*\mu} N^\nu|^2 + |f_{\mu\nu}(\omega(1,\bhn)) M^{\mu} N^\nu|^2 \big]
\,.
\end{aligned}
\label{PowerAmplitudeLO}
\end{equation}
This relation is the avatar of the relation between the energy of the wave and the squared amplitude of the wave, the only difference being that here we are measuring the momentum emitted in a given direction at a large distance
$r$ from the source.
The emitted radiation observable provides information about the 
magnitude of the observed
messenger wave, but not about its phase. The direct derivation in 
previous sections adds that information. 

A recently proposed generalization of a standard event shape
is sensitive to amplitude phases~\cite{Korchemsky:2021okt}.
It would be interesting to explore a possible connection to the 
waveform.

\section{Conclusions}
\label{ConclusionsSection}

In this paper, we have developed an observables-based
formalism for computing classical waves from quantum
scattering amplitudes.  We have shown how to incorporate
both outgoing and incoming narrowly sampled waves, via 
the ``local'' observables needed for the former, and
scattering of waves needed for the latter.

Waveforms measured at gravitational wave observatories are ``local'' measurements, in the sense that 
the passing gravitational wave train is sampled only at the (small) spatial location of the 
observatory relative to the (very large) spatial extent of the gravitational wave. In this paper, our first major
focus was on developing a quantum-field theoretic formalism to describe this kind of classical, local 
measurement.  This is in contrast to previous
work~\cite{KMOC,Maybee:2019jus} on classical observables in quantum field theory, which discussed 
``global'' observables, such as the total
amount of energy-momentum radiated in a scattering event. 
Our formalism is very general, 
though in our explicit discussions
we focused on the case of electromagnetic radiation, which has the pedagogical benefit of being 
slightly easier to work with. We look forward
to applications of our formalism in gravity.

Scattering amplitudes are remarkably simple objects which can be computed efficiently. For this 
reason, it seems very promising that waveforms
can be computed so directly in terms of amplitudes. In particular, it is clear from our work that 
there is no obstacle to using the double copy
to compute gravitational waveforms (sourced by a scattering event) to any order of perturbation 
theory. It may be worth emphasizing that we do 
not need the BCJ formulation~\cite{Doublecopy1} of the double copy at loop level, which remains 
conjectural, to perform such a computation. The 
gravitational waveform at higher orders could be computed using the unitarity method, 
with only tree-level gravitational amplitudes required as inputs.  
For those amplitudes, the BCJ relations are proven.
The insensitivity of the classical waveform 
to delta-function contributions localized on the worldlines of the particles 
offers another, potentially significant simplification: only a subset of all possible 
quantum factorization channels needs to be computed. 
The possibility of computing leading-order gravitational
radiation using amplitudes and the double copy was previously 
discussed by Luna \textit{et al.}~\cite{Donal8}, 
building on a leading-order worldline
treatment by Goldberger and Ridgway~\cite{Doublecopy2}.
The formalism presented here makes this computation possible to any order. 
Shen~\cite{Doublecopy3} has already
computed the next-to-leading order waveform;
it will be interesting to compare the efficiency of 
our methods, using the conventional double
copy of amplitudes, with Shen's ingenious world-line implementation of the double copy.

One important theme in the calculation and exploration 
of scattering amplitudes is the search for the simplest forms in
which to cast them.
An early realization came through the focus
on helicity amplitudes rather than covariant forms (in terms of polarization vectors
and momenta).  The former contain all physical information, and are simpler.  This
is especially true when they are expressed in terms of spinorial variables.
The translation comes through a spinor-helicity formalism; historically, that
of Xu, Zhang, and Chang~\cite{XZC} played an important role.

Remarkably, the same phenomenon occurs in classical field theory. 
Newman--Penrose (NP) scalars~\cite{Newman:1961qr} are classical
analogs of 
helicity amplitudes; indeed, as we have seen, the NP scalar $\Phi_2$ is an integral 
over a helicity amplitude. The NP scalars
can be defined by contracting tensorial quantities, 
such as the electromagnetic field strength 
$F^{\mu\nu}$, with a basis of four null
vectors. This basis is a direct analog of the momentum of a particle, 
along with its two possible polarization 
vectors, and a gauge choice. 
Alternatively,
the NP scalars can be constructed directly by passing from the tensorial field strength to its 
spinorial equivalent. In this formulation, a natural
basis of spinors occurs classically, in an exact analogue of the spinor-helicity method in scattering 
amplitudes. It seems likely that further study
will reveal more close connections between sophisticated approaches to classical physics and the 
methods of scattering amplitudes.

As a concrete application of our formalism, we computed a simple
waveform: the electromagnetic radiation emitted as two charges
scatter. We extracted the asymptotic spectral functions, as
well as the relevant asymptotic Newman-Penrose scalar.
At leading order, these quantities are closely related to five-point amplitudes.
In the Fourier domain, they are built out of modified Bessel functions.
At higher orders, the connection to five-point amplitudes will persist.
We expect that an interesting class of functions, generalizing Bessel functions,
will appear. In the time domain, the functions were simpler; we suspect that
this may be an accident of low orders.
 
Our second major focus in this paper has been developing
a quantum field-theoretic description of 
massless classical waves readily amenable 
to calculations using scattering amplitudes. 
Coherent states are key tools in extracting classical 
behavior from quantum field
theory~\cite{Yaffe:1981vf}, so it is
no surprise that we found them to be very helpful. Indeed, they 
mesh very naturally with 
amplitudes, and especially with the transition operator
$T$ whose matrix elements are the amplitudes. The reason is that 
the $T$ matrix can be written out in 
terms of amplitudes and of creation and
annihilation operators. These operators, in turn, act very simply 
on coherent states. 

As an application of massless waves, we studied the scattering of a massless electromagnetic wave off 
a classical charge. We showed that
the resulting outgoing wave is determined, at leading order, by the classical limit of the Compton 
four-point amplitude. We expect this final state 
to also be coherent. In appendix~\ref{app:opFactorisation} we provide
evidence in favor of the coherence of this radiation.

Throughout our paper, the focus has been on scattering events. These are very naturally described using 
amplitudes. Scattering events in 
general relativity are interesting in themselves
given the possibility that the tightly-bound compact binaries 
observed by the LIGO and Virgo 
collaborations are created after a scattering event with a third object~\cite{LIGO3}. Of course,
a major goal for the future will be to 
understand how 
gravitational waveforms from classically bound objects can also be computed using amplitudes. This will 
need a new understanding,
perhaps building on the work~\cite{Kalin:2019rwq,Kalin:2019inp} of K\"alin and Porto in the context of 
conservative classical dynamics. Yet 
even without  such a direct connection, it seems clear that our work can be used in the context of 
bound state physics by developing an 
effective action to enable the transfer of know-how from
unbound to bound cases. 
The reader may also be interested in forthcoming work by Bautista, Guevara,
Kavanagh and Vines~\cite{BGKV} on 
related subjects.

The future for gravitational wave physics is data-rich and high precision. We will need every good 
idea we can find to calculate waveform
templates at the necessary precision. By now it is clear that amplitudes and the double copy will be a 
useful tool. The double copy, at least
in its BCJ form, was a theoretical discovery which was a
by-product of the drive for precision theory for LHC physics. 
New theoretical discoveries may well
await us as we develop our understanding of 
gravitational amplitudes in the drive for precision
gravitational-wave physics.

\acknowledgments
We thank Enrico Herrmann, Ben Maybee, Alex Ochirov,
Alasdair Ross, and Matteo Sergola
for helpful discussions and comments,
and Alfredo Guevara for sharing a preliminary draft of
ref.~\cite{BGKV}. This project has received funding from the European Union’s Horizon 2020 research and innovation program under the Marie Skłodowska-Curie grant agreement No. 764850 “SAGEX”.
DAK's work was supported in part by the French 
\textit{Agence 
Nationale pour la
 Recherche\/}, under grant ANR--17--CE31--0001--01,
 and in part by the European Research Council, under
 grant ERC--AdG--885414. 
 DOC is supported in part by the STFC consolidated grant `Particle Physics at the Higgs Centre'. Several of our figures were produced with the help of the Inkscape~\cite{Inkscape}.

\appendix

\section{Beam Spreading}
\label{BeamSpreadAppendix}

Let us obtain a more refined picture of the time
dependence of the classical wave in \eqn{WavesFieldStepI}.
Expand the square root in the exponent in that expression,
keeping the next-to-leading term in the expansion,
\begin{equation}
\sqrt{\omega^2 + 
    (\kb^\xcomponentlabel)^2 + (\kb^\ycomponentlabel)^2} = 
\omega + \frac{(\kb^\xcomponentlabel)^2 
       + (\kb^\ycomponentlabel)^2}{2\omega} 
       + \cdots\,.
\label{BroadenedDeltaFunctionExpansionNLO}
\end{equation}
Substituting this into \eqn{WavesFieldStepI}, we obtain,
\begin{equation}
\Acl^\mu(x)=\sqrt{2}\ANorm   \Re \beampolv^{\mu}(\kbbeam) 
e^{-i \omega(t-z)} \mathcal{I}(\omega,\ell_{\perp}) \,,
\label{WavesFieldStepIInlo}
\end{equation}
where we have introduced the following scalar integral
(recall that $\sigma_\perp=\ell_\perp^{-1}$),
\begin{equation}
\mathcal{I}(\omega,\ell_{\perp}) =
\int d^2 \kb_\perp \, 
\deltasm{\sigma_\perp}{\kb^x} \deltasm{\sigma_\perp}{\kb^y} 
e^{i \kb^x \, x}e^{i \kb^y \, y} e^{-it\kb_x^2/(2\omega)} e^{-it\kb_y^2/(2\omega)}
\end{equation}
Integrating over the angular variable, we find,
\begin{equation}
\mathcal{I}(\omega,\ell_{\perp}) =
2\ell^2_{\perp} \int_{0}^{\infty} d k \, k\, 
J_{0}(\sqrt{x^2+y^2}\, k) e^{-k^2 [\ell^2_{\perp}+i t/(2\omega)]}
\,.
\end{equation}
Performing the integral, we obtain,
\begin{equation}
\mathcal{I}(\omega,l_{\perp}) = \frac{e^{-\frac{(x^2+y^2)}{4l^2_{\perp}}\big[1+i\frac{t}{2\omega l^2_{\perp}} \big]^{-1}}}
{1+\frac{it}{2\omega l^2_{\perp}}}\,.
\label{Iomegaperp}
\end{equation}
Yet higher-order contributions may be computed by noticing that the electromagnetic field can be expressed --- without expanding the square root in \eqn{BroadenedDeltaFunctionExpansionNLO} --- as,
\begin{equation}
\Acl^\mu(x)=\sqrt{2}\ANorm   \Re \beampolv^{\mu}(\kbbeam)
\; \; e^{i\omega z-it\hat{H}(\omega)}
\Bigl[e^{-\frac{(x^2+y^2)}{4l^2_{\perp}}}\Bigr]\,,
\end{equation}
where we have introduced the operator $\hat{H}(\omega)=\sqrt{\omega^2-\nabla^{2}_{(x,y)}}$. 
In this reformulation, the problem is now equivalent to computing the time evolution --- for a relativistic Hamiltonian with effective mass $\omega$ --- of a Gaussian wavepacket. Restricting the time evolution to the nonrelativistic limit, we obtain the 
well-known result for the spread of a Gaussian wavepacket in two dimensions, in agreement with \eqn{Iomegaperp}. In a similar way, we can easily generalize the computation by adding contributions from the expansion of the polarization vectors in the integrand as in \eqn{WavesFieldNLO}.

\section{Factorization and Unitarity in the Classical Limit}
\label{app:opFactorisation}

Our framework allows the computation of classical phenomena such as the electromagnetic field generated by the scattering of an incoming beam of light with a massive particle.  In this appendix, we address the question of whether the final state is coherent, in the context of a perturbative calculation. For coherence to hold, we must show that the mean value of the electromagnetic field operator on the final state factorizes. 
The final state is given by the evolution of the initial
state, 
\begin{equation}
\ket{\psi}_{\textrm{out}}= \int d \Phi(p)\: \phi(p)\: e^{i b \cdot {p}/{\hbar}}\: S\ket{p \: \alpha^{+}}_{\textrm{in}} \; . 
\end{equation}
We say that the final state is coherent if the following correlation function vanishes in the classical limit,
\begin{equation}
\label{correlator}
\Delta=  {}_{\textrm{out}}\langle \psi| \mathbb{F}^{\mu \nu}(x) \mathbb{F}^{\alpha \beta}(y)  | \psi \rangle_{\textrm{out}}-{}_{\textrm{out}}\langle \psi| \mathbb{F}^{\mu \nu}(x)  | \psi \rangle_{\textrm{out}}\, {}_{\textrm{out}}\langle \psi| \mathbb{F}^{\alpha \beta}(y)  | \psi \rangle_{\textrm{out}}
\end{equation}
where the electromagnetic field operator is given by
\eqn{eqn:fieldstrengthdn}.
Let us prove that the previous correlation function vanishes at the first nontrivial order in the coupling $g$. The second term in \eqn{correlator} is already known to this order as it matches the value of the electromagnetic field in Thomson scattering times its free counterpart. What is left is to compute is the first term. We can safely disregard contributions quadratic in the transfer matrix,
leaving us to compute the classical limit of,
\begin{equation}
\label{correFsquared}
{}_{\textrm{out}}\langle \psi| \mathbb{F}^{\mu \nu}(x) \mathbb{F}^{\alpha \beta}(y)   | \psi \rangle_{\textrm{out}}=F^{\mu \nu,(0)}(x)F^{\alpha \beta,(0)}(y)+i \: {}_{\textrm{in}}\!\bra{\psi} [\mathbb{F}^{\mu \nu}(x) \mathbb{F}^{\alpha \beta}(y), T ] \ket{\psi}_{\textrm{in}}\,,
\end{equation}
where $F^{(0)}_{\mu \nu}(x)$ denotes the free field. Expanding the electromagnetic field operator in terms of annihilation and creation operators,
\begin{equation}
\begin{aligned}
\mathbb{F}^{\mu \nu}(x) \mathbb{F}^{\alpha \beta}(y)=
\hspace*{25mm}&
\\ 
-\frac{4}{\hbar^3} \sum_{\eta_1,\eta_2}\int \df(k_{1}) \df(k_{2}) \Bigl[& \annihilationh{\eta_1}(k_1) \annihilationh{\eta_2}(k_2) 
k_{1}^{[\mu} \pol^{(\eta_1) \nu] *} \:  k_{2}^{[\alpha} \pol^{(\eta_2) \beta] *} 
e^{-i (k_{1} \cdot x+k_{2} \cdot y)/\hbar}   \: 
\\&\,
+ \creationh{\eta_1}(k_1) \creationh{\eta_2}(k_2)  
   k_{1}^{[\mu} \pol^{(\eta_1) \nu]} \:  k_{2}^{[\alpha} \pol^{(\eta_2) \beta]}
      e^{i (k_{1} \cdot x+k_{2} \cdot y)/\hbar}
\:
\\&\,
- \creationh{\eta_2}(k_2) \annihilationh{\eta_1}(k_1) 
   k_{1}^{[\mu} \pol^{(\eta_1) \nu]*} \:  k_{2}^{[\alpha} \pol^{(\eta_2) \beta]}
      e^{-i (k_{1} \cdot x- k_{2} \cdot y)/\hbar} 
\:
\\&\,
- \creationh{\eta_1}(k_1) \annihilationh{\eta_2}(k_2) 
   k_{1}^{[\mu} \pol^{(\eta_1) \nu]} \:  k_{2}^{[\alpha} \pol^{(\eta_2) \beta]*}
      e^{i (k_{1} \cdot x-k_{2} \cdot y)/\hbar} 
\:
\\&\,
- \delta_{\Phi}(k_1-k_2) k_{1}^{[\mu} \pol^{(\eta_1) \nu]*} \:  k_{2}^{[\alpha} \pol^{(\eta_2) \beta]}
   e^{i (k_{1} \cdot x-k_{2} \cdot y)/\hbar} 
\: \Bigr]\,.
\label{eq:FF_factorization}
\end{aligned}
\end{equation}
At leading order in the coupling, the $T$ matrix reads
\[
T = \sum_{\hellabel, \hellabel'} \int \df(\tilde k', \tilde k, \tilde p', \tilde p) \; \langle \tilde k'^{\hellabel'} \tilde p' | T | \tilde k^{\hellabel} \tilde p \rangle \;
\creationh{\hellabel'}(\tilde k') \creation{}(\tilde p')\, \annihilationh{\hellabel}(\tilde k) \annihilation{}(\tilde p) + \cdots \,,
\] 
We can now evaluate the correlation function. The first two terms inside the bracket in \eqn{eq:FF_factorization} can contribute only at higher order in the coupling, and can be safely neglected in the evaluation of the correlation function. As for the last term in \eqn{eq:FF_factorization}, we can see it is similar to \eqref{eq:towardsClassicalFactorization}, providing a quantum contribution at leading order in the coupling which will disappear in the classical limit. We are left with the following,
\begin{equation}
\begin{aligned} 
[a^{\dagger}_{(\eta_1)}(k_1) a_{(\eta_2)}(k_2), T ] &=\sum_{\hellabel} \int \df(\tilde p,\tilde p',\tilde k) \bra{k_2^{\eta_2} \tilde p'} T\ket{\tilde{k}^{\eta} \tilde p} \creationh{\eta_1}(k_1) \creation{}(\tilde p')\, \annihilationh{\eta}(\tilde k) \annihilation{}(\tilde p) \\
&- \sum_{\hellabel'}\int \df(\tilde p,\tilde p',\tilde k') \bra{\tilde{k}'^{\hellabel'} \tilde p'} T\ket{k_1^{\eta_1} \tilde p} \creationh{\hellabel'}(\tilde k') \creation{}(\tilde p')\, \annihilationh{\eta_2}(k_2) \annihilation{}(\tilde p)  \\
[a^{\dagger}_{(\eta_2)}(k_2) a_{(\eta_1)}(k_1), T ] &= \sum_{\hellabel}\int \df(\tilde p,\tilde p',\tilde k) \bra{k_1^{\eta_1} \tilde p'} T\ket{\tilde{k}^{\eta} \tilde p} \creationh{\eta_2}(k_2) \creation{}(\tilde p')\, \annihilationh{\eta}(\tilde k) \annihilation{}(\tilde p) \\
&- \sum_{\hellabel'}\int \df(\tilde p,\tilde p',\tilde k') \bra{\tilde{k}'^{\hellabel'} \tilde p'} T\ket{k_2^{\eta_2} \tilde p} \creationh{\hellabel'}(\tilde k') \creation{}(\tilde p')\, \annihilationh{\eta_1}(k_1) \annihilation{}(\tilde p) \,.
\hspace*{-8mm}
\end{aligned}
\label{intcalc0}
\end{equation}
These results imply that,
\begin{equation}
\begin{aligned} 
{}_{\textrm{out}}&\langle \psi| \mathbb{F}^{\mu \nu}(x) \mathbb{F}^{\alpha \beta}(y)   | \psi \rangle_{\textrm{out}}=
\\ &F^{\mu \nu,(0)}(x)F^{\alpha \beta,(0)}(y) \\
&+\frac{8}{\hbar^3} \Re \sum_{\eta, \eta_1,\eta_2} \int \df(k_1,k_2,\tilde k,p,p') \phi(p) \phi^*(p') \\
&\hspace{5mm}\times
\left[i \bra{k_1^{\eta_1} p'} T\ket{\tilde{k}^{\eta} p} 
\bra{\alpha^+} \creationh{\eta_2}(k_2) \annihilationh{\eta}(\tilde k) \ket{\alpha^+}  
k_{1}^{[\mu} \pol^{(\eta_1) \nu]*} \:  k_{2}^{[\alpha} \pol^{(\eta_2) \beta]}
 e^{-i(k_{1} \cdot x-k_{2} \cdot y)/\hbar}\right] \\
&+\frac{8}{\hbar^3}  \Re \sum_{\eta, \eta_1,\eta_2} \int \df(k_1,k_2,\tilde k,p,p') \phi(p) \phi^*(p')  \\
&\hspace{5mm}\times 
\left[i \bra{k_2^{\eta_2} p'} T\ket{\tilde{k}^{\eta} p} \bra{\alpha^+} \creationh{\eta_1}(k_1)  \annihilationh{\eta}(\tilde k) \ket{\alpha^+}  k_{1}^{[\mu} \pol^{(\eta_1) \nu]} \:  k_{2}^{[\alpha} \pol^{(\eta_2) \beta]*}
 e^{i (k_{1} \cdot x-k_{2} \cdot y)/\hbar}\right] \,.
\end{aligned}
\end{equation}
After some simple algebra, we find
\begin{equation}
\begin{aligned} 
\label{fsquared2}
{}_{\textrm{out}}&\langle \psi| \mathbb{F}^{\mu \nu}(x) \mathbb{F}^{\alpha \beta}(y)   | \psi \rangle_{\textrm{out}}=
\\&F^{\mu \nu,(0)}(x)F^{\alpha \beta,(0)}(y)  \\
&+\frac{8}{\hbar^3} \Re \sum_{\eta_1,\eta_2}\int \df(k_1,\tilde k,p,p') \phi(p) \phi^*(p')  \\
&\hspace{5mm}\times\left[i \bra{k_1^{\eta_1} p'} T\ket{\tilde{k}^{\eta} p} \alpha(\tilde k)  k_{1}^{[\mu} \pol^{(\eta_1) \nu]*} 
e^{-i k_{1} \cdot x/\hbar} \: 
\int \df(k_2) \, \alpha^*(k_2) 
 k_{2}^{[\alpha} \pol^{(\eta_2) \beta]} 
 e^{i k_{2} \cdot y/\hbar}\right]
\\
&+\frac{8}{\hbar^3}  \Re \sum_{\eta_1,\eta_2}\int \df(k_1,k_2,\tilde k,p,p') \phi(p) \phi^*(p')   \\
&\hspace{5mm}\times \left[i \bra{k_2^{\eta_2} p'} T\ket{\tilde{k}^{\eta} p} \alpha(\tilde k)  \: k_{2}^{[\alpha} \pol^{(\eta_2) \beta]*} 
e^{-i k_{2} \cdot y/\hbar} 
\int \df(k_1) \, \alpha^*(k_1) k_{1}^{[\mu} \pol^{(\eta_1) \nu]} \:  e^{i k_{1} \cdot x/\hbar}\right]; 
\end{aligned}
\end{equation}
reorganizing the terms we then obtain, as expected,
\begin{equation}
\begin{aligned}
{}_{\textrm{out}}\langle \psi| &\mathbb{F}^{\mu \nu}(x) \mathbb{F}^{\alpha \beta}(y)   | \psi \rangle_{\textrm{out}} = 
\\&F^{\mu \nu,(0)}(x)F^{\alpha \beta,(0)}(y)
\\&
 +F^{\alpha \beta,(0)}(y) 
 \frac{4}{\hbar^{3/2}} 
 \Re \sum_{\eta_1}\int \df(k_1,\tilde k,p,p') \phi(p) \phi^*(p') 
 \\&\hspace*{50mm}\times
 \left[i \bra{k_1^{\eta_1} p'} T\ket{\tilde{k}^{\eta} p} \alpha(\tilde k)  k_{1}^{[\mu} \pol^{(\eta_1) \nu]*} 
 e^{-i k_{1} \cdot x/\hbar} \right] 
 \\&
+ F^{\mu \nu,(0)}(x) 
\frac{4}{\hbar^{3/2}} \Re \sum_{\eta_2}\int \df(k_2,\tilde k,p,p') \phi(p) \phi^*(p') 
 \\&\hspace*{50mm}\times
\left[i \bra{k_2^{\eta_2} p'} T\ket{\tilde{k}^{\eta} p} \alpha(\tilde k)  k_{2}^{[\mu} \pol^{(\eta_2) \nu]*} e^{-i k_{2} \cdot y/\hbar} \right] 
\,.
\end{aligned}
\label{finalcorr}
\end{equation}
From this result we conclude that,
\begin{equation}
\Delta\big|_{g^2}=0\,.
\end{equation}
This demonstrates that the semiclassical state generated in Thomson
scattering is a coherent state to this nontrivial order 
in the coupling.

\section{Integrals}
\label{IntegralsAppendix}

We require explicit expressions for the integrals appearing 
in the leading-order radiation kernel, eq.~(5.46) of ref.~\cite{KMOC}.
The integral is,
\begin{equation}
\begin{aligned}
\RadKerCl(\kb;b) &= {4} \int \! \dd^4 \xferb_1 \dd^4 \xferb_2 \;
\del(2p_1\cdot\xferb_1) \del(2p_2\cdot\xferb_2) \del^{(4)}(\kb - \xferb_1 - \xferb_2) \,
 e^{i\xferb_1 \cdot b} 
\\&\hphantom{=}\hspace*{10mm}
\times \biggl\{ \frac{Q_1^2Q_2^{\vphantom{2}}}{\xferb_2^2} 
\biggl[-p_2\cdot\pol + \frac{(p_1\cdot p_2)(\xferb_2\cdot\pol)}{p_1\cdot\kb} 
+ \frac{(p_2\cdot\bar{k})(p_1\cdot\pol)}{p_1\cdot\kb} 
\\&\hspace*{35mm} 
 - \frac{(\kb\cdot\xferb_2)(p_1\cdot p_2)(p_1\cdot\pol)}{(p_1\cdot\kb)^2}\biggr] 
\vphantom{\frac{(p_1\cdot p_2)(\xferb_2\cdot\pol_h)}{p_1\cdot\bar{k}}}
+ (1 \leftrightarrow 2)\biggr\} \,.
\label{RadiationKernel}
\end{aligned}
\end{equation}
We replace $p_i^\mu$ by $m_i u_i^\mu$, and introduce a fourth basis vector,
\begin{equation}
v_\mu = 4\epsilon_{\mu\nu\lambda\rho} u_1^\nu u_2^\lambda b^\rho\,.
\end{equation}
Its square is given by,
\begin{equation}
v^2 = -2 G(u_1,u_2,b)\,,
\end{equation}
where $G$ is the Gram determinant
\begin{equation}
G(\{p_i\}) = \det(2 p_i\cdot p_j)\,.
\end{equation}

The only nontrivial Lorentz invariants that can be built out of the $u_i^\mu$,
$b^\mu$, and $v^\mu$ are,
\begin{equation}
\begin{aligned}
\gamma &= u_1\cdot u_2\,, 
\end{aligned}
\end{equation}
and $b^2$, as $u_i^2 = 1$.

We note that,
\begin{equation}
v^2 = 16 b^2 (\gamma^2-1)\,.
\end{equation}
It is convenient to introduce two rescaled four-vectors,
\begin{equation}
\begin{aligned}
\tb^\mu &= b^\mu/\sqrt{-b^2}\,,
\\ \tv^\mu &= v^\mu/\sqrt{-v^2} = v^\mu/(4\sqrt{-b^2(\gamma^2-1)})\,.
\end{aligned}
\end{equation}

\def\zc#1{z^{[#1]}}
Let us also introduce as well new coordinates $\zc{i}_{1,2,b,v}$,
\begin{equation}
\xferb_i^\mu = \zc{i}_1 u_1^\mu + \zc{i}_2 u_2^\mu + \zc{i}_b \tb^\mu + \zc{i}_v \tv^\mu\,.
\label{ChangeVar1}
\end{equation}
The Jacobian from the change of variables in each $\xferb_i$ is,
\begin{equation}
|\epsilon_{\mu\nu\lambda\rho} \tv^\mu u_1^\nu u_2^\lambda \tb^\rho| =  
-\frac{v^2}{4\sqrt{-v^2}\sqrt{-b^2}} = \sqrt{\gamma^2-1}\,.
\end{equation}
We also have the following expression for each square,
\begin{equation}
\xferb_i^2 = (\zc{i}_1){}^2 + 2\gamma \zc{i}_1 \zc{i}_2+(\zc{i}_2){}^2 
-(\zc{i}_b){}^2 -(\zc{i}_v){}^2\,.
\end{equation}

There are four elementary integrals we need to evaluate,
\begin{equation}
\begin{aligned}
I_1 &= \int \! \dd^4 \xferb_1 \dd^4 \xferb_2 \;
\del(u_1\cdot\xferb_1) \del(u_2\cdot\xferb_2) \del^{(4)}(\kb - \xferb_1 - \xferb_2) \,
 \frac{e^{i\xferb_1 \cdot b}}{\xferb_1^2}\,,
\\ I_2^\mu &= \int \! \dd^4 \xferb_1 \dd^4 \xferb_2 \;
\del(u_1\cdot\xferb_1) \del(u_2\cdot\xferb_2) \del^{(4)}(\kb - \xferb_1 - \xferb_2) \,
 \frac{e^{i\xferb_1 \cdot b} \xferb_1^\mu}{\xferb_1^2}\,,
\\ I_3 &= \int \! \dd^4 \xferb_1 \dd^4 \xferb_2 \;
\del(u_1\cdot\xferb_1) \del(u_2\cdot\xferb_2) \del^{(4)}(\kb - \xferb_1 - \xferb_2) \,
 \frac{e^{i\xferb_1 \cdot b}}{\xferb_2^2}\,,
\\ I_4^\mu &= \int \! \dd^4 \xferb_1 \dd^4 \xferb_2 \;
\del(u_1\cdot\xferb_1) \del(u_2\cdot\xferb_2) \del^{(4)}(\kb - \xferb_1 - \xferb_2) \,
 \frac{e^{i\xferb_1 \cdot b} \xferb_2^\mu}{\xferb_2^2}\,.
\end{aligned}
\end{equation}

Start evaluating $I_1$ by using the four-fold delta function to evaluate the $\xferb_2$
integral,
\begin{equation}
\begin{aligned}
I_1 &= \int \! \dd^4 \xferb_1 \;
\del(u_1\cdot\xferb_1) \del(u_2\cdot\xferb_1-u_2\cdot \kb) \,
 \frac{e^{i\xferb_1 \cdot b}}{\xferb_1^2}\,,
\end{aligned}
\end{equation}
and then make the change of variables~(\ref{ChangeVar1}),
\begin{equation}
\begin{aligned}
&\frac{\sqrt{\gamma^2-1}}{(2\pi)^2}
\int \! d\zc1_1 d\zc1_2 d\zc1_b d\zc1_v \;
\delta(\zc1_1 + \gamma \zc1_2) \delta(\gamma\zc1_1+\zc1_2-u_2\cdot \kb) \,
 \\ &\hspace*{30mm}\times
 \frac{e^{-i \zc1_b \sqrt{-b^2}}}
   {(\zc1_1){}^2 + 2\gamma \zc1_1 \zc1_2+(\zc1_2){}^2 
     - (\zc1_b){}^2 -(\zc1_v){}^2}\,.
\end{aligned}
\end{equation}
Use the delta functions to perform the $\zc1_{1,2}$ integrals,
\begin{equation}
\frac{1}{(2\pi)^2\,\sqrt{\gamma^2-1}}
\int \! d\zc1_b d\zc1_v \;
 \frac{e^{-i \zc1_b \sqrt{-b^2}}}
   {-(u_2\cdot \kb)^2/(\gamma^2-1) - (\zc1_b){}^2 -(\zc1_v){}^2}\,.
\end{equation}
Perform the $\zc1_v$ integral to obtain,
\begin{equation}
-\frac{1}{4\pi\,\sqrt{\gamma^2-1}}
\int \! d\zc1_b\;
 \frac{e^{-i \zc1_b \sqrt{-b^2}}}
   {\sqrt{(\zc1_b){}^2+(u_2\cdot \kb)^2/(\gamma^2-1)}}\,.
\end{equation}
This can be evaluated as a Fourier transform,
\begin{equation}
I_1 = -\frac{1}{2\pi\,\sqrt{\gamma^2-1}} 
  K_0\bigl(\sqrt{-b^2}\,u_2\cdot \kb/\sqrt{\gamma^2-1}\bigr)\,,
\end{equation}
where $K_0$ is a modified Bessel function of the second kind.

The first two steps are the same for $I_2^\mu$,
\begin{equation}
\begin{aligned}
I_2^\mu &= \int \! \dd^4 \xferb_1 \;
\del(u_1\cdot\xferb_1) \del(u_2\cdot\xferb_1-u_2\cdot \kb) \,
 \frac{e^{i\xferb_1 \cdot b}\xferb_1^\mu}{\xferb_1^2}
\\ &= \frac{\sqrt{\gamma^2-1}}{(2\pi)^2}
\int \! d\zc1_1 d\zc1_2 d\zc1_b d\zc1_v \;
\delta(\zc1_1 + \gamma \zc1_2) \delta(\gamma\zc1_1+\zc1_2-u_2\cdot \kb) \,
 \\ &\hspace*{30mm}\times
 \frac{e^{-i \zc1_b \sqrt{-b^2}} 
   (\zc1_1 u_1^\mu + \zc1_2 u_2^\mu + \zc1_b \tb^\mu + \zc1_v \tv^\mu)}
   {(\zc1_1){}^2 + 2\gamma \zc1_1 \zc1_2+(\zc1_2){}^2 
     -(\zc1_b){}^2 -(\zc1_v){}^2}\,.
\end{aligned}
\end{equation}
The $\tv^\mu$ term will vanish because of the antisymmetry in $\zc1_v$;
the $u_{1,2}^\mu$ terms will yield a result proportional to $I_1$,
\begin{equation}
\begin{aligned}
I_{2a}^\mu &= \frac{u_2\cdot\kb}{\gamma^2-1} \bigl(\gamma u_1^\mu-u_2^\mu) I_1
\\ &= -\frac{u_2\cdot\kb}{2\pi\,({\gamma^2-1})^{\expfrac32}}
\bigl(\gamma u_1^\mu-u_2^\mu)
  K_0\bigl(\sqrt{-b^2}\,u_2\cdot \kb/\sqrt{\gamma^2-1}\bigr)\,.
\end{aligned}
\end{equation}
The remaining ($\tb^\mu$) term is,
\begin{equation}
\begin{aligned}
I_{2b}^\mu &= 
-\frac{\tb^\mu}{4\pi\,\sqrt{\gamma^2-1}}
\int \! d\zc1_b\;
 \frac{e^{-i \zc1_b \sqrt{-b^2}} \zc1_b}
   {\sqrt{(\zc1_b){}^2+(u_2\cdot \kb)^2/(\gamma^2-1)}}\,.
\\ &=
\frac{i u_2\cdot\kb\,\tb^\mu}{2\pi\,({\gamma^2-1})} 
  K_1\bigl(\sqrt{-b^2}\,u_2\cdot \kb/\sqrt{\gamma^2-1}\bigr)\,,
\end{aligned}
\end{equation}
where we have dropped a delta-function contribution.  The total is,
\begin{equation}
I_2^\mu = I_{2a}^\mu + I_{2b}^\mu\,.
\end{equation}

In $I_3$, start by using the four-fold delta function to integrate out $\xferb_1$,
\begin{equation}
\begin{aligned}
I_3 &= e^{i b\cdot \kb}\int \! \dd^4 \xferb_2 \;
\del(u_1\cdot\xferb_2-u_1\cdot \kb) \del(u_2\cdot\xferb_2)\,
 \frac{e^{i\xferb_2 \cdot b}}{\xferb_2^2}\,.
\end{aligned}
\end{equation}
This is proportional to $I_1$, with the exchange $u_1\leftrightarrow u_2$,
\begin{equation}
I_3 = -\frac{e^{i b\cdot \kb}}{2\pi\,\sqrt{\gamma^2-1}} 
  K_0\bigl(\sqrt{-b^2}\,u_1\cdot \kb/\sqrt{\gamma^2-1}\bigr)\,.
\end{equation}
Similarly for $I_4^\mu$, 
\begin{equation}
I_4^\mu = I_{4a}^\mu + I_{4b}^\mu\,,
\end{equation}
with,
\begin{equation}
\begin{aligned}
I_{4a}^\mu &= -\frac{u_1\cdot\kb}{\gamma^2-1} 
 \bigl(u_1^\mu-\gamma u_2^\mu) I_3
\\ &= \frac{u_1\cdot\kb\,e^{i b\cdot \kb}}{2\pi\,({\gamma^2-1})^{\expfrac32}} 
 \bigl(u_1^\mu-\gamma u_2^\mu)
  K_0\bigl(\sqrt{-b^2}\,u_1\cdot \kb/\sqrt{\gamma^2-1}\bigr)\,,
\\ I_{4b}^\mu &= \frac{i u_1\cdot\kb\,e^{i b\cdot \kb}\,\tb^\mu}{2\pi\,({\gamma^2-1})} 
  K_1\bigl(\sqrt{-b^2}\,u_1\cdot \kb/\sqrt{\gamma^2-1}\bigr)\,.
\end{aligned}
\end{equation}

\nocite{Newton:1982qc}\nocite{Low:1997ut}

\end{document}